\PassOptionsToPackage{prologue,dvipsnames, table}{xcolor}
\documentclass[review]{fcs}

\title{\revise{}{Towards spatial computing: recent advances in multimodal natural interaction for XR headsets}}
\shorttitle{Recent advances in multimodal natural interaction for XR headsets}
\author[1]{Zhimin Wang} \author[1]{Maohang Rao} \author[1]{Shanghua Ye} \author[2]{Weitao Song} \author*[1]{Feng Lu}
\address[1]{State Key Laboratory of VR Technology and Systems, School of CSE, Beihang University, Beijing, China}
\address[2]{School of Optics and Photonics, Beijing Institute of Technology, Beijing, China}
\fcssetup{
  received       = {month dd, yyyy},
  accepted       = {month dd, yyyy},
  corr-email     = {lufeng@buaa.edu.cn},
}
\usepackage{threeparttable}
\usepackage{multirow}
\usepackage{makecell}    
\usepackage{multicol}
\usepackage[table]{xcolor}

\newcommand{\revise}[2]{\textcolor[rgb]{0,0,0}{#2}}

\definecolor{oBlue}{rgb}{0.607, 0.75, 0.851}
\definecolor{oYellow}{rgb}{0.976, 0.858, 0.384}
\definecolor{oRed}{rgb}{0.8, 0.8, 0.8}
\definecolor{oPink}{rgb}{0.964, 0.827, 0.858}

\begin{abstract}
With the widespread adoption of Extended Reality (XR) headsets, spatial computing technologies are gaining increasing attention. Spatial computing enables interaction with virtual elements through natural input methods such as eye tracking, hand gestures, and voice commands, thus placing natural human-computer interaction at its core. While previous surveys have reviewed conventional XR interaction techniques, recent advancements in natural interaction, particularly driven by artificial intelligence (AI) and large language models (LLMs), have introduced new paradigms and technologies. In this paper, we review research on multimodal natural interaction for wearable XR, focusing on papers published between 2022 and 2024 in six top venues: ACM CHI, UIST, \revise{}{IMWUT (Ubicomp)}, IEEE VR, ISMAR, and TVCG.
We classify and analyze these studies based on application scenarios, operation types, and interaction modalities. This analysis provides a structured framework for understanding how researchers are designing advanced natural interaction techniques in XR. Based on these findings, we discuss the challenges in natural interaction techniques and suggest potential directions for future research. This review provides valuable insights for researchers aiming to design natural and efficient interaction systems for XR, ultimately contributing to the advancement of spatial computing.
\end{abstract}

\begin{document}

\section{Introduction}

Extended Reality (XR), which includes Virtual Reality (VR), Augmented Reality (AR), and Mixed Reality (MR), merges the virtual and physical worlds to provide immersive experiences. 
In recent years, XR technologies have developed rapidly. This has led to the widespread adoption of headsets such as the Microsoft HoloLens 2 \cite{HoloLens} and Meta Quest 3 \cite{Meta}, across various fields, 
including industrial maintenance, remote collaboration, online education and entertainment \cite{DBLP:conf/vr/JingLB22, DBLP:conf/vr/QuereMJWW24, Barteit2021, an2023arcosmetics}. These developments highlight XR's potential for diverse applications and its promising market prospects.

In 2024, Apple introduced a new XR headset called the Vision Pro, which reignited public enthusiasm for XR and marked the beginning of a new era in spatial computing \cite{Apple}. 
Spatial computing leverages advanced technologies to perceive and digitize the surrounding physical environment. It seamlessly integrates this environment with computer-generated virtual content, enabling natural interactions between humans and digital systems \cite{spatialcomputing}. 
\revise{The advent of spatial computing is expected to initiate a transformative 'iPhone moment' for XR \cite{hackl2024spatial}.}{The advent of spatial computing is expected to bring a transformative redefinition to XR devices, often referred to as the 'iPhone moment' for XR \cite{hackl2024spatial}.}


\revise{At the core of spatial computing is human computer interaction (HCI)}{The core of spatial computing is human-computer interaction (HCI)}, with a particular focus on developing natural and intuitive interaction techniques \cite{Apple}. Traditional methods, such as keyboards and handheld controllers, are inadequate for delivering the immersive experiences \cite{10108465, DBLP:conf/interact/BerardIBEBC09}. 
\revise{In response, research has increasingly focused on direct human inputs as interaction channels, including eye gaze, hand gestures, and voice commands}{In light of this, research has increasingly focused on direct human input as interaction channels, including eye gaze, hand gestures, and voice commands} \cite{DBLP:conf/vr/MatthewsTIS22, DBLP:conf/vr/GiunchiNGS24, chai2022speech, ding2016survey, huang2024matching, DBLP:conf/ismar/WangGL23}. 
Although these modalities provide more intuitive interfaces, each faces significant limitations.
\revise{For example, gaze-based interaction encounters the Midas Touch problem \cite{DBLP:conf/ismar/MohanGFY18, velichkovsky1997towards} and often lacks accuracy \cite{10108465}. Gesture-based interaction can cause arm fatigue after prolonged use \cite{DBLP:conf/vr/ChaconasH18,DBLP:conf/chi/Hincapie-RamosGMI14}, and speech-based interaction increases cognitive load due to the need to remember complex commands \cite{DBLP:conf/ismar/ChenGFCL23}. Thus, the efficiency of unimodal natural interactions requires further improvement.}{For example, gesture-based interaction can cause arm fatigue after prolonged use \cite{DBLP:conf/vr/ChaconasH18,DBLP:conf/chi/Hincapie-RamosGMI14}. The performance of unimodal natural interactions requires further improvement.}

\begin{figure*}
    \begin{center}
    \includegraphics[width=0.8\linewidth]{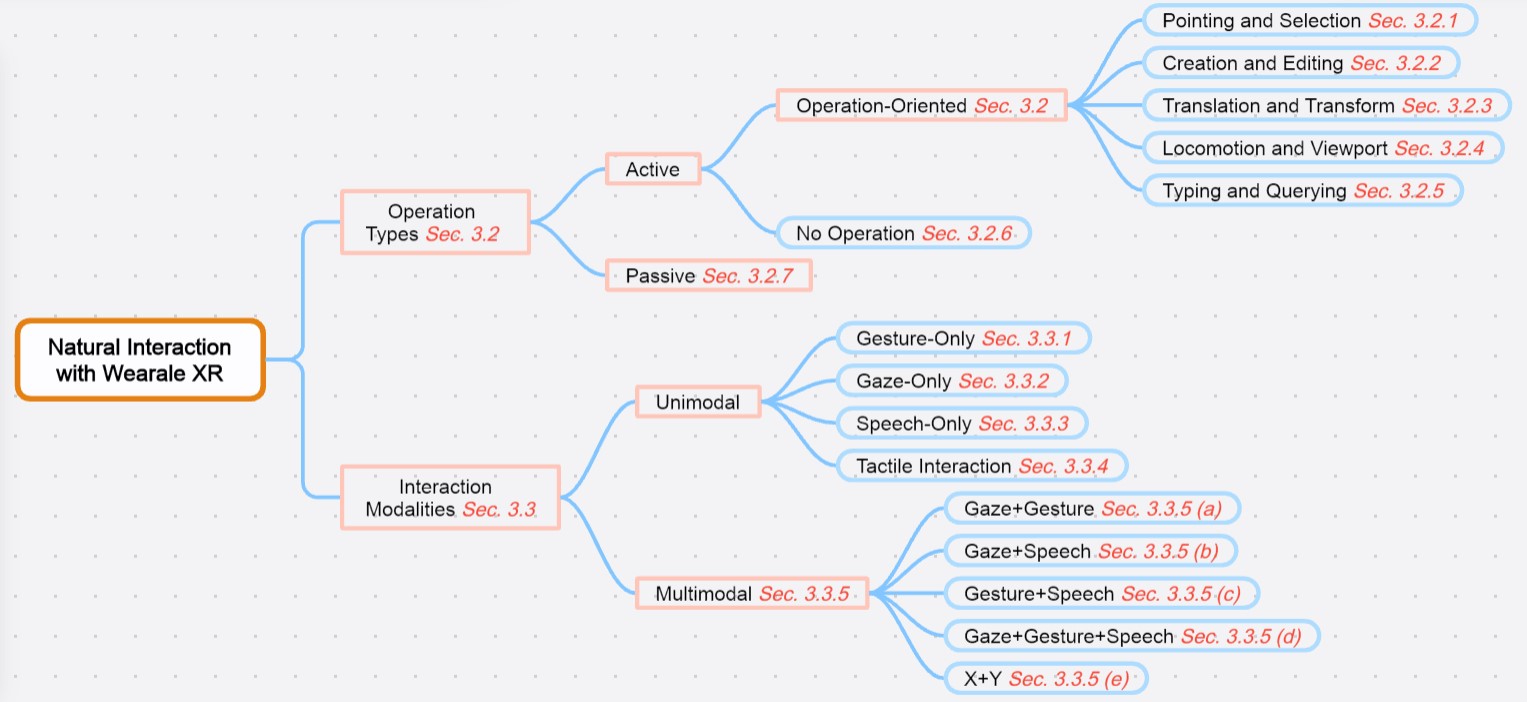}
    \end{center}
    \caption{
    \revise{}{We classify operation types into seven categories based on whether users actively input or passively receive feedback. Additionally, we divide interaction modalities into nine categories, distinguishing between unimodal and multimodal natural interactions. (The overlap between Operation Types and Interaction Modalities is shown in table. 1)}
    }
    \label{fig:taxonomy}
\end{figure*}

To address these challenges, multimodal interaction has emerged as a promising solution by combining the strengths of individual modalities. \revise{Typically, such systems integrate two modalities, \textit{e.g.}, Gesture + Speech \cite{DBLP:conf/chi/CaoKWAX24, DBLP:conf/chi/TorreFHBFL24}, Gaze + Gesture \cite{DBLP:journals/tvcg/SidenmarkP0CGWG22, 10.1145/3530886}, Gaze + Speech \cite{10.1145/3613904.3642068}, or Gaze + Electroencephalography \cite{DBLP:conf/embc/WangDCS15}.}{Such systems integrate typically two modalities, \textit{e.g.}, Gesture + Speech \cite{DBLP:conf/chi/CaoKWAX24, DBLP:conf/chi/TorreFHBFL24}, Gaze + Gesture \cite{DBLP:journals/tvcg/SidenmarkP0CGWG22, 10.1145/3530886}, Gaze + Speech \cite{10.1145/3613904.3642068}, and Gaze + Electroencephalography \cite{DBLP:conf/embc/WangDCS15}.}
\revise{In these systems, each modality is assigned a specific task.}{Each modality in these systems is assigned a specific task.} For example, a user might select an object using the eye gaze and trigger an action with a hand gesture \cite{10108465}. These approaches enhance the XR experience, providing users with a more natural and efficient way to interact with digital environments.

In recent years, researchers have conducted extensive reviews on various aspects of XR interaction techniques. These reviews have typically focused on specific areas such as AR environments \cite{DBLP:journals/tvcg/TranBWBP23, DBLP:conf/ismar/HertelKSBSS21}, VR environments \cite{pirker2021potential, katona2021review}, or immersive environments involving handheld displays \cite{DBLP:journals/vi/ZhangWZST23, DBLP:journals/tvcg/SpittlePC023}, highlighting the advantages, challenges, and emerging trends in each domain \cite{DBLP:journals/tvcg/TranBWBP23, DBLP:journals/tvcg/SpittlePC023}. However, the rapid advancements in natural interaction, particularly driven by artificial intelligence (AI) and large language models (LLMs) \cite{DBLP:conf/vr/GiunchiNGS24, DBLP:conf/chi/WangYWJ024, DBLP:conf/vr/YangQCSBLL24}, have introduced new interaction paradigms and technologies. It underscores the need for updated reviews that synthesize and evaluate the latest developments in the field.

In this paper, we aim to capture the latest trends by providing a comprehensive review of multimodal natural interaction techniques for wearable XR. \revise{}{Specifically, our objective is to determine and answer the following research questions (RQs):}

\begin{enumerate}
    \item \revise{}{What novel interaction paradigms and techniques have emerged over the past three years? (answered in Section \ref{Interaction techniques})}

    \item \revise{}{What are the key evolving trends of multimodal natural interaction in XR over the past three years? (answered in Section \ref{4.1} and \ref{4.2})}

    \item \revise{}{How have recent advancements in AI and LLMs been leveraged to enhance natural interaction in XR environments? (answered in Section \ref{4.3})}
\end{enumerate}

\revise{We focus on papers published between 2022 and 2024 across six top venues: ACM CHI, UIST, IMWUT(UbiComp), IEEE VR, ISMAR, and TVCG.}{In order to answer these questions, we investigate papers published between 2022 and 2024 across six top venues: ACM CHI, UIST, IMWUT, IEEE VR, ISMAR, and TVCG.} We categorize the reviewed literature based on application context, operation types, and interaction modalities. \revise{Specifically, operation types are divided into seven categories}{Particularly, operation types are divided into seven categories}, distinguishing between active and passive interactions, as shown in Fig. \ref{fig:taxonomy}. Interaction modalities are discussed across nine distinct types.
\revise{Additionally, we present statistical results into advanced natural interaction techniques}{Additionally, we present statistical results on advanced natural interaction techniques.}
\revise{Based on these findings, we discuss the current challenges of natural interaction techniques and suggest potential future research directions.}{Based on these findings, we discuss the current challenges of natural interaction techniques and propose potential directions for future research.} Our review offers valuable insights for researchers and promotes the further development of multimodal natural interaction for XR headsets.


The contributions of this review are threefold:

\begin{enumerate}
\item We provide a systematic review of the latest developments in multimodal interaction techniques for wearable XR, drawing from six top venues with papers published between 2022 and 2024.

\item We categorize these papers based on application contexts, operation types, performance measures, and interaction modalities, and present statistical insights into advanced natural interaction techniques.

\item We identify the current challenges of natural interaction techniques and propose potential directions for future research to improve the effectiveness and usability of multimodal interactions in XR.
\end{enumerate}


\revise{}{\textbf{Relationship to other reviews.} Several prior reviews have explored immersive interaction techniques, but with distinct scopes and focuses compared to our work. Hertel \textit{et al.} \cite{DBLP:conf/ismar/HertelKSBSS21} proposed a taxonomy for AR interaction techniques, focusing on task and modality dimensions based on works from 2016 to 2021. Spittle \textit{et al.} \cite{DBLP:journals/tvcg/SpittlePC023} reviewed AR/VR interaction techniques from 2013 to 2020, focusing on display type, study type, input methods, and tasks. 
Pirker \textit{et al.} \cite{pirker2021potential} analyzed the educational applications of 360° videos and real VR from 2010 to 2020, including language learning, teacher education, history and social studies, \textit{etc}. Zhang \textit{et al.} \cite{DBLP:journals/vi/ZhangWZST23} focused on immersive visualization from 2014 to 2023, detailing multimodal perception and interaction techniques, particularly sensory modalities including vision, touch, and olfaction, while also analyzing collaborative analysis and hardware devices. 
Ghamandi \textit{et al.} \cite{DBLP:conf/ismar/GhamandiHNGKTL23} reviewed 30 years of collaborative XR tasks (1993–2023) and proposed a taxonomy that classifies tasks based on actions (\textit{e.g.}, manipulation and navigation) and properties (\textit{e.g.}, temporal state and dependency). Their work focuses on understanding task structures and collaboration dynamics across the mixed reality spectrum.
In contrast, our review focuses on recent developments (2022–2024) in multimodal natural interaction for wearable XR headsets, emphasizing the integration of AI and LLMs to enable novel interaction paradigms and providing a structured framework for advancing spatial computing.}

The remainder of this paper is organized as follows. \revise{Section 2 outlines the methodology used for selecting the reviewed literature and provides an analysis of the selected papers.}{Section 2 outlines the methodology to select the reviewed literature and provides an statistic analysis of the selected papers.} \revise{Section 3 presents a detailed review and analysis of natural interaction techniques in XR, categorized based on the previously mentioned criteria}{Section 3 presents a detailed review and analysis of natural interaction techniques in XR, categorized based on the criteria mentioned above}. Section 4 discusses key challenges and offers recommendations for future research on natural interaction techniques in XR. Finally, Section 5 describes the limitations of this study, and Section 6 concludes the paper.


\renewcommand{\dblfloatpagefraction}{.9}
\begin{table*}
\renewcommand\arraystretch{1.2}
  \centering
  \caption{The reviewed literature is categorized based on six operation types, nine interaction modalities, and year of publication \revise{}{(``X+Y'' represents other modality combinations, where ``X'' denotes a modality from gesture, speech, or gaze, and ``Y'' denotes any other modality except these three (\textit{e.g.}, gaze + Head pose))}}
  \scalebox{0.97}{
  \begin{tabular}{m{1.6cm}<{\centering}|m{0.7cm}<{\centering}|m{1.29cm}<{\centering}|m{1.29cm}<{\centering}|m{1.29cm}<{\centering}|m{1.2cm}<{\centering}|m{1.29cm}<{\centering}|m{1.29cm}<{\centering}|m{1.29cm}<{\centering}|m{1.2cm}<{\centering}|m{1.2cm}<{\centering}}
    \hline
    
     \cellcolor{oBlue!80}  & \cellcolor{oPink!80} & \multicolumn{4}{c|}{ \cellcolor{oYellow!90}{Unimodal Interactions}} & \multicolumn{5}{c}{\cellcolor{oRed!100}{Multimodal Interactions}} \\
\cline{3-11}    \multirow{-2}{1.6cm}{\centering \cellcolor{oBlue!80} Operation Types}      &  \multirow{-3}{0.7cm}{\centering \cellcolor{oPink!80} Years}   & \cellcolor{oYellow!20}{GEST only} & \cellcolor{oYellow!20}{GAZE only} & \cellcolor{oYellow!20}{SPCH only} & \cellcolor{oYellow!20}{Tactile} & \cellcolor{oRed!20}{GAZE + GEST} & \cellcolor{oRed!20}{GAZE + SPCH} & \cellcolor{oRed!20}{GEST + SPCH} & \cellcolor{oRed!20}{GAZE + GEST + SPCH} & \cellcolor{oRed!20}{X + Y} \\
    \hline
       \cellcolor{oBlue!30} & \cellcolor{oPink!30} 2022  & \cite{DBLP:conf/vr/MatthewsTIS22}\cite{DBLP:conf/ismar/ChowdhuryUPIH22}\cite{DBLP:conf/chi/SchmitzGS022}\cite{DBLP:conf/ismar/BanMNK22}\cite{DBLP:conf/ismar/YuZDV022}  & \cite{DBLP:conf/ismar/LeeHM22}\cite{DBLP:conf/chi/ChoiSO22}\cite{kim2022lattice}\cite{DBLP:conf/vr/0001LC00S22}\cite{DBLP:journals/tvcg/WangZ022}  &   & \cite{DBLP:journals/imwut/ChenLYZ22}  & \cite{DBLP:journals/tvcg/SidenmarkP0CGWG22}\cite{10.1145/3530886}\cite{DBLP:conf/uist/SendhilnathanZL22}  & \cite{DBLP:conf/vr/JingLB22}  & \cite{DBLP:conf/uist/LiaoKJKS22}  &   & \cite{DBLP:conf/uist/0001QTFLS22}\cite{DBLP:conf/ismar/XuMYSL22}\cite{DBLP:journals/imwut/ShenYYS22}\cite{DBLP:conf/ismar/MengXL22}\\
\cline{2-11}      \cellcolor{oBlue!30}   &  \cellcolor{oPink!30} 2023  & \cite{DBLP:conf/ismar/DasNH23}\cite{DBLP:conf/ismar/ZhuSSG23}\cite{DBLP:journals/tvcg/SongDK23}\cite{DBLP:conf/chi/0003HLG23}  & \cite{DBLP:conf/ismar/ChenHTHH23}\cite{DBLP:conf/chi/SidenmarkCNLPG23}  &   & \cite{DBLP:conf/ismar/DasNH23}  & \cite{10.1145/3544548.3581423}\cite{10.1145/3591129}\cite{10108465}\cite{DBLP:conf/ismar/CailletGN23}  &   &   &  \cite{DBLP:conf/ismar/ChenGFCL23}  & \cite{DBLP:conf/chi/WeiSYW0YL23}\cite{DBLP:conf/chi/HouNSKBG23}\cite{10049667} \\
\cline{2-11}         \multirow{-5}{1.6cm}{\centering \cellcolor{oBlue!30} Pointing and Selection} &  \cellcolor{oPink!30} 2024  & \cite{DBLP:conf/vr/QuereMJWW24}\cite{DBLP:conf/vr/SindhupathirajaUDH24}\cite{DBLP:conf/chi/DupreARSP24}  & \cite{DBLP:conf/vr/OrloskyLSSM24}\cite{DBLP:conf/chi/ZhangCSS24}\cite{DBLP:conf/chi/TurkmenGBSASPM24}  &   &   & \cite{DBLP:conf/chi/ZennerKFAK24}\cite{10.1145/3613904.3642758}  &   &   &   & \cite{DBLP:conf/vr/LaiSL24}\cite{marquardt2024selection} \\
    \hline
    \cellcolor{oBlue!30} & \cellcolor{oPink!30} 2022  & \cite{DBLP:conf/ismar/YuZDV022}\cite{DBLP:conf/chi/PeiCLZ22}  &   &   & \cite{DBLP:journals/imwut/ChenLYZ22}\cite{DBLP:conf/chi/SatriadiSECCLYD22}  &   &   & \cite{DBLP:conf/uist/LiaoKJKS22}  &   &  \\
\cline{2-11}    \cellcolor{oBlue!30}     & \cellcolor{oPink!30} 2023  & \cite{DBLP:journals/tvcg/XuZSFY23}\cite{DBLP:journals/tvcg/SongDK23} &   &   & \cite{DBLP:journals/imwut/ZhanXZGCGLQ23}  &   &   &   & \cite{DBLP:conf/ismar/ChenGFCL23}  &  \\
\cline{2-11}     \multirow{-3}{1.6cm}{\centering \cellcolor{oBlue!30} Creation and Editing}   & \cellcolor{oPink!30} 2024  & \cite{DBLP:conf/vr/QuereMJWW24}\cite{DBLP:conf/chi/DupreARSP24}  & \cite{DBLP:conf/chi/TurkmenGBSASPM24}  & \cite{DBLP:conf/chi/TorreFHBFL24}\cite{DBLP:conf/vr/GiunchiNGS24}  &   &   &   & \cite{DBLP:conf/chi/CaoKWAX24}\cite{10.1145/3613904.3642758}  &   &  \\
    \hline
    \cellcolor{oBlue!30} & \cellcolor{oPink!30} 2022  & \cite{DBLP:conf/chi/PeiCLZ22}\cite{DBLP:conf/ismar/BanMNK22}\cite{DBLP:conf/ismar/YuZDV022}   &   &   & \cite{DBLP:conf/chi/SatriadiSECCLYD22}  &   &   &\cite{9873984}   &   &  \\
\cline{2-11}    \cellcolor{oBlue!30}     & \cellcolor{oPink!30} 2023  & \cite{DBLP:journals/tvcg/SongDK23}\cite{DBLP:journals/tvcg/XuZSFY23}\cite{DBLP:conf/chi/0003HLG23}  &   &   &   & \cite{10108465}  &   &   &   & \cite{DBLP:conf/chi/HouNSKBG23} \\
\cline{2-11}  \multirow{-4}{1.6cm}{\centering \cellcolor{oBlue!30} Translation and Transform}       & \cellcolor{oPink!30} 2024  & \cite{DBLP:journals/ijhci/DengSZK24}  &   & \cite{DBLP:conf/vr/GiunchiNGS24}  &   &   &   & \cite{DBLP:conf/chi/CaoKWAX24}  &   &  \\
    \hline
    \cellcolor{oBlue!30} & \cellcolor{oPink!30} 2022  & \cite{DBLP:conf/ismar/YuZDV022}\cite{DBLP:conf/ismar/ChowdhuryUPIH22}  &   &   &   &   &   &   &   &  \\
\cline{2-11}    \cellcolor{oBlue!30}     & \cellcolor{oPink!30} 2023  &   &   & \cite{DBLP:conf/vr/HombeckVHDL23}  & \cite{DBLP:conf/vr/MortezapoorVVK23}  &   &   &   &   &  \\
\cline{2-11}     \multirow{-4}{1.6cm}{\centering \cellcolor{oBlue!30} Locomotion and Viewport}    & \cellcolor{oPink!30} 2024  & \cite{DBLP:conf/vr/SindhupathirajaUDH24}\cite{DBLP:conf/vr/SinJLLLN24}  &   & \cite{DBLP:conf/chi/WangYWJ024}  &   & \cite{10.1145/3613904.3642147}  &   &   &   & \cite{DBLP:conf/chi/LeeWSG24} \\
    \hline
    \cellcolor{oBlue!30} & \cellcolor{oPink!30}2022  & \cite{10.1145/3491102.3517682}\cite{DBLP:conf/ismar/SongDK22}  &   &   &   & \cite{DBLP:conf/chi/HeLP22}  &   & \cite{DBLP:conf/uist/LiaoKJKS22}  &   &  \\
\cline{2-11}     \cellcolor{oBlue!30}    & \cellcolor{oPink!30}2023  & \cite{DBLP:journals/tvcg/ShenDK23}  & \cite{DBLP:conf/iui/ZhaoPTZWJBG23}\cite{cui2023glancewriter}  & \cite{DBLP:conf/chi/ZhangLHWLGZ23}  & \cite{DBLP:journals/imwut/ZhanXZGCGLQ23}  &   &   &   &   &  \\
\cline{2-11}    \multirow{-3}{1.6cm}{\centering \cellcolor{oBlue!30} Typing and Querying}      & \cellcolor{oPink!30} 2024  & \cite{shen2024ringgesture}  & \cite{DBLP:conf/vr/HuDK24}  & \cite{DBLP:conf/chi/TorreFHBFL24}\cite{DBLP:journals/tvcg/CaiML24}\cite{DBLP:conf/chi/WangYWJ024}  &   & \cite{10474330}& \cite{10.1145/3613904.3642068}\cite{DBLP:journals/corr/abs-2405-18537}\cite{DBLP:journals/imwut/WangSWYYWJXY24}  &   & \cite{DBLP:conf/chi/0005WBCRF24}  &  \\

 \hline
    \multicolumn{2}{c|}{\cellcolor{oBlue!30} No Operation} & \cite{DBLP:conf/chi/WangLZ24}\cite{DBLP:journals/tvcg/ShenDMK22}\cite{DBLP:conf/chi/LeeZAYGLKYDLSGZ24}\cite{DBLP:conf/vr/RuppGBK24}  &   & \cite{10522613}\cite{DBLP:journals/tvcg/CaiML24} & \cite{DBLP:conf/chi/XuZKN23}\cite{DBLP:conf/iswc/KitamuraYS23}\cite{DBLP:conf/ismar/LiLMHLS22}  &   &   &   &   &  \\
    \hline
    \end{tabular}%
    }
  \label{tab:table1}%
\end{table*}%

\begin{table}[t]
    \centering
    \renewcommand\arraystretch{1.2}
    \caption{The literature related to passive interaction}
    \begin{tabular}{m{2.5cm}<{\centering}|m{4cm}<{\centering}}
    \hline
        \cellcolor{oBlue!80} Passive Interaction & \cellcolor{oPink!80} Literature \\ \hline
        \cellcolor{oBlue!30} Visual & \cite{DBLP:conf/chi/Pohlmann0MMB23}\cite{DBLP:conf/chi/MedlarLG24}\cite{DBLP:conf/chi/FeickR0K22}\cite{DBLP:conf/chi/WuQQCRS24}\cite{DBLP:conf/vr/YangQCSBLL24}\cite{DBLP:conf/chi/ElsharkawyAYAHK24}\cite{DBLP:conf/vr/LiLYTFX24}\cite{DBLP:conf/chi/RaschRS023}\cite{DBLP:conf/chi/TanX0SZHH24}\cite{DBLP:conf/vr/WangZF24}\cite{DBLP:conf/uist/Tao022}\cite{10462901}\cite{wang2024tasks} \\ \hline
        \cellcolor{oBlue!30} Acoustic & \cite{DBLP:conf/vr/YangQCSBLL24}\cite{DBLP:conf/chi/Pohlmann0MMB23}\cite{DBLP:conf/chi/ElsharkawyAYAHK24} \\ \hline
        \cellcolor{oBlue!30} Haptic & \cite{DBLP:conf/chi/ShenS022}\cite{DBLP:conf/chi/TatzgernDWCEDGH22}\cite{DBLP:conf/chi/0001OPSB24}\cite{DBLP:conf/vr/YamazakiH23}\cite{DBLP:conf/uist/ShenRM0S23}\cite{DBLP:conf/uist/JinguWS23}\cite{DBLP:conf/vr/SaintAubertAMPAL23} \\ \hline
        \cellcolor{oBlue!30} Hybrid  & \cite{DBLP:conf/chi/FeickR0K22}\cite{DBLP:conf/chi/Pohlmann0MMB23}\cite{DBLP:conf/chi/0001OPSB24} \\ \hline
    \end{tabular}
    \label{tab:passivetable}
\end{table}

\section{Survey Methodology}

We conducted a systematic review and analysis of natural interaction techniques in wearable XR. To this end, we surveyed relevant papers from the top conferences and journals in the field, focusing on recent publications. Additionally, we categorized and coded these papers based on various aspects of natural interaction techniques in XR.

\subsection{Selection Criteria}
\label{criteria}
This review focuses on the technologies and applications of these interaction techniques within XR. \revise{We selected relevant papers based on the following criteria}{Relevant papers are selected based on the following criteria}:

1. \textbf{\revise{Wearable XR (including VR, AR, and MR).}{Involve wearable XR (including VR, AR, and MR).}} We focus on wearable XR because head-mounted displays (HMDs) offer superior mobility and support more natural interaction methods, such as gaze, gesture, and voice control, which ultimately enhance the sense of immersion.

2. \textbf{\revise{Considering natural interaction techniques.}{Consider natural interaction techniques.}} \revise{We examined both unimodal and multimodal interactions.}{We take both unimodal and multimodal interactions into consideration.}  \revise{Unimodal interactions include hand gestures, eye gaze, speech, and tactile inputs, while multimodal interactions combine these modalities.}{Unimodal interactions include hand gesture, eye gaze, speech, and tactile input, while multimodal interactions combine these modalities.} \revise{Among the unimodal types, hand-based interactions are the most common in everyday life, such as pouring water or typing on a keyboard, and they allow for precise operations.}{Among the unimodal interactions, hand-based ones allow precise operations and are most commonly used in daily life, such as pressing buttons and typing on a keyboard.} \revise{With the widespread adoption of smartphones, voice interaction has also gained significant use, with systems like Apple Siri, Microsoft Cortana, and Xiaomi XiaoAI improving input efficiency.}{With the widespread adoption of smartphones, voice interactions have also become progressively prevalent, with systems like Apple Siri, Microsoft Cortana, and Xiaomi XiaoAI enhancing input efficiency.} In recent years, advances in eye-tracking technology have drawn increasing attention to gaze-based interaction. For tactile interactions, we focus on lightweight, wearable, and technologically advanced input devices.

\begin{figure*}
    \begin{center}
    \includegraphics[width=1\linewidth]{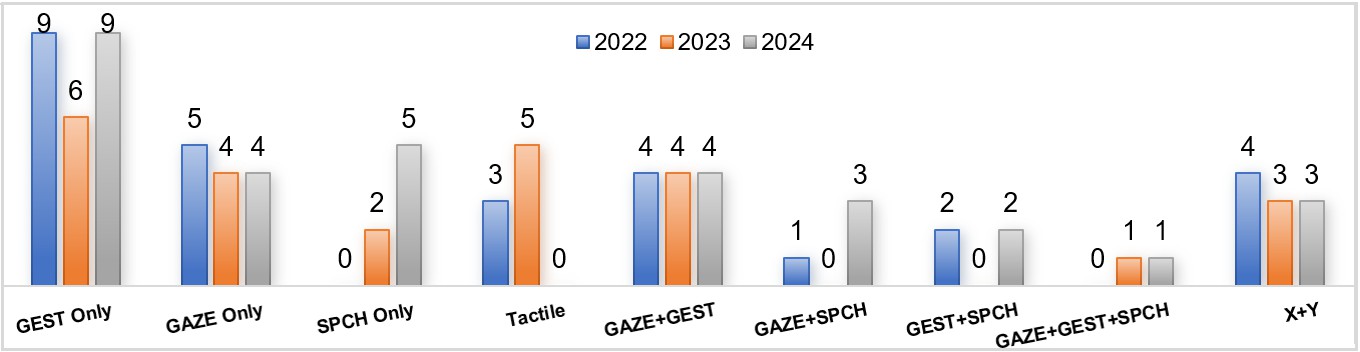}
    \end{center}
    \caption{
The bar charts display interaction modality data, categorized by nine modalities and three years. This figure is consistent with the literature recorded in Table \ref{tab:table1}.
    }
    \label{fig:bymodality}
\end{figure*}


3. \textbf{\revise{Focusing on top-tier conferences and journals.}{Focus on top-tier conferences and journals.}} \revise{We concentrated on research published in leading venues such as ACM CHI, UIST, IMWUT,  IEEE VR, ISMAR, and IEEE TVCG.}{We concentrate on research published in leading venues such as ACM CHI, UIST, IMWUT (UbiComp),  IEEE VR, ISMAR, and IEEE TVCG.} These publications typically feature pioneering work, often marked by innovative contributions.

4. \textbf{\revise{Focusing on research from 2022, 2023, and 2024}{Focus on research from 2022, 2023, and 2024}\footnote{Our final search was conducted on 16th October 2024.}}. \revise{The release of XR devices such as Microsoft HoloLens 2, Apple Vision Pro, HTC Vive, Meta Quest, and Pico series has provided significant technical support to the field of XR research.}{The release of XR devices, such as Microsoft HoloLens 2, Apple Vision Pro, HTC Vive, Meta Quest, and Pico series, has provided significant technical support to the field of XR research.} We believe that recent studies offer valuable insights for researchers in this area.


\subsection{Data Collection}

We search keywords in Google Scholar such as ``virtual reality'', ``augmented reality", ``extended reality", ``multimodal interaction'', ``eye gaze", ``hand", ``speech" and ``tactile". 
Article retrieval is primarily conducted following the standards outlined in Section \ref{criteria}. In addition, we identify some articles from other high-level journals and conferences that also meet criteria 1, 2, and 4 from Section \ref{criteria}. Although smaller in number, this portion of the literature serves as a valuable supplement to the collected data.

In summary, we gather a total of 104 research papers. These include 40 from CHI, 20 from IEEE VR, 14 from ISMAR, 14 from TVCG, 6 from UIST, 4 from IMWUT, 5 from other conferences and journals (IJHCS, ETRA, IUI, ISWC), and 1 from ArXiv.


\subsection{Data Analysis} 

The collected literature covers a range of interaction modalities and their combinations, with a focus on various application contexts, performance measures and operation types. These publications were classified and analyzed across several dimensions, as outlined in Section \ref{section3}. 
In this section, we perform a statistical analysis of the reviewed literature from various perspectives. This high-level analysis explores the quantity and proportion of attention given to different categories of interaction research in recent years. These statistical results also support the taxonomy proposed in Section \ref{section3}.


Based on whether humans act as initiators or receivers of interaction, the literature is categorized into \textbf{Active} interaction (84 papers) and \textbf{Passive} interaction (20 papers). In the Active interaction literature, the modalities include \textit{Gesture} (24), \textit{Gaze} (13), \textit{Speech} (7), \textit{Tactile} (8), and various combinations of these modalities, such as \textit{Gaze+Speech} (12), \textit{Gaze+Gesture} (4), \textit{Gesture+Speech} (4), \textit{Gaze+Gesture+Speech} (2), and other combinations (\textit{X+Y}) (10). Notably, in some publications that examined multiple single or multimodal interaction schemes, the baseline methods were excluded from classification, focusing instead on the newly proposed interaction schemes. The proportion of different modalities in research from 2022 to 2024 is shown in Fig. \ref{fig:bymodality}.
Studies on Gesture and Gaze, including their combined modalities, are the most prevalent. A notable increase in research on Speech-only interaction is observed in 2024, accompanied by a rise in Speech-related multimodal studies, likely driven by recent advancements in LLMs. Tactile interaction-related literature is comparatively less common. In the analysis of Passive interaction modalities, tactile interaction is frequently studied as a feedback mechanism rather than as an active interaction method.


\begin{figure}
    \begin{center}
    \includegraphics[width=1\linewidth]{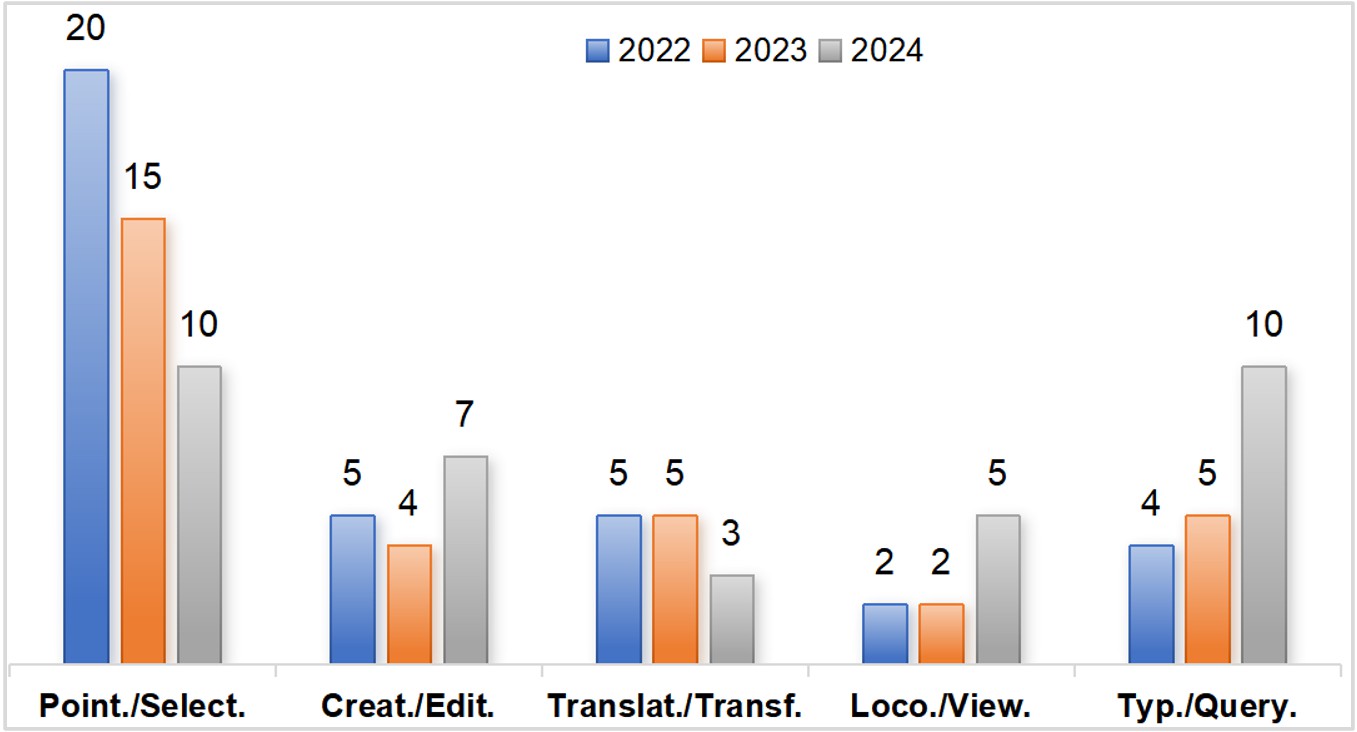}
    \end{center}
    \caption{
The bar charts display the yearly statistics for the five main operation types. This figure is consistent with the literature recorded in Table \ref{tab:table1}.
    }
    \label{fig:bytype}
    \vspace{-6mm}
\end{figure}

The number of publications for various operation types in the collected literature is as follows: \textit{Pointing and Selection} (45), \textit{Creation and Editing} (16), \textit{Translation and Transform} (13), \textit{Locomotion and Viewport} (9), \textit{Typing and Querying} (19). The proportions of research from 2022 to 2024 are illustrated in Fig. \ref{fig:bytype}. Pointing and Selection remains the most focused area, although the number of studies has been decreasing annually. This decline may be due to advancements in more precise multimodal selection techniques.
Significant increases in research on Locomotion and Viewport and Typing and Querying were observed in 2024. This rise may be attributed to growing attention on users' subjective experiences and the development of LLMs. These models enable users to communicate more diverse semantic information to XR systems. Further analysis of the technological developments behind these trends will be provided in Section \ref{section3}.

Passive interaction types are categorized as Visual, Acoustic, Haptic, and Hybrid. Specific details are available in Table \ref{tab:passivetable}. 
\revise{}{Passive interaction is separated into a new table  rather than include it in Table \ref{tab:table1} for the following reasons.
Firstly, there are inherent differences between the modalities of active and passive interactions, and they are not directly equivalent. For example, while active interaction involving human's eyes is typically referred to as \textit{gaze} (or \textit{eye gaze}) interaction, passive interaction involving the eyes is generally categorized as \textit{visual feedback}. 
Besides, Table \ref{tab:table1} is primarily designed to illustrate the relationship between modalities and operation types of the reviewed papers. Passive interactions, however, do not engage with the concept of `operation' in the same way. Thus, it is not appropriate to include passive interactions in Table \ref{tab:table1}, which focuses on active interactions.
}

\revise{}{Recent studies have increasingly focused on haptic feedback}, likely due to its ability to significantly enhance users' immersive experiences. \revise{Combined Visual and Acoustic feedback is relatively common,}{It is relatively commmon to combine visual and acoustic feedback,} while standalone acoustic feedback is rarely studied. Some research also explores visual feedback alone. Moreover, the integration of the three primary sensory feedback modalities (visual, acoustic and haptic) has emerged as a major research direction.

%


\revise{}{There are some devices that are preferred by researchers to conduct their experiment. The situation is illustrated in Table \ref{devices}. Microsoft Hololens 2 and HTV Vive Pro / Vive Pro Eye are the most popular devices used for experiment in the recently 2 years, while Hololens 2 is more preferred in 2024. Meta Quest 2 (Oculus Quest 2) is less used after 2022 but still popular.}

\section{A Taxonomy and Analysis of Natural Interaction Techniques}
\label{section3}

\revise{}{Previous XR survey studies have primarily categorized research based on \textit{display type}, \textit{study type}, \textit{use case}, \textit{input technique}, and \textit{task type} \cite{DBLP:journals/vi/ZhangWZST23,DBLP:conf/ismar/HertelKSBSS21,DBLP:journals/tvcg/SpittlePC023}. Since this paper focuses on head-mounted displays, there is only one display type.  
The \textit{study type} refers to the type of user evaluation conducted (\textit{e.g.}, assessment or comparison), while the \textit{use case} examines the conditions under which studies are conducted (\textit{e.g.}, static or in motion). Given the focus of RQ1 and RQ2, this paper primarily explores novel interaction paradigms and techniques, as well as evolving trends in multimodal natural interaction. Therefore, we do not discuss \textit{study type} or \textit{use case} in this work but retain \textit{input technique}. 
Regarding \textit{task type}, existing research often considers \textit{application scenario} and \textit{operation type} as part of task classification and discusses them together \cite{DBLP:conf/ismar/HertelKSBSS21,DBLP:journals/tvcg/SpittlePC023}. In this paper, we separate these two aspects.}  

\revise{}{As a result, we retain three classification dimensions. Our categorization follows the order of \textit{application scenario} → \textit{operation type} → \textit{interaction technique}. This structure facilitates the analysis of inherent developmental trends in interaction modalities through the lens of application requirements. It also progresses from a broad, high-level perspective to a more detailed and specific one.}



\revise{}{The subsequent sections are organized as follows: Section \ref{Appl.} presents a detailed discussion of natural interaction applications, Section \ref{operationtypes} provides an analysis of seven distinct operation types, and Section \ref{Interaction techniques} examines the implementation methodologies of various interaction modalities and the design considerations employed by researchers in developing these interaction systems.}

\begin{table}[t]
    \centering
    \renewcommand\arraystretch{1.2}
    \caption{\revise{}{The literature of different application scenarios}}
    \begin{tabular}{m{2.5cm}<{\centering}|m{4cm}<{\centering}}
    \hline
        \cellcolor{oBlue!80} Application Scenarios & \cellcolor{oPink!80} Literature \\ \hline
        \cellcolor{oBlue!30} drawing and sketching & \cite{DBLP:conf/ismar/ChenGFCL23, DBLP:conf/chi/TurkmenGBSASPM24, DBLP:journals/tvcg/XuZSFY23, 10.1145/3613904.3642758, DBLP:journals/imwut/ChenLYZ22, DBLP:journals/tvcg/SongDK23}
        \\ \hline
        \cellcolor{oBlue!30} smart assistants & \cite{DBLP:conf/vr/GiunchiNGS24, DBLP:conf/chi/WangYWJ024, DBLP:conf/vr/YangQCSBLL24, DBLP:conf/chi/0005WBCRF24, 10.1145/3613904.3642068, DBLP:conf/chi/WuQQCRS24, DBLP:journals/imwut/WangSWYYWJXY24}
        \\ \hline
        \cellcolor{oBlue!30} virtual meetings & \cite{DBLP:conf/vr/SaintAubertAMPAL23, 10462901, DBLP:conf/uist/LiaoKJKS22, DBLP:conf/chi/CaoKWAX24, 10049667, DBLP:conf/vr/WangZF24}
        \\ \hline
        \cellcolor{oBlue!30} VR/AR navigation  & \cite{DBLP:conf/vr/QuereMJWW24, DBLP:conf/chi/WangYWJ024, DBLP:conf/vr/YangQCSBLL24}
        \\ \hline
        \cellcolor{oBlue!30} reading  & \cite{DBLP:conf/ismar/LeeHM22, DBLP:conf/ismar/MengXL22, DBLP:conf/ismar/XuMYSL22}
        \\ \hline
        \cellcolor{oBlue!30} furniture assembly/maintenance  & \cite{DBLP:conf/vr/YangQCSBLL24,DBLP:conf/vr/QuereMJWW24}
          \\ \hline
        \cellcolor{oBlue!30} remote collaboration  & \cite{DBLP:conf/vr/JingLB22, DBLP:journals/corr/abs-2405-18537}
         \\ \hline
        \cellcolor{oBlue!30} autonomous driving  & \cite{DBLP:conf/chi/ElsharkawyAYAHK24}
        \\ \hline
        \cellcolor{oBlue!30} enter passwords  & \cite{DBLP:conf/ismar/SongDK22, DBLP:conf/vr/RuppGBK24}
        \\ \hline
    \end{tabular}
    \label{scene}
\end{table}

\begin{table}[t]
    \centering
    \renewcommand\arraystretch{1.2}
    \caption{\revise{}{Devices chosen by researchers (by years)}}
    \begin{tabular}{m{2.5cm}<{\centering}|m{1.5cm}<{\centering}|m{1.5cm}<{\centering}|m{1.5cm}<{\centering}}
    \hline
         Device Name &  2022 & 2023 & 2024
        \\ \hline
         Microsoft Hololens 2 & 7 & 6 & 13
        \\ \hline
        HTC Vive Pro/Vive Pro Eye & 9 & 8 & 7
        \\ \hline
        Meta Quest 2 (Oculus Quest 2) & 8 & 5 & 6
        \\ \hline

    \end{tabular}
    \label{devices}
\end{table}

\subsection{Application Scenarios}  
\label{Appl.}

\begin{figure}
    \begin{center}
    \includegraphics[width=1\linewidth]{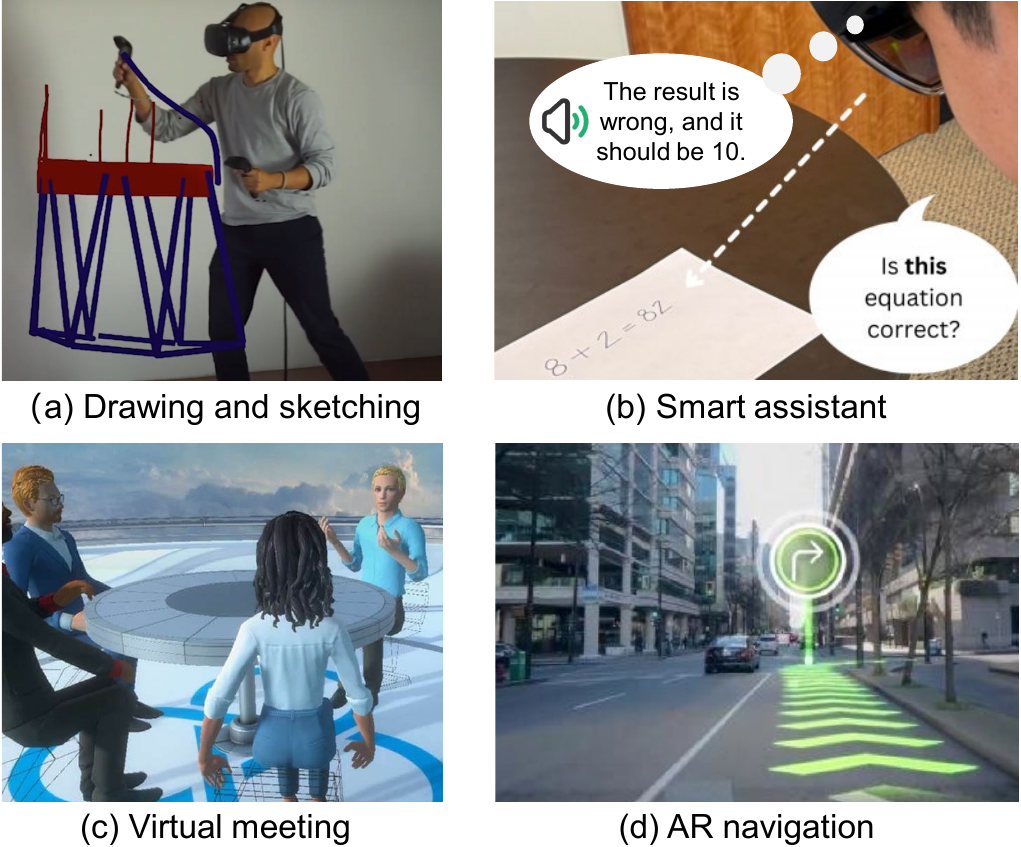}
    \end{center}
    \caption{
\revise{}{The illustrations of XR application scenarios: (a) drawing and sketching; (b) smart assistant \cite{DBLP:conf/chi/0005WBCRF24}; (c) virtual meeting; (d) AR navigation. Image courtesy of \cite{DBLP:conf/chi/0005WBCRF24}.}
    }
    \label{fig:application}
\end{figure}

In the reviewed literature, nearly 70\% of the research focuses on interaction design for general scenarios, without specifying particular application contexts. For example, some studies design gaze vergence control techniques \cite{DBLP:conf/chi/SidenmarkCNLPG23, DBLP:journals/tvcg/WangZ022, DBLP:conf/chi/ZhangCSS24}, create more than 10 different hand gestures for interaction \cite{DBLP:conf/chi/PeiCLZ22, DBLP:journals/tvcg/SongDK23}, and develop robust voice keyword detection methods \cite{DBLP:journals/tvcg/CaiML24, DBLP:conf/chi/ZhangLHWLGZ23}.
However, we argue that applying multimodal natural interaction to specific application scenarios is crucial. 
Implementing these techniques in real-world contexts not only enhances user experience but also increase user exposure to multimodal interaction technologies.
In turn, this can accelerate the adoption of these technologies and promote their use across a wider range of fields.
Therefore, in this section, we discuss studies that explore specific application scenarios. \revise{}{Several exemplary application scenarios are illustrated in Fig. \ref{fig:application}. Meanwhile, we also provide an overview of the interaction types and operation types utilized.}


We found 6 papers applied natural interaction techniques to \textbf{drawing and sketching} \cite{DBLP:conf/ismar/ChenGFCL23, DBLP:conf/chi/TurkmenGBSASPM24, DBLP:journals/tvcg/XuZSFY23, 10.1145/3613904.3642758, DBLP:journals/imwut/ChenLYZ22, DBLP:journals/tvcg/SongDK23}. In these studies, users primarily engage in sketching through gestures (bare hands, pen, touch devices, or controllers), while using eye-tracking, voice commands or gestures for menu selection, such as switching brush colors. Eye-tracking is also employed to control 3D grids, allowing users to perceive depth more intuitively \cite{DBLP:conf/chi/TurkmenGBSASPM24}. Sketching is considered a relatively complex task, involving operations like pointing, selection, creation, and editing. 2 papers allow users to manipulate different geometric objects, including actions such as scaling, translation, and rotation \cite{DBLP:journals/tvcg/SongDK23, DBLP:journals/tvcg/XuZSFY23}.

7 papers explored the application of \textbf{smart assistants} \cite{DBLP:conf/vr/GiunchiNGS24, DBLP:conf/chi/WangYWJ024, DBLP:conf/vr/YangQCSBLL24, DBLP:conf/chi/0005WBCRF24, 10.1145/3613904.3642068, DBLP:conf/chi/WuQQCRS24, DBLP:journals/imwut/WangSWYYWJXY24}. Among these, 6 papers utilize LLMs as assistants, benefiting from their superior comprehension and reasoning abilities. For example, Giunchi \textit{et al.} allowed users to directly edit virtual objects through voice commands, such as creating or moving them \cite{DBLP:conf/vr/GiunchiNGS24}. Wang \textit{et al.} provided intelligent guidance for virtual tours by recognizing user speech and environmental information, then delivering multimodal feedback, such as avatars, voice, text windows, and minimaps \cite{DBLP:conf/chi/WangYWJ024}. 
Lee \textit{et al.} offered assistance with daily activities by using LLMs to answer questions related to objects of interest such as food calories or offer book recommendations, as identified by the user’s gaze and speech \cite{DBLP:conf/chi/0005WBCRF24}.

6 papers explored the application of natural interaction techniques in \textbf{video conferencing} or \textbf{virtual meetings} \cite{DBLP:conf/vr/SaintAubertAMPAL23, 10462901, DBLP:conf/uist/LiaoKJKS22, DBLP:conf/chi/CaoKWAX24, 10049667, DBLP:conf/vr/WangZF24}.
For example, Saint-Aubert \textit{et al.} focused on verbal communication with avatars in VR, introducing synchronized haptic feedback based on speech content to enhance immersion \cite{DBLP:conf/vr/SaintAubertAMPAL23}. Lee \textit{et al.} proposed using visual cues (lighting effects) and auditory cues (spatial audio) in VR social interactions to guide users' attention to new speakers \cite{10462901}.
Liao \textit{et al.} and Cao \textit{et al.} explored techniques for enhancing speaker presentations in video conferences \cite{DBLP:conf/uist/LiaoKJKS22, DBLP:conf/chi/CaoKWAX24}. They extracted keywords from users' speech and matched them with predefined images and animations, which were then synchronized with hand movements to improve expression.

3 papers investigated the application of natural interaction techniques in \textbf{VR/AR navigation} \cite{DBLP:conf/vr/QuereMJWW24, DBLP:conf/chi/WangYWJ024, DBLP:conf/vr/YangQCSBLL24}. 
For example, Quere \textit{et al.} used a hand-menu system to create virtual annotations for a large campus hosting a reception, which helped direct participants to meeting rooms \cite{DBLP:conf/vr/QuereMJWW24}. Wang \textit{et al.} used LLMs to generate guidance for virtual tours through avatars, voice, and text windows \cite{DBLP:conf/chi/WangYWJ024}.
It is important to note that our definition of navigation refers to guiding the user to quickly locate their position, which differs from the definition of virtual navigation in related studies that focus on controlling travel speed and direction of users \cite{10.1145/3613904.3642147, DBLP:conf/vr/SinJLLLN24, DBLP:conf/vr/SindhupathirajaUDH24}.

3 papers explored the application of natural interaction in \textbf{reading} \cite{DBLP:conf/ismar/LeeHM22, DBLP:conf/ismar/MengXL22, DBLP:conf/ismar/XuMYSL22}. For example, Lee \textit{et al.} designed a set of gaze-based interaction strategies to select text, zoom in on specific areas, and scroll through content, thereby enhancing the reading experience \cite{DBLP:conf/ismar/LeeHM22}. Meng \textit{et al.} investigated head-based pointing combined with three hands-free selection mechanisms, \textit{i.e.}, dwell, eye blinks, and voice (hum), to select text during reading \cite{DBLP:conf/ismar/MengXL22}.

1 paper discussed the use of turn-by-turn animations and text instructions displayed on AR glasses for \textbf{furniture assembly} \cite{DBLP:conf/vr/YangQCSBLL24}. 1 paper explored the application of AR in \textbf{maintenance}, such as annotating a broken video projector to guide other users in understanding how to operate it \cite{DBLP:conf/vr/QuereMJWW24}. 2 paper focused on \textbf{remote collaboration}, proposing specific methods to quickly guide collaborators to follow an expert's line of sight \cite{DBLP:conf/vr/JingLB22, DBLP:journals/corr/abs-2405-18537}. 
1 paper explores the application of VR-based interactive feedback in \textbf{autonomous driving} \cite{DBLP:conf/chi/ElsharkawyAYAHK24}.
Additionally, 2 papers examined how gestures can be used to quickly \textbf{enter passwords} in VR \cite{DBLP:conf/ismar/SongDK22, DBLP:conf/vr/RuppGBK24}.

To summarize, current research on XR natural interaction techniques has explored 10 types of applications, offering a relatively diverse range of studies. However, these papers account for only 30\% of the reviewed literature, indicating that further exploration is needed on how to apply natural interaction techniques to practical use cases. Additionally, the range of application types should be expanded to include fields such as medicine, industrial training, and education, all of which hold significant potential for impactful applications. \revise{}{Fig. \ref{scene} show the papers with specific scenarios.} Further discussion can be found in Section \ref{4.3}.

\subsection{Operation Types}
\label{operationtypes}
The applications mentioned above are built on specific natural interaction operations. Researchers have developed various operations and explored diverse interaction modalities, resulting in a complex mapping between operation forms and interaction modalities \cite{DBLP:conf/ismar/HertelKSBSS21}. This complexity underscores the importance of identifying similarities among these operations. In this section, we introduce a rational classification that reveals the relationships between different operations. This taxonomy helps to identify the core issues each type of operation must address and suggests future development trends. The paper categorizes XR operations into seven main classes, as explained below.

Before interacting with an object, the initial step is to select it. Even when creating new objects, users must first preselect a location. The ability to quickly, accurately, and stably select desired objects is essential in nearly all scenarios \cite{marquardt2024selection}. This process is categorized in this paper as \textbf{Pointing and Selection}.
XR enhances human capabilities by providing greater control over virtual environments, particularly with respect to object manipulation \cite{DBLP:conf/chi/WangLZ24}. The ability to create objects and modify their properties is a central feature that human-centered XR should offer. This paper classifies these actions as \textbf{Creation and Editing}.
Once objects are selected, users can perform various interactions, depending on the scenario, including controlling spatial position, size, and orientation. These actions are referred to in this paper as translation, scaling, and rotation, with the latter two collectively termed \textit{Transform}. Therefore, these operations are grouped under a single category: \textbf{Translation and Transform}.

The aforementioned categories relate to user interactions with objects. However, users also need to navigate and observe within the virtual environment, which involves movement through space ({Locomotion}) and changes in viewpoint ({Viewport}). This paper combines these activities under the category of \textbf{Locomotion and Viewport}.

With the continued advancement of XR and AI technologies, users are presented with increasingly rich information within virtual environments. This development necessitates more frequent exchanges of information between users and systems \cite{DBLP:journals/imwut/WangSWYYWJXY24}. Users must input task-related information into the system, while the system must accurately interpret user intentions and provide appropriate feedback. Based on recent research, we classify information interaction into two categories: \textbf{Typing and Querying}. The former refers to the primary method of text input, while the latter includes various forms of information retrieval for users.




\begin{figure}[t]
    \centering
    \includegraphics[width=\linewidth]{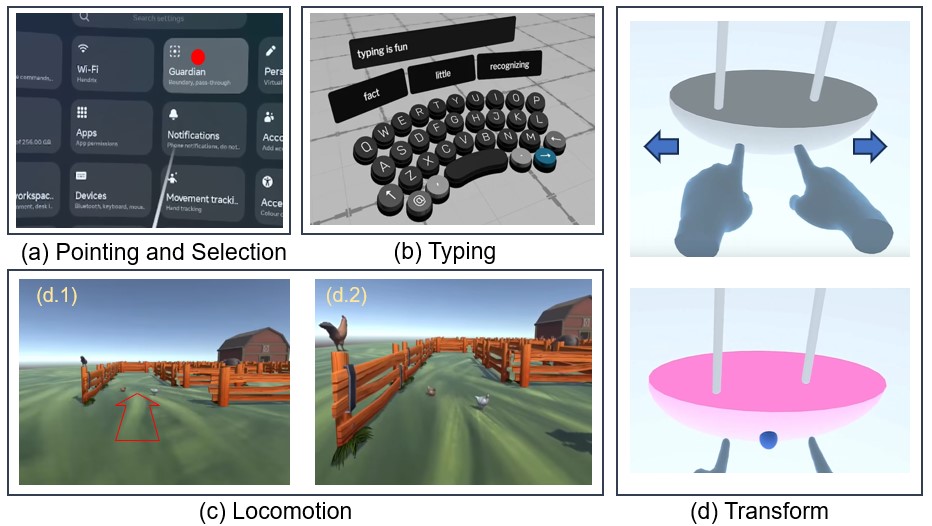}
    \caption{\revise{}{The illustrations of  operation types: (a) pointing and selection; (b) typing; (c) locomotion; (d) transform \cite{DBLP:journals/tvcg/SongDK23}. Image courtesy of \cite{DBLP:journals/tvcg/SongDK23}.}}
    \label{fig:Operation}
\end{figure}

The aforementioned interaction types involve active en-
gagement of users with the environment, where users input
information into the system. We categorize these opera-
tions as Active Interactions. Fig. gives an illuatration of these operations.
Beyond user input, receiving feedback from the XR system is another critical component of natural interaction, as it enhances user immersion. Research that investigates methods for delivering feedback to users is categorized as \textbf{Passive Interactions}.

In our literature review, we also identify a distinct category of research that does not focus on specific operations but rather on improving the recognition accuracy of input modalities. We classify this body of work under the \textbf{No Operation} category.
 
This classification offers a structured framework for understanding and analyzing XR interactions, capturing the core ways in which users engage with and manipulate virtual content. It encompasses both the creation and modification of virtual objects as well as user movement and perception within the XR environment, providing a comprehensive approach to the study of XR operations.

\subsubsection{Pointing and Selection}
Pointing-and-Selection is the most common and fundamental operation in XR, essential for interacting with any object. Selection operations are often influenced by scene factors such as object size, density, and depth distribution \cite{marquardt2024selection}. \revise{In recent researches, 2D menu layouts are prevalent in selection operation.}{Among recent research, 2D menu layouts are prevalent in selection operation.} \revise{These researches focuse on designing task-specific menu layouts for different application contexts to achieve better selection}{These studies focus on designing task-specific menu layouts for different application contexts to obtain better performance of selection} \cite{DBLP:conf/ismar/ChenGFCL23, DBLP:conf/ismar/LeeHM22, DBLP:conf/vr/OrloskyLSSM24, kim2022lattice, DBLP:conf/vr/0001LC00S22, DBLP:conf/vr/QuereMJWW24, 10.1145/3544548.3581423, DBLP:conf/vr/LaiSL24}. \revise{Researches focusing on selection in 3D space primarily addresses high-occlusion scenarios}{Research focusing on selection in the 3D space primarily aims to address the problem of accuracy in high-occlusion scenarios}  \cite{DBLP:conf/chi/SidenmarkCNLPG23, DBLP:conf/chi/WeiSYW0YL23, DBLP:conf/uist/0001QTFLS22}. Some studies also discuss user experience in text selection \cite{DBLP:conf/ismar/LeeHM22, DBLP:conf/ismar/MengXL22} and selection across different depth planes \cite{DBLP:conf/chi/ZhangCSS24, DBLP:conf/uist/0001QTFLS22, DBLP:journals/tvcg/WangZ022}.

\textbf{Performance measures.} These papers share a common set of evaluation metrics. Objective metrics predominate, including \textit{Target Delay}, which describes the time for the pointing modality to align with the target object; \textit{Selection Time}, measuring the total duration from object appearance to selection; \textit{False Positive Rate}, indicating the frequency of false touch occurrences; \textit{Selection Accuracy}, describing the overall accuracy of a selection method; \textit{Error and Abort Rate}, assessing the difficulty of using a particular method; and \textit{Hand or Controller Movement}, quantifying the required hand motion, closely related to arm fatigue. Common subjective metrics primarily evaluate \textit{Cognitive Load} and \textit{Subjective Preferences}.

\textbf{Modalities.} Eye tracking is the most commonly used modality for pointing and selection because the human eye naturally focuses on objects of interest, as demonstrated in Table \ref{tab:table1}. However, gaze estimation methods still have limitations in precision, as sampled gaze points represent a spatial distribution rather than a fixed point due to  tremors or
involuntary saccades \cite{DBLP:conf/chi/SidenmarkCNLPG23}. Pure eye-tracking pointing and selection often incorporate special interface designs or vergence control to address these issues \cite{DBLP:conf/chi/SidenmarkCNLPG23,DBLP:conf/ismar/LeeHM22,DBLP:conf/chi/WeiSYW0YL23,DBLP:conf/vr/OrloskyLSSM24,DBLP:conf/chi/ZhangCSS24,DBLP:conf/chi/ChoiSO22,kim2022lattice,DBLP:conf/vr/0001LC00S22,DBLP:journals/tvcg/WangZ022}. Additionally, introducing secondary modalities for fine-tuning or confirming selections is a common approach \cite{DBLP:journals/tvcg/SidenmarkP0CGWG22,10.1145/3544548.3581423,DBLP:conf/chi/HouNSKBG23,10.1145/3591129,10.1145/3530886}. A few studies explore the reliability of gesture-based selection, typically applied in scenarios where gestures align more closely with human intuition \cite{DBLP:conf/vr/QuereMJWW24,DBLP:conf/chi/SchmitzGS022,DBLP:journals/tvcg/SongDK23}.

\subsubsection{Creation and Editing}
As XR technology advances, allowing users to freely create and modify objects in XR significantly expands their interaction freedom with the environment \cite{DBLP:conf/chi/TorreFHBFL24}. This set of operations primarily involves Creation and Editing operations. The current mainstream Creation mode uses predefined patterns, where users or developers predefine a set of virtual components that can be created by users through specific interaction modalities (\textit{e.g.}, \cite{DBLP:conf/uist/LiaoKJKS22,DBLP:conf/chi/CaoKWAX24,DBLP:conf/vr/QuereMJWW24}), mostly triggered by voice \cite{DBLP:conf/uist/LiaoKJKS22} or gestures \cite{DBLP:conf/chi/CaoKWAX24}. In this mode, designing easily memorable and manageable Creation methods for users is a primary concern\cite{DBLP:conf/chi/PeiCLZ22,DBLP:conf/ismar/ChenGFCL23}. Additionally, some studies explore methods for users to freely create non-predefined objects \cite{DBLP:journals/tvcg/XuZSFY23,DBLP:conf/vr/GiunchiNGS24,DBLP:journals/tvcg/SongDK23,DBLP:conf/chi/PeiCLZ22}. They focus on diversifying creatable content to provide users with more freedom. Editing operations act on these created objects. As virtual elements can take various forms such as text, images, windows, maps, etc., there are multiple editable attributes like size, texture, and color\cite{DBLP:conf/chi/WangYWJ024}.

\textbf{Performance measures.}
The primary concern in Editing operations is whether users can easily remember and use the relevant interaction methods, which are highly application-specific. Common metrics include both objective and subjective measures. Objective metrics include \textit{Completion Time} and \textit{Success Rate}. Subjective metrics mainly cover \textit{Engagement}, \textit{Confidence}, \textit{Expressiveness}, \textit{Learnability}, and \textit{Easiness of Use}.

\textbf{Modalities.}
Hand gestures are the most common modality for Creation. Six degree-of-freedom (DoF) gestures have attracted significant research attention due to their extensive design space and the diversity of creatable objects. Numerous studies have designed intuitive creation gestures for various objects \cite{DBLP:conf/chi/PeiCLZ22,DBLP:conf/ismar/ChenGFCL23,DBLP:journals/tvcg/XuZSFY23,DBLP:conf/chi/0003HLG23}. Given the rich semantics of human language, coupled with advanced NLP technologies, speech interaction can provide more precise information about the objects to be created and their attributes \cite{DBLP:conf/chi/CaoKWAX24,DBLP:conf/uist/LiaoKJKS22}. A few studies have incorporated eye tracking to select creation positions or specify menu items \cite{DBLP:conf/ismar/ChenGFCL23,10.1145/3613904.3642758,DBLP:conf/ismar/MengXL22}.

\subsubsection{Translation and Transform}
This operation corresponds to users' fundamental control capabilities on objects, which include transform, rotation, and scaling. Recent studies have defined richer interactions for precise object manipulation, extending beyond intuitive hand gestures. As XR technology advances, virtual objects have varied, necessitating simultaneous multiple basic operations on single objects. 
Research have focused on enhancing transformation richness and flexibility through customized gestures \cite{DBLP:conf/chi/CaoKWAX24,DBLP:journals/tvcg/XuZSFY23} and intuitive designs for various objects \cite{DBLP:conf/chi/PeiCLZ22,DBLP:journals/tvcg/SongDK23}. Studies also aim to extend user reach for larger-scale transformations in limited spaces \cite{DBLP:journals/ijhci/DengSZK24,DBLP:conf/chi/0003HLG23}, and improve interaction feedback, including haptic responses.

\textbf{Performance measures.}
Evaluation in these works combined objective and subjective metrics. Objective metrics include \textit{Accuracy}, \textit{Speed}, \textit{Stability} (for measuring consistency during jittery movements and across multiple operations), and \textit{Precision} (for positional accuracy). Subjective metrics focus on \textit{Learnability} and \textit{Fatigue}.

\textbf{Modalities.}
This operation typically employ intuitive hand gestures, such as pinch and grab for moving objects \cite{DBLP:conf/vr/QuereMJWW24}, and two-handed pull for scaling \cite{DBLP:conf/chi/CaoKWAX24}. Studies have explored varied implementation details, including customizable gestures \cite{DBLP:conf/chi/CaoKWAX24} and gestures inspired by everyday tool use \cite{DBLP:conf/chi/PeiCLZ22}. Some research examines learning abilities and operational stability for specific gesture-based transformations \cite{DBLP:conf/chi/CaoKWAX24}.
However, gesture detection inevitably involves errors \cite{DBLP:conf/chi/CaoKWAX24}. Some studies incorporate voice input for improved stability \cite{DBLP:conf/chi/CaoKWAX24,DBLP:conf/uist/LiaoKJKS22}. Gaze interaction, being quicker and less fatiguing, has been utilized for object transformation, with gestures often used for fine-tuning and 3DoF rotation control \cite{DBLP:conf/chi/HouNSKBG23}. A few studies have explored hands-free approaches using eye tracking and head-gaze for transformations \cite{DBLP:conf/chi/HouNSKBG23}.

\subsubsection{Locomotion and Viewport}
This operation includes both \textit{Locomotion} and \textit{Viewport}, which are essential for user exploration and movement in virtual environments, significantly influencing user comfort and immersion \cite{DBLP:conf/vr/SinJLLLN24}. Traditional locomotion research has primarily focused on 2D planar movement, with \textit{teleportation} and \textit{steering} being the two most common modes. The former rarely causes discomfort, while the latter offers greater immersion \cite{DBLP:conf/vr/HombeckVHDL23}. Recent studies have extended this focus to 3D space exploration \cite{DBLP:conf/vr/SinJLLLN24,DBLP:conf/vr/SindhupathirajaUDH24}, aiming to improve the overall user experience. In collaborative VR scenarios, enhanced locomotion techniques help users better understand each other's movement intentions, preventing disconnection or separation \cite{DBLP:conf/chi/RaschRS023}. However, traditional controller-based rotation and directed steering can detract from immersion, as they fail to provide natural and continuous turning experiences \cite{DBLP:conf/vr/HombeckVHDL23}.
Improvements to the \textit{Viewport} are equally important for enhancing user comfort and overall experience. One study introduced three gaze-controlled viewport methods that enable hands-free and controller-free interaction \cite{DBLP:conf/chi/LeeWSG24}.

\textbf{Performance measures.} Measures for locomotion primarily utilize subjective metrics, including \textit{Presence}, \textit{Workload}, \textit{Cybersickness}, \textit{Preference}, and \textit{Overall User Experience}, while the main objective metric is \textit{Task Completion Time}. For viewport evaluation, researchers employ similar subjective measures as locomotion, with the addition of \textit{Error Rate} as an objective metric to quantify the ease of view manipulation.

\textbf{Modalities.} In steering mode, except from body-leaning, recent studies have focused on gesture-based movement \cite{DBLP:conf/vr/SinJLLLN24} and gaze-directed locomotion with speed and turning control using gestures\cite{10.1145/3613904.3642147}. Voice commands have also proven to be a viable option for steering control \cite{DBLP:conf/vr/HombeckVHDL23}. In teleportation mode, research has explored voice-based destination matching \cite{DBLP:conf/vr/HombeckVHDL23} and gesture-based alternatives to controllers \cite{DBLP:conf/vr/SindhupathirajaUDH24}. Some studies focus on teleportation in collaborative VR scenarios, addressing the challenge of communicating teleportation intentions between users\cite{DBLP:conf/chi/RaschRS023}.

\subsubsection{Typing and Querying}
This kind of operation includes typing and querying, crucial for inputting information into and retrieving feedback from XR systems. Typing efficiency significantly impacts productivity \cite{DBLP:conf/chi/HeLP22}. Recent XR interaction advancements offer diverse modalities for information retrieval\cite{DBLP:journals/imwut/WangSWYYWJXY24}. These modalities of information and their combinations have been leveraged in recent works\cite{DBLP:journals/imwut/WangSWYYWJXY24, DBLP:conf/chi/0005WBCRF24}.  User immersion in XR querying is enhanced when input methods closely resemble everyday communication, requiring systems to discern user intent \cite{DBLP:conf/chi/0005WBCRF24}.

\textbf{Performance measures.} Evaluation for typing primarily use objective metrics such as \textit{Words per Minute (WPM)}, \textit{Error Rate}, \textit{Deletion Count}, and \textit{Prediction Count}. Subjective metrics include cognitive, visual, and hand fatigue levels. Evaluation of Querying employs both subjective metrics (\textit{Simplicity}, \textit{Naturalness}, \textit{Human-likeness}, \textit{Personal Preferences}) and objective metrics (\textit{Task Time}, \textit{Times of Attempts}).

\textbf{Modalities.} Recent studies often retain traditional keyboard layouts while incorporating new modalities to improve speed and experience \cite{DBLP:conf/iui/ZhaoPTZWJBG23,DBLP:conf/chi/HeLP22,DBLP:journals/tvcg/ShenDK23,DBLP:conf/ismar/SongDK22,DBLP:conf/chi/Hedeshy0MS21,10474330}. Gaze-tracking is used to predict characters of interest, accelerating cursor movement or providing visual cues\cite{DBLP:conf/iui/ZhaoPTZWJBG23,10474330}. He \textit{et al.} proposed a gaze-selecting method combined with language models for error-tolerant blind typing \cite{DBLP:conf/chi/HeLP22}. Some studies achieve faster input speeds compared to midair-tapping using pure gaze input\cite{cui2023glancewriter,DBLP:conf/vr/HuDK24}. Gestures are used for keyboard control \cite{DBLP:conf/ismar/SongDK22}, and new 3D decoding techniques enable precise interpretation of aerial gestures \cite{DBLP:journals/tvcg/ShenDK23}. For querying, speech-based interaction is fundamental, with lip-reading and other auxiliary methods improving robustness \cite{DBLP:conf/chi/ZhangLHWLGZ23,DBLP:journals/tvcg/CaiML24}. Eye-tracking data and gesture information can also help resolve pronoun ambiguities in speech \cite{DBLP:conf/chi/0005WBCRF24,DBLP:journals/imwut/WangSWYYWJXY24}. One study enables gaze-only querying through an AR interface design \cite{DBLP:conf/uist/LiaoKJKS22}.

\subsubsection{No Operation Type}

In our literature review, we identified nine studies that did not explicitly focus on specific interaction types. Instead, these studies provided optimizations aimed at improving the recognition accuracy of input modalities. These studies primarily aimed to enhance the system's ability to accurately interpret and respond to users' active interaction intents. Since these papers are not discussed elsewhere in this work, we provide a detailed description here.

The majority of these studies centered around gesture and touch-based interactions. Xu \textit{et al.} proposed a fine-grained hand gesture detection method that leverages both visual and auditory sensors, enabling users to trigger interactions with smaller movements \cite{DBLP:conf/chi/XuZKN23}. Kitamura \textit{et al.} developed a contact-based wearable device capable of accurately recognizing micro-gestures, including continuous gesture changes and pressure inputs \cite{DBLP:conf/iswc/KitamuraYS23}. Lee \textit{et al.} introduced a wrist-worn device that employs active acoustics to continuously capture hand movements and interactions with objects \cite{DBLP:conf/chi/LeeZAYGLKYDLSGZ24}. Liu \textit{et al.} designed a gesture recognition method using audio signals, enabling the recognition of various hand gestures on different material surfaces, thus expanding the possibilities of gesture interaction\cite{10522613}. Li \textit{et al.} developed a finger-ring-based micro-gesture recognition device equipped with a miniature camera, allowing for gesture interaction on various surfaces \cite{DBLP:conf/ismar/LiLMHLS22}. Rupp \textit{et al.} proposed a set of gesture authentication schemes that achieve the same entropy as PIN codes \cite{DBLP:conf/vr/RuppGBK24}. Shen \textit{et al.} introduced a Key Gesture Spotting architecture to assist developers in rapidly developing gesture recognition systems, while simultaneously reducing gesture detection latency for a better user experience.

1 other study leverage visual cues for preventing user
 \cite{DBLP:journals/tvcg/ShenDMK22}. Wang \textit{et al.} designed a visual guidance mechanism for bare-hand interaction to help users perform interaction actions more consistently, reducing recognition errors \cite{DBLP:conf/chi/WangLZ24}. Cai \textit{et al.} developed a dual-mode keyword detection system using both speech and echo signals, enabling accurate keyword recognition in a wider range of environments 
 \cite{DBLP:journals/tvcg/CaiML24}.

\subsubsection{Passive Interaction}
\begin{figure}[t]
    \centering
    \includegraphics[width=0.85\linewidth]{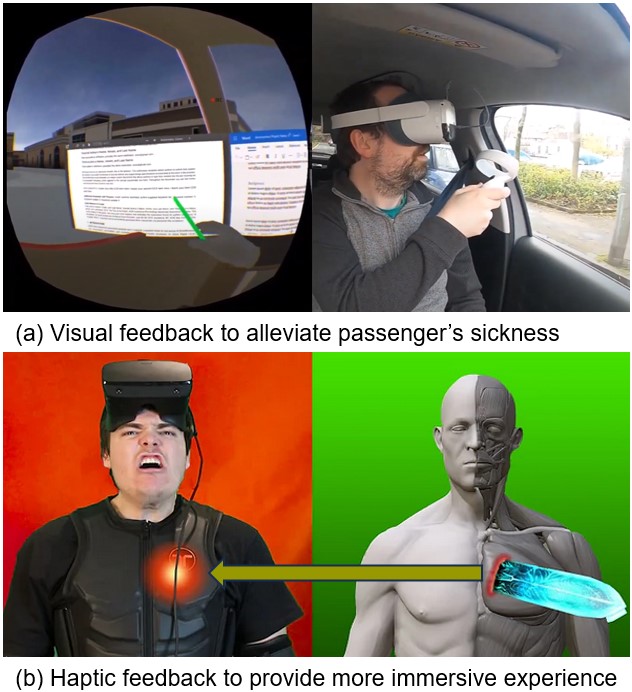}
    \caption{\revise{}{The illustrations of passive interaction: (a) visual feedback \cite{DBLP:conf/chi/Pohlmann0MMB23}; (b) haptic feedback. Image courtesy of \cite{DBLP:conf/chi/Pohlmann0MMB23}}.}
    \label{img:passive}
\end{figure}

In addition to the previously mentioned {Active Operations} and {No Operation}, we also examine {Passive Interactions}. In recent years, many studies have focused on this topic. Based on the types of feedback provided by XR environments, we categorize these studies into four groups: \textit{Visual}, \textit{Acoustic}, \textit{Haptic}, and \textit{Hybrid}. For more detailed information on this classification, please refer to Table \ref{img:passive}. Fig \ref{img:passive} illustrates passive interactions.

\textbf{Performance measures}.
Passive Interaction primarily focuses on users' perception of the environment, making subjective metrics the primary evaluation criteria. Data in various studies is typically collected through questionnaires designed by the authors. The most commonly assessed indicators include \textit{Realism}, \textit{Immersion}, and \textit{Confidence}. In haptic-related research, \textit{Visuo-Haptic Match} is also a key metric \cite{DBLP:conf/chi/ShenS022}. Additionally, some studies use task completion scores in XR environments as an evaluation method \cite{DBLP:conf/vr/LiLYTFX24,DBLP:conf/chi/Pohlmann0MMB23}.

\textbf{Modalities}.
The acoustic modality is not observed in isolation in the collected articles. It is typically studied in conjunction with other modalities. Visual feedback represents one of the primary feedback modalities. The impact of singular visual feedback on users' self-position perception during teleportation has been investigated \cite{DBLP:conf/chi/MedlarLG24}. Visual guidance mechanisms are employed in 3 articles to enhance communication efficiency among multiple users in XR environments \cite{10462901,DBLP:conf/vr/WangZF24,DBLP:conf/chi/RaschRS023}. 2 articles dynamically alter environmental parameters based on users' task stages in the XR environment to improve task completion efficiency \cite{DBLP:conf/chi/WuQQCRS24,DBLP:conf/vr/LiLYTFX24}. 
3 papers design feedback in VR environments based on real-world distractions, aiming to reduce the impact of distractions on user immersion and comfort \cite{DBLP:conf/chi/ElsharkawyAYAHK24,DBLP:conf/uist/Tao022,DBLP:conf/chi/Pohlmann0MMB23}.

The haptic modality has garnered significant attention from researchers in recent years as a passive interaction modality. 5 articles utilize auxiliary wearable hardware to provide additional tactile feedback, enabling users to experience a wider range of sensations in XR\cite{DBLP:conf/chi/ShenS022,DBLP:conf/vr/YamazakiH23,DBLP:conf/uist/JinguWS23,DBLP:conf/chi/TatzgernDWCEDGH22,DBLP:conf/uist/ShenRM0S23}, such as wind\cite{DBLP:conf/chi/ShenS022}. Visuo-Haptic illusion with Proxies is examined in 2 articles, offering users perception of objects' size, weight, and motion trajectories \cite{DBLP:conf/chi/FeickR0K22,DBLP:conf/chi/0001OPSB24}.

\subsection{Interaction Techniques}
\label{Interaction techniques}

The aforementioned operations are supported by specific interaction modalities. This section focuses on the hardware and algorithmic implementations of these modalities, as well as the design of concrete interaction techniques. Section \ref{GestureOnly} discusses the hardware and software implementations of gesture-only interactions, provides a summary of existing research, and analyzes the strengths and limitations of this modality. Section \ref{GazeOnly} covers the same aspects for gaze-only interactions, while Section \ref{SpeechOnly} addresses speech-only interactions. Section \ref{Tactile} explores tactile interaction in a similar manner. Section \ref{MultiInteraction} focuses on multimodal interaction techniques, specifically covering \textit{Gaze + Gesture}, \textit{Gaze + Speech}, \textit{Gesture + Speech}, and a triple-modal technique (\textit{Gaze + Gesture + Speech}). Lastly, we discuss various \textit{X + Y} interactions, which compare different multimodal interaction techniques in these studies.

\begin{table}[t]
\centering
\renewcommand\arraystretch{1.2}
\caption{\revise{}{Comparison of hand tracking performance across devices}}
\begin{threeparttable} 
\begin{tabular}{
     >{\centering\arraybackslash}m{2.8cm} 
     >{\centering\arraybackslash}m{2cm} 
     >{\centering\arraybackslash}m{1.2cm} 
     >{\centering\arraybackslash}m{1cm} 
}
\hline
\textbf{Device Name}  & \textbf{Accuracy} & \textbf{Latency} & \textbf{Sampling Rate}  \\ \hline
Microsoft HoloLens 2  & around 15 mm & -\tnote{1} & -  \\ \hline
Meta Quest 2           & around 11 mm & 45 ms & 60 Hz  \\ \hline
HTC Vive Pro              & around 37 mm & - & -  \\ \hline
Lee \textit{et al.}, 2024 \cite{DBLP:conf/chi/LeeZAYGLKYDLSGZ24} & 4.81 mm & 500 ms & - \\ \hline
\end{tabular}

\begin{tablenotes}
\item[1] \revise{}{- indicates that no reports are found regarding this item.}
\end{tablenotes}

\end{threeparttable}
\label{tab:hand_tracking_comparison}
\end{table}

\subsubsection{Gesture Only}
\label{GestureOnly}

We begin by introducing the implementation methods for hand gestures.
\textbf{Hardware}. Currently, hand tracking mainly relies on sensors integrated into VR/AR HMDs, primarily utilizing infrared (IR) cameras. The latest devices, such as the Meta Quest 3, additionally use two RGB cameras for enhanced tracking \cite{Meta}. 
\revise{}{The hand-tracking performance of XR devices used in recent research is summarized in Table \ref{tab:hand_tracking_comparison}. Among these, the HTC Vive Pro shows lower accuracy compared to the HoloLens 2 and Meta Quest 2 \cite{10.1145/3485279.3485283}. Although the latency and sampling rate of the HoloLens 2 and HTC Vive Pro have not been documented, users have not reported experiencing any noticeable delays during use. Typically, users rely on the HTC Vive Pro's hand controller for gesture-based interactions.}
\revise{2 paper explores low-power sensing modules, such as speakers and microphones, to detect hand-reflected sound waves as an alternative to image-based tracking \cite{DBLP:conf/chi/LeeZAYGLKYDLSGZ24, 10522613}.}{2 papers \cite{DBLP:conf/chi/LeeZAYGLKYDLSGZ24, 10522613} explore low-power sensing modules, such as speakers and microphones, to detect hand-reflected sound waves as an alternative to image-based tracking.}
\textbf{Algorithm}. Hand tracking in VR/AR HMDs typically employs computer vision techniques to output hand skeleton data \cite{cai2020generalizing, cai2018desktop}. 
\revise{4 papers aimed to classify a larger variety of gestures (\textit{e.g}., 10 or more) using neural networks or machine learning \cite{DBLP:journals/tvcg/SongDK23, DBLP:conf/chi/PeiCLZ22, DBLP:journals/tvcg/ShenDMK22,DBLP:conf/chi/LeeZAYGLKYDLSGZ24}.}{4 papers \cite{DBLP:journals/tvcg/SongDK23, DBLP:conf/chi/PeiCLZ22, DBLP:journals/tvcg/ShenDMK22,DBLP:conf/chi/LeeZAYGLKYDLSGZ24} aim to classify a larger variety of gestures (\textit{e.g}., 10 or more) using neural networks or machine learning.} 
\revise{Other studies focused on simpler gestures, like pinching or clicking, which require only distance detection between the index finger and thumb.}{Other studies focus on simpler gestures like pinching or clicking, which only require distance detection between the index finger and thumb.}

\begin{figure}[t]
    \centering
    \includegraphics[width=1\linewidth]{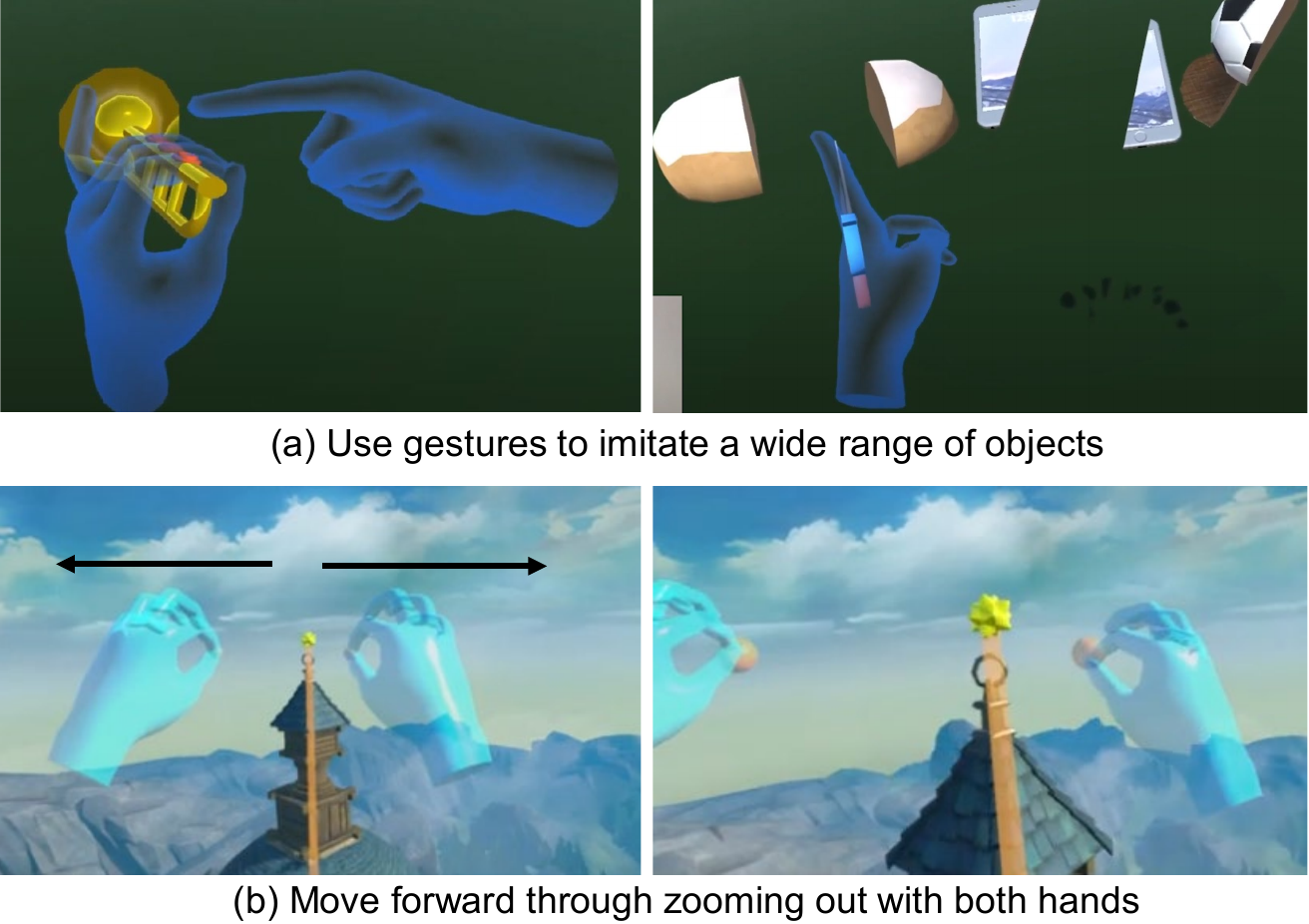}
    \caption{\revise{}{ The illustrations of hand gesture only interaction. Images courtesy of \cite{DBLP:conf/chi/PeiCLZ22, DBLP:conf/vr/SinJLLLN24}.}}
    \label{Hand}
\end{figure}

\textbf{Summary of current research}. In the reviewed literature, gesture only interaction is the most frequently studied modality for spatial computing, \revise{with 24 papers focused on this topic (see Fig. \ref{fig:bymodality})}{with 24 papers focusing on this topic, shown in Fig. \ref{fig:bymodality}}. 
\revise{Hands are a natural and intuitive interface for interacting with objects in daily life.}{The hands are natural and intuitive interfaces for interacting with objects in daily life.} For example, 20 papers explored using hand gestures for object selection or translation. Hand movements in 6DoF have been used to create complex 3D scenes \cite{DBLP:journals/tvcg/XuZSFY23, 10.1145/3491102.3517682} and for VR locomotion \cite{DBLP:conf/vr/SinJLLLN24, DBLP:conf/vr/SindhupathirajaUDH24}. Gestures also enable fine-grained control of small objects. 3 papers investigated text input in VR, either via direct virtual keyboard interactions \cite{DBLP:conf/vr/RuppGBK24, DBLP:journals/tvcg/ShenDK23} or gesture-to-text conversions \cite{DBLP:conf/ismar/SongDK22}.
Gesture transformations can convey complex semantic information. For instance, 2 papers used 10 or more gestures to imitate a wide range of objects, such as a telescope, scissors, or camera, \textit{etc} \cite{DBLP:conf/chi/PeiCLZ22, DBLP:journals/tvcg/SongDK23}. Additionally, 3 papers explored hand redirection techniques, leveraging the perceptual phenomenon of change blindness in hand and arm movements \cite{DBLP:conf/vr/MatthewsTIS22, DBLP:conf/chi/ZennerKFAK24, DBLP:conf/ismar/BanMNK22}. This allows users to operate within a limited physical space while enabling broader interactions in a virtual environment.

\textbf{Advantages and disadvantages of hand gesture only interaction}. Hand-gesture only interaction offers several advantages. It is natural and intuitive, as gestures align with everyday interactions. Gestures also allow for complex semantic representation and providing fine-grained control for precise manipulation of small objects. However, there are notable drawbacks. 
For the use of complex gestures, the cognitive load increases as users need considerable time to master them \cite{DBLP:conf/chi/PeiCLZ22}. 
Gesture recognition accuracy still requires improvement, particularly in cases of occlusion, as mentioned in 2 papers \cite{DBLP:conf/chi/PeiCLZ22, DBLP:journals/tvcg/SongDK23}. Additionally, prolonged use of gestures can lead to arm fatigue \cite{DBLP:conf/vr/SindhupathirajaUDH24, DBLP:conf/vr/QuereMJWW24}, and the lack of tangible support means users do not receive physical feedback \cite{DBLP:conf/chi/SchmitzGS022, DBLP:journals/tvcg/ShenDK23}. 
Social acceptance can also pose a challenge, as obvious hand gesture interaction may not be appropriate in public \cite{DBLP:conf/ismar/LiLMHLS22}. 
To summarize, hand gesture interactions often require users to adapt to complex paradigms defined by researchers, which increases the cognitive load. Future developments should focus on improving ease of use, making hand gesture interaction more user-friendly for novices, and minimizing tracking accuracy requirements.

\subsubsection{Gaze Only}
\label{GazeOnly}

Gaze-based interaction is a key focus in extended reality, leveraging humans' intuitive eye movements. Eye gaze rapidly reaches objects of interest, making it ideal for pointing and selection tasks \cite{cheng2024appearance}. Gaze-only methods eliminate the need for body movements, benefiting users in various situations, including those with disabilities or in socially awkward scenarios \cite{DBLP:conf/chi/LeeWSG24}.

\begin{table}[t]
\centering
\renewcommand\arraystretch{1.2}
\caption{\revise{}{Comparison of eye-tracking performance across XR devices}}
\begin{threeparttable} 
\begin{tabular}{
     >{\centering\arraybackslash}m{1.7cm} 
     >{\centering\arraybackslash}m{1.2cm} 
     >{\centering\arraybackslash}m{1.3cm} 
     >{\centering\arraybackslash}m{1cm} 
     >{\centering\arraybackslash}m{1.4cm} 
}
\hline
\textbf{Device Name}  & \textbf{Accuracy} & \textbf{Latency} & \textbf{Sampling Rate} & \textbf{Slippage-Robust} \\ \hline
Microsoft HoloLens 2  & $1.5^\circ$ & - & 30 Hz & \checkmark \\ \hline
HTC Vive Pro Eye           & $0.5^\circ$-$1.1^\circ$ & 50 ms & 120 Hz & \checkmark \\ \hline
Meta Quest Pro              & $1.6^\circ$ & 58 ms & 90 Hz & \checkmark \\ \hline
Microsoft HoloLens 1\tnote{1}  & $1^\circ$ & 8.5 ms & 120 Hz & $\times$ \\ \hline
\end{tabular}

\begin{tablenotes}
\item[1] \revise{}{Microsoft HoloLens 1 is integrated with Pupil Labs' eye tracker.}
\end{tablenotes}

\end{threeparttable}
\label{tab:eye_tracking_comparison}
\end{table}

\textbf{Hardware and algorithm.} There are two main eye-tracking strategies in XR environments \cite{DBLP:conf/ismar/WangZLL21}. The first is the Pupil Center Corneal Reflection (PCCR) method \cite{DBLP:journals/tbe/GuestrinE06, DBLP:journals/corr/abs-2003-08806}. This approach uses 1-2 near-infrared cameras positioned close to the eye, along with multiple near-infrared light sources (typically 6-12) directed at the eye. Based on the corneal reflection patterns, the system reconstructs an accurate eye model. The advantage of this method lies in its high precision and its ability to support slippage detection and compensation for device slippage \cite{DBLP:conf/etra/SantiniNK19}. However, it has drawbacks, including higher hardware costs and a more complex process for calibrating the optical positions of the light sources and cameras \cite{6949395}.
\revise{}{
Currently, commercial XR devices such as Microsoft HoloLens 2, HTC Vive Pro Eye, and Meta Quest Pro utilize the PCCR-based eye-tracking approach, as shown in Table \ref{tab:eye_tracking_comparison}. These devices typically achieve eye-tracking accuracy within the range of 0.5°–1.6°. Furthermore, because they are robust against device slippage, the differences in eye-tracking performance during interaction can be negligible. The latency falls within an acceptable range for interaction purposes, and a sampling rate of 30 Hz is generally sufficient for gaze-based interaction. Overall, the eye-tracking performance of these XR devices is comparable.}

The second method is glint-free, which relies on temporal pupil information rather than corneal reflections \cite{DBLP:conf/etra/DierkesKB19, DBLP:conf/chi/KytoEPLB18}. This technique typically uses a single camera and a single light source primarily for illumination. By analyzing pupil data over multiple frames, it reconstructs the eye model. The advantage of this method is its simplicity and lower hardware requirements, although the accuracy of the eye model reconstruction is generally lower compared to PCCR \cite{DBLP:journals/tvcg/WangZ022}.
\revise{}{Early implementations, such as the Microsoft HoloLens 1 integrated with Pupil Labs' eye-tracking system \cite{DBLP:conf/chi/KytoEPLB18}, used this glint-free approach. While its accuracy was comparable to the PCCR method under ideal conditions, it was highly sensitive to device slippage, resulting in a rapid deterioration of gaze accuracy over time.}


\textbf{Summary of current research.} Gaze-only interactions with virtual user interfaces (UIs) often encounter \textbf{the Midas touch problem} due to lack of confirmation modalities. 
To address this, researchers have explored peripheral vision areas and auxiliary UIs. 
\revise{Choi \textit{et al.} proposed the Kuiper Belt concept \cite{DBLP:conf/chi/ChoiSO22}, while Yi \textit{et al.} studied optimal virtual menu layouts \cite{DBLP:conf/vr/0001LC00S22}.}{Choi \textit{et al.} \cite{DBLP:conf/chi/ChoiSO22} proposed the Kuiper Belt concept, while Yi \textit{et al.} \cite{DBLP:conf/vr/0001LC00S22} studied optimal virtual menu layouts.} Orlosky \textit{et al.} \cite{DBLP:conf/vr/OrloskyLSSM24} and Kim \textit{et al.} \cite{kim2022lattice} designed auxiliary interfaces to enhance pure gaze selection operations.
Besides, several techniques have been developed for \textbf{heavily-occluded scenarios}. 
\revise{Sidenmark \textit{et al.} matched object depth motion with gaze vergence changes \cite{DBLP:conf/chi/SidenmarkCNLPG23}, while Wei \textit{et al.} used probabilistic models based on head and gaze endpoints \cite{DBLP:conf/chi/WeiSYW0YL23}. Yi \textit{et al.} combined planar and depth information analysis \cite{DBLP:conf/uist/0001QTFLS22}.}{Sidenmark \textit{et al.} \cite{DBLP:conf/chi/SidenmarkCNLPG23} matched object depth motion with gaze vergence changes, while Wei \textit{et al.} \cite{DBLP:conf/chi/WeiSYW0YL23} used probabilistic models based on head and gaze endpoints. Yi \textit{et al.} \cite{DBLP:conf/uist/0001QTFLS22} combined planar and depth information analysis.}
Vergence estimation technology has introduced \textbf{vergence control} as a novel gaze-only interaction mode. 
\revise{Zhang \textit{et al.} designed visual depth control methods for object selection \cite{DBLP:conf/chi/ZhangCSS24}, while Wang \textit{et al.} developed depth control schemes for see-through vision.}{Zhang \textit{et al.} \cite{DBLP:conf/chi/ZhangCSS24} designed visual depth control methods for object selection, while Wang \textit{et al.} \cite{DBLP:journals/tvcg/WangZ022} developed depth control schemes for see-through vision.}
In \textbf{typing} applications, pure eye movement input methods have shown promising results. 
\revise{Cui \textit{et al.} designed a word prediction algorithm based on eye movement trajectories \cite{cui2023glancewriter}.}{Cui \textit{et al.} \cite{cui2023glancewriter} designed a word prediction algorithm based on eye movement trajectories.} Similar systems integrated with a LLM were also proposed by Hu \textit{et al.} \cite{DBLP:conf/vr/HuDK24}.

\begin{figure}[t]
    \centering
    \includegraphics[width=1\linewidth]{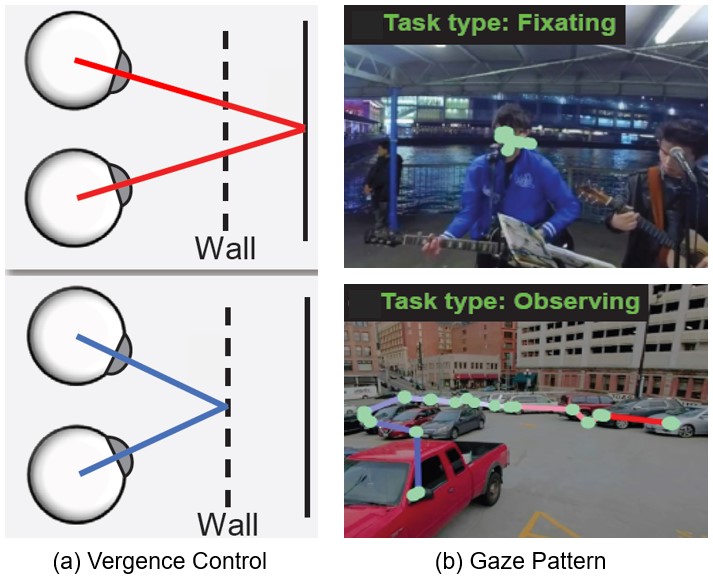}
    \caption{\revise{}{The illustrations of gaze only interaction: (a) gaze vergence control \cite{DBLP:journals/tvcg/WangZ022}; (b) gaze patterns \cite{wang2024tasks}. Images courtesy of \cite{DBLP:journals/tvcg/WangZ022, wang2024tasks}}.}
    \label{Gaze}
\end{figure}

Gaze-based interactions have also been applied to \textbf{enhance user experience} in various contexts.
\revise{Chen \textit{et al.} explored activating hidden objects in virtual films \cite{DBLP:conf/ismar/ChenHTHH23}, Turkmen \textit{et al.} investigated gaze-activated auxiliary grids in virtual sketching \cite{DBLP:conf/chi/TurkmenGBSASPM24}, and Lee \textit{et al.} introduced gaze-activated magnification in VR reading \cite{DBLP:conf/ismar/LeeHM22}.}{Chen \textit{et al.} \cite{DBLP:conf/ismar/ChenHTHH23} explored activating hidden objects in virtual films, Turkmen \textit{et al.} \cite{DBLP:conf/chi/TurkmenGBSASPM24} investigated gaze-activated auxiliary grids in virtual sketching, and Lee \textit{et al.} \cite{DBLP:conf/ismar/LeeHM22} introduced gaze-activated magnification in VR reading.}
Additional applications include \textbf{perspective switching control} using different eye movement modes\cite{DBLP:conf/chi/LeeWSG24} and text operations such as Selection-and-Snap and Gaze Scroll (Lee \textit{et al.} \cite{DBLP:conf/ismar/LeeHM22}). These diverse applications demonstrate the potential of gaze-based interactions to significantly enhance user experiences across various XR scenarios. \revise{}{Fig. \ref{Gaze} illustrates gaze-only interactions.}

\textbf{Advantages and disadvantages of gaze-only interaction.}
The main advantages of eye movement interaction include speed, ease of use, and intuitiveness. Its maximum utility is demonstrated when large-scale physical movements are impossible or when both hands are occupied with tasks. Furthermore, eye movements possess dual selection capabilities in both 2D planes and depth directions. These capabilities can be combined to achieve more diverse interactions. Richer semantic information can also be extracted from human eye movement patterns. Pure eye movement modality has the potential to enable more powerful interaction functions.
In eye movement interaction, the human eye serves as both the medium for initiating interactive behaviors and the primary sensory organ for observation in XR. Consequently, the Midas touch problem persists. This issue remains a primary consideration for researchers in subsequent studies. Similarly, many eye movement interactions involve the design of new interaction interfaces and visual cues to assist users in completing interactions. The visual disturbances and cognitive load imposed on users by these new interfaces are significant issues that cannot be overlooked.

\subsubsection{Speech Only} 
\label{SpeechOnly}

\begin{figure}[t]
    \begin{center}
    \includegraphics[width=1\linewidth]{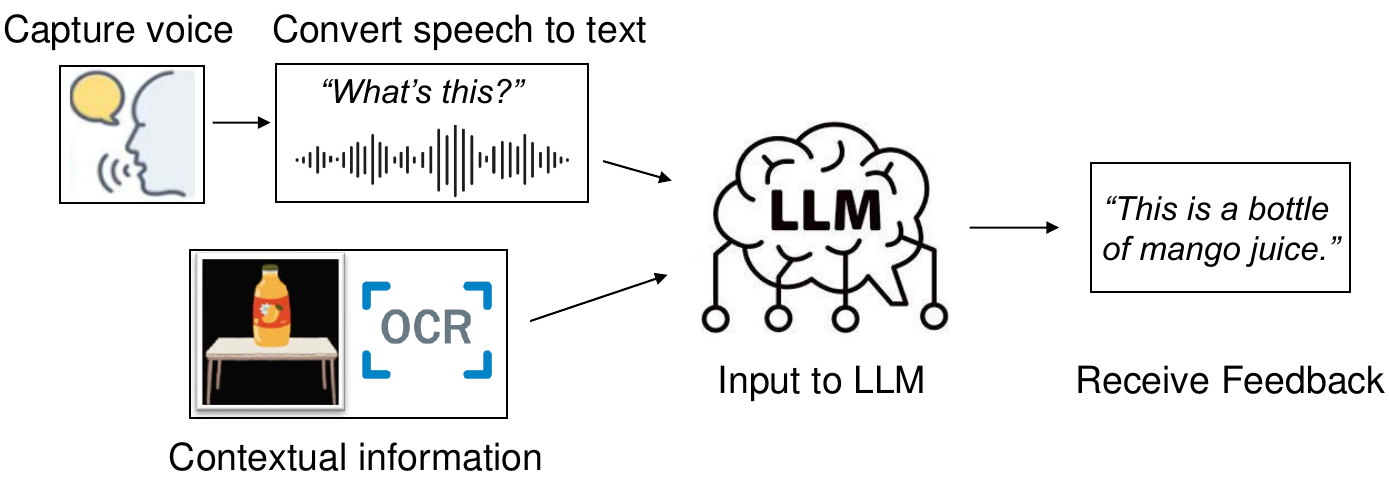}
    \end{center}
    \caption{
        \revise{}{A common process in LLM-based speech interactions. The user's voice is captured and transcribed into text, which is then combined with contextual information to form prompts for the LLM. The LLM's feedback is subsequently provided to the user or processed further by the system.}
    }
    \label{fig:LLMSpeech}
\end{figure}

\textbf{Hardware and algorithm.} Speech-based interactions in XR environments are typically enabled by a single microphone. While some researchers utilize an external microphone \cite{DBLP:conf/vr/SaintAubertAMPAL23, DBLP:conf/chi/Hedeshy0MS21}, the majority depend on the built-in microphone of VR/AR HMDs, such as Hololens 2 \cite{DBLP:conf/ismar/ChenGFCL23}, HTC Vive Pro Eye \cite{DBLP:conf/vr/JingLB22, DBLP:conf/ismar/MengXL22}, and Valve Index \cite{DBLP:conf/vr/HombeckVHDL23}. For speech recognition, most researchers rely on established speech-to-text systems or APIs, such as Windows DictationRecognizer \cite{DBLP:conf/vr/JingLB22} and WebSpeech \cite{DBLP:conf/uist/LiaoKJKS22}. To summarize, these devices and methods have been well established in recent years.

In addition to traditional method above, there is a growing interest in \textbf{silent speech recognition}, which enables speech detection without audible vocalization. This approach often involves additional devices, such as those used in active acoustic sensing.
\revise{For example, EchoSpeech leverages inaudible echo wave that emitted by two speakers and received by two microphones, mounted on a glass-frame \cite{DBLP:conf/chi/ZhangLHWLGZ23}.}{For example, EchoSpeech\cite{DBLP:conf/chi/ZhangLHWLGZ23} leverages inaudible echo wave that emitted by two speakers and received by two microphones, mounted on a glass-frame.} 
\revise{Cai \textit{et al.} further enhanced this by combining both vocal and echoic modalities for more robust speech recognition across various scenarios \cite{DBLP:journals/tvcg/CaiML24}.}{Cai \textit{et al.} \cite{DBLP:journals/tvcg/CaiML24} further enhanced this by combining both vocal and echoic modalities for more robust speech recognition across various scenarios.}
Depth cameras have also been explored for silent speech recognition. 
\revise{Wang \textit{et al.} used TrueDepth camera at three different locations to capture lip movement depth data, enabling silent speech recognition via point cloud video analysis \cite{DBLP:conf/chi/WangSRZ24}.}{Wang \textit{et al.}\cite{DBLP:conf/chi/WangSRZ24} used TrueDepth camera at three different locations to capture lip movement depth data, enabling silent speech recognition via point cloud video analysis.}

\textbf{Summery of current research.} Recently, speech-based interactions can be generally classified into two categories: keyword-based and LLM-based. 
\textbf{Keyword-based} speech interactions identify keywords provided by users as interaction cues or commands, typically including pronouns (\textit{e.g.}, ``this" and ``here") \cite{DBLP:conf/vr/JingLB22}, commands (\textit{e.g.}, ``teleport" and ``select") \cite{DBLP:conf/vr/HombeckVHDL23, DBLP:conf/ismar/ChenGFCL23}, and user-defined keywords \cite{DBLP:conf/uist/LiaoKJKS22, DBLP:conf/chi/CaoKWAX24}. For instance, Hombeck \textit{et al.} \cite{DBLP:conf/vr/HombeckVHDL23} introduced three speech-only interaction techniques for locomotion, by keywords of direction (\textit{e.g.}, ``\textit{left}"), landmark (\textit{e.g.}, ``jump to \textit{bed}") or grid number (\textit{e.g.}, ``teleport to \textit{fifteen}"). However, keyword-based interaction imposes significant limitations on the vocabulary available for interaction \cite{DBLP:conf/vr/HombeckVHDL23}.

In contrast, \textbf{LLM-based} speech interactions overcome this restriction. With the increasing sophistication of large language models (LLMs), more researchers are leveraging LLMs as intelligent agents for interaction, allowing users to issue commands without vocabulary constraints. 
\revise{In this approach, the entire speech input is transcribed into text, which is then combined with prompt words or other contextual information as input to the LLM for further inference\cite{DBLP:conf/chi/WangYWJ024, DBLP:conf/vr/GiunchiNGS24, DBLP:conf/chi/0005WBCRF24, DBLP:conf/chi/TorreFHBFL24}.}{In this approach, the entire speech input is transcribed into text, which is then combined with prompt words or other contextual information as input to the LLM for further inference\cite{DBLP:conf/chi/WangYWJ024, DBLP:conf/vr/GiunchiNGS24, DBLP:conf/chi/0005WBCRF24, DBLP:conf/chi/TorreFHBFL24}, as shown in Fig. \ref{fig:LLMSpeech}.} 
\revise{DreamCodeVR exemplifies the use of LLMs to translate spoken language into code, enabling users to modify the behaviour of a running VR application irrespective of their programming skills \cite{DBLP:conf/vr/GiunchiNGS24}.}{DreamCodeVR \cite{DBLP:conf/vr/GiunchiNGS24} exemplifies the use of LLMs to translate spoken language into code, enabling users to modify the behaviour of a running VR application irrespective of their programming skills.} LLM-based speech interactions have significantly expanded the scope of operations and applications of speech-based interaction.

\textbf{Advantages and disadvantages of speech only interaction.} Speech only interaction is widely acknowledged for its minimal physical effort, as speaking in natural language requires little exertion\cite{DBLP:conf/vr/HombeckVHDL23}. 
\revise{Moreover, speech interactions for discrete selection are robust and require no physical movements or gestures \cite{DBLP:conf/ismar/ChenGFCL23}.For text input, speech aided by speech-to-text technology is much quicker and more intuitive than keyboard input \cite{DBLP:conf/vr/HombeckVHDL23}.}{Moreover, speech interactions for discrete selection are robust and do not require physical movements or gestures \cite{DBLP:conf/ismar/ChenGFCL23}. For text input task, speech input aided by speech-to-text technology is much quicker and more intuitive than keyboard input \cite{DBLP:conf/vr/HombeckVHDL23}.} However, speech-only interaction has notable drawbacks.\revise{The most frequently cited weaknesses are latency and inaccuracy.}{The weaknesses cited most frequently are high latency and inaccuracy.} 
\revise{Several studies reported that delays or recognition errors have impacted the user experience during the interaction\cite{DBLP:conf/vr/HombeckVHDL23, DBLP:conf/vr/SaintAubertAMPAL23, DBLP:conf/vr/JingLB22, DBLP:conf/ismar/MengXL22, DBLP:conf/uist/LiaoKJKS22}.}{Several studies\cite{DBLP:conf/vr/HombeckVHDL23, DBLP:conf/vr/SaintAubertAMPAL23, DBLP:conf/vr/JingLB22, DBLP:conf/ismar/MengXL22, DBLP:conf/uist/LiaoKJKS22} reported that delays or recognition errors have impacted the user experience during interaction.} Additionally, speech interactions, especially keyword-based interactions, often increase cognitive load and raises learning curve due to the requirement of memorizing specific commands \cite{DBLP:conf/ismar/ChenGFCL23}. Furthermore, speech interactions lack subtlety and may face social acceptance challenges \cite{DBLP:conf/chi/WangYWJ024, DBLP:conf/ismar/MengXL22}. Fortunately, recent advancements mentioned above in silent speech recognition offer potential solutions of this problem.


\begin{figure}
    \begin{center}
    \includegraphics[width=1\linewidth]{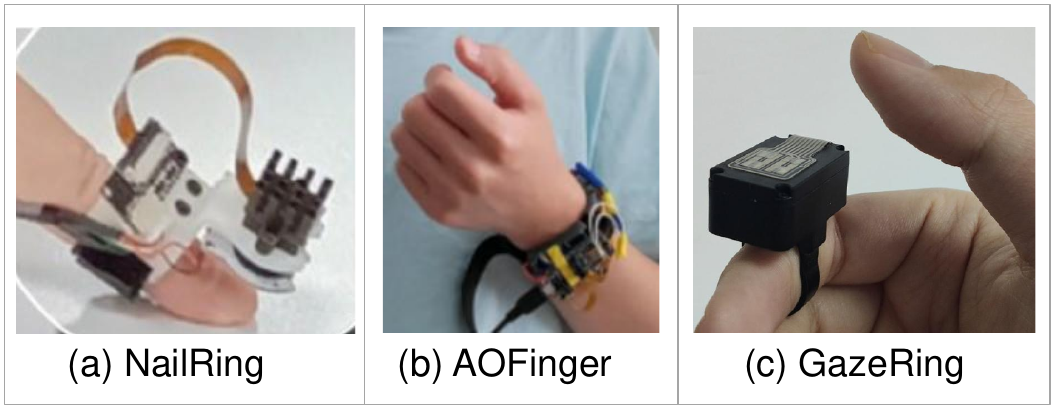}
    \end{center}
    \caption{
        \revise{}{Different devices and locations of tactile interactions: (a) NailRing \cite{DBLP:conf/ismar/LiLMHLS22}, (b) AO-Finger \cite{DBLP:conf/chi/XuZKN23}, (c) GazeRing \cite{DBLP:conf/ismar/WangSHRS024}. Images courtesy of \cite{DBLP:conf/ismar/LiLMHLS22}, \cite{DBLP:conf/chi/XuZKN23} and \cite{DBLP:conf/ismar/WangSHRS024}.}
    }
    \label{fig:TactileInteractions}
\end{figure}

\subsubsection{Tactile Interaction}
\label{Tactile}

\revise{In discussing ``Tactile Interactions", we refer to interactions with additional input devices that are light-weight, wearable and technologically advanced.}{In discussing ``Tactile Interaction", we refer to interactions with additional input devices that are light-weight, wearable and technologically advanced, for example those illustrated in Fig. \ref{fig:TactileInteractions}.}
As such, interactions involving controllers or cumbersome tangible objects are excluded. This section focuses on two primary interaction techniques: finger gesture interaction and touch-based interaction. Additionally, studies investigating tactile feedback to enhance user immersion are also presented.

\textbf{Finger gesture interactions} have been extensively studied in recent years. Researchers various input modalities, such as cameras \cite{DBLP:conf/ismar/LiLMHLS22}, optical sensors \cite{DBLP:conf/iswc/KitamuraYS23, DBLP:conf/chi/XuZKN23} and electric field sensors \cite{DBLP:journals/imwut/ChenLYZ22}, to recognize predefined finger gestures. 
\revise{The positioning of input devices also varies: NailRing \cite{DBLP:conf/ismar/LiLMHLS22} and TouchLog \cite{DBLP:conf/iswc/KitamuraYS23} placed their devices above the user's index fingernail.}{The positioning of input devices also differs: NailRing \cite{DBLP:conf/ismar/LiLMHLS22} and TouchLog \cite{DBLP:conf/iswc/KitamuraYS23} place their devices above the user's index fingernail, whereas AO-Finger \cite{DBLP:conf/chi/XuZKN23} positions its devices around the user's wrist}. 
\revise{NailRing utilized a micro close-focus camera to capture changes of color in the nail and finger \cite{DBLP:conf/ismar/LiLMHLS22}.}{NailRing\cite{DBLP:conf/ismar/LiLMHLS22} utilizes a micro close-focus camera to capture changes of color in the nail and finger.} 
\revise{Rather than cameras, TouchLog employed a nail-type device with photo-reflective sensors for privacy, to detect the skin deformation of the fingertip during gestures \cite{DBLP:conf/iswc/KitamuraYS23}.}{Rather than cameras, TouchLog \cite{DBLP:conf/iswc/KitamuraYS23} employs a nail-type device with photo-reflective sensors for privacy, to detect the skin deformation of the fingertip during gestures.} 
\revise{Beyond discrete gestures, EFRing explored continuous 1D finger micro-movement tracking using a ring-shaped device through electric-field sensing \cite{DBLP:journals/imwut/ChenLYZ22}.}{Beyond discrete gestures, EFRing \cite{DBLP:journals/imwut/ChenLYZ22} explores continuous 1D finger micro-movement tracking using a ring-shaped device through electric-field sensing.}

Finger gesture interactions offer several advantages. Their primary advantage lies in their subtlety and greater social acceptance \cite{DBLP:conf/ismar/LiLMHLS22, DBLP:conf/iswc/KitamuraYS23, DBLP:conf/chi/XuZKN23, DBLP:journals/imwut/ChenLYZ22}. Notably, the pre-defined finger gestures in recent works typically involve no more than two fingers (thumb and index). Thus, these techniques are much more effortless and light-weight compared to hand gesture and other methods \cite{DBLP:conf/chi/XuZKN23}. However, finger gesture interactions face challenges with generalization accuracy due to individual differences, as reported by \cite{DBLP:conf/iswc/KitamuraYS23, DBLP:conf/ismar/LiLMHLS22}. Both studies suggest that individual calibration before use can mitigate this issue. Additionally, due to the complexity of recognition algorithms and the computational limitations of mobile devices, finger gesture recognition exceeds the capabilities of the minimal input devices and thus often confined to PC platforms \cite{DBLP:conf/chi/XuZKN23, DBLP:journals/imwut/ChenLYZ22}.

\textbf{Touch-based interactions} involve works with the need for a surface or device to be physically touched. Researchers proposed numerous devices for touch-based interactions, such as flexible sensors \cite{DBLP:journals/imwut/ZhanXZGCGLQ23},  packaged microphones \cite{10522613}, and even robots \cite{DBLP:conf/vr/MortezapoorVVK23}.
\revise{For text editing task in speech-unfriendly AR environments, TouchEditor equipped the user's arm with a flexible touchpad composed of flexible pressure sensors \cite{DBLP:journals/imwut/ZhanXZGCGLQ23}.}{For text editing task in speech-unfriendly AR environments, TouchEditor \cite{DBLP:journals/imwut/ZhanXZGCGLQ23} equipped the user's arm with a flexible touchpad composed of flexible pressure sensors.} They designed numerous operations including text selection, cursor positioning, text retyping and editing commands. 
\revise{Similarly focusing on text entry in AR, TapGazer explored typing with TapStrap (finger-worn accelerometers), touch-sensitive gloves or touchpads, supported by gaze and language model \cite{DBLP:conf/chi/HeLP22}.}{Similarly focusing on text entry in AR, TapGazer \cite{DBLP:conf/chi/HeLP22} explored typing with TapStrap (finger-worn accelerometers), touch-sensitive gloves or touchpads, supported by gaze and language model.} Buttons can also be wearable to offer a touch-based solution.
\revise{FingerButton allows users to seamlessly transition between real world and VR with a finger-worn button device \cite{DBLP:conf/ismar/DasNH23}. Satriadi \textit{et al.} demonstrated greater creativity by leveraging tangible globes to explore the design of immersive data visualization in AR \cite{DBLP:conf/chi/SatriadiSECCLYD22}.}{FingerButton \cite{DBLP:conf/ismar/DasNH23} allows users to seamlessly transition between the real world and VR with a finger-worn button device. Satriadi \textit{et al.} \cite{DBLP:conf/chi/SatriadiSECCLYD22} demonstrated greater creativity by leveraging tangible globes to explore the design of immersive data visualization in AR.}

Touch-based interactions seem to represent a compromise between traditional device (\textit{e.g.}, controller and keyboard) and finger or hand gesture. Their performance of speed and accuracy is comparable to traditional input devices like keyboards \cite{DBLP:conf/chi/HeLP22}, while offering greater convenience and subtlety akin to finger or hand gestures \cite{DBLP:journals/imwut/ZhanXZGCGLQ23, DBLP:conf/ismar/DasNH23}. Moreover, they cause less arm and hand fatigue compared to gesture-based interactions. On the other hand, they still require additional hardware \cite{DBLP:conf/chi/HeLP22} and face the challenges related to the larger form factors of these devices \cite{DBLP:journals/imwut/ZhanXZGCGLQ23}. Furthermore, touch-based interactions are dependent on surfaces (\textit{e.g.}, arm) for input.

\textbf{Tactile feedback.} 
Feedback from the interactive device is necessary to enhance the interactive experience \cite{DBLP:conf/ismar/LiLMHLS22}. Since wearable tactile devices is always available to access \cite{DBLP:conf/ismar/DasNH23}, several studies have investigated these devices to provide tactile feedback. 
\revise{For example, Saint-Aubert \textit{et al.} utilized a HapCoil-Plus actuator to generate speech-based tactile vibrations, aiming to enhance users' persuasiveness, co-presence, and leadership \cite{DBLP:conf/vr/SaintAubertAMPAL23}. Jingu \textit{et al.} introduced an electrotactile device with a thin and flexible form factor to enable double-sided simulation feedback within pinched fingerpads \cite{DBLP:conf/uist/JinguWS23}.}{For example, Saint-Aubert \textit{et al.} \cite{DBLP:conf/vr/SaintAubertAMPAL23} utilized a HapCoil-Plus actuator to generate speech-based tactile vibrations, aiming to enhance users' persuasiveness, co-presence, and leadership. Jingu \textit{et al.} \cite{DBLP:conf/uist/JinguWS23} introduced an electrotactile device with a thin and flexible form factor to enable double-sided simulation feedback within pinched fingerpads.}

\subsubsection{Multimodal Interaction}
\label{MultiInteraction}

Unimodal interactions, as discussed above, have been explored across various spatial computing scenarios, but each comes with its own limitations.
\revise{By leveraging the complementary strengths of these modalities, multimodal interactions seeks to enhance usability. Wang \textit{et al.} argued that eye gaze-based interaction is more suitable for primary target pointing, hand gestures provide capabilities for transformation and editing, and descriptive voice input improves system controllability, \textit{e.g.}, commands such as open/close or up/down \cite{DBLP:journals/thms/WangWYL21}.}{By leveraging the complementary strengths of these modalities, multimodal interactions seek to enhance usability. Wang \textit{et al.} \cite{DBLP:journals/thms/WangWYL21} argued that eye gaze-based interaction is more suitable for primary target pointing, hand gestures provide capabilities for transformation and editing, and descriptive voice input improves system controllability, \textit{e.g.}, commands such as open/close or up/down.} Inspired by these insights, recent research has developed various multimodal interactions for different XR tasks \cite{DBLP:conf/ismar/ChenGFCL23, DBLP:conf/vr/JingLB22}. In this paper, we primarily focus on dual-modal techniques (\textit{e.g.}, \textit{Gaze + Gesture}, \textit{Gaze + Speech}, \textit{Gesture + Speech}) and a triple-modal technique (\textit{Gaze + Gesture + Speech}).

\textbf{(a) Gaze + Gesture}. Many studies adhere to the principle of “gaze selects, hand manipulates” \cite{10108465}, as eye gaze, which can rapidly locate a target, complements the fine-grained control provided by hand gestures. Apple Vision Pro identified this dual-modal interaction as the primary access method for spatial computing \cite{Apple}. In the reviewed literature, two papers focused on text input tasks in XR environments \cite{10474330, DBLP:conf/chi/HeLP22}.
\revise{Ren \textit{et al.} utilized the fact that gaze reaches the target before the hand, predicting the next likely key and enlarging and highlighting it, allowing users to press it more quickly and accurately \cite{10474330}. He \textit{et al.} enabled users to input text by tapping the approximate area of a character on a keyboard, without needing to see their hands or the keyboard. Additionally, users can resolve ambiguities by selecting the intended word via eye movement \cite{DBLP:conf/chi/HeLP22}.}{Ren \textit{et al.} \cite{10474330} found that gaze reaches the target before the hand, and leveraged this to enlarge and highlight the possible following key, providing users quicker and more accurate selection. He \textit{et al.} \cite{DBLP:conf/chi/HeLP22} enabled users to input text by tapping the approximate area of a character on a keyboard, without looking at their hands or the keyboard. Additionally, users can resolve ambiguities by selecting the intended word via eye movement.}
Researchers also explored using gaze as a directional reference, combined with hand movements along the gaze ray or swipes to navigate in VR environments \cite{10.1145/3613904.3642147}.

Besides above tasks, many studies have examined menu selection and 3D object manipulation in XR \cite{10.1145/3544548.3581423, 10.1145/3530886, 10.1145/3591129, 10.1145/3613904.3642758, 10108465, DBLP:conf/ismar/CailletGN23}.
\revise{Wagner \textit{et al.} designed a Fitts’ Law study to evaluate the efficiency of two different gaze-hand alignment techniques for target selection \cite{10.1145/3544548.3581423, 10.1145/3530886}. Shi \textit{et al.} used eye gaze and hand gestures for region selection in AR \cite{10.1145/3591129}. Rodríguez \textit{et al.} enabled artists to draw VR sketches using this dual-modal interaction \cite{10.1145/3613904.3642758}. Bao \textit{et al.} explored how hand-eye coordination can facilitate object selection and manipulation in occluded environments \cite{10108465}. Caillet \textit{et al.} designed a two-phase interaction technique with gaze and hand gesture for reduce fatigue in 3D selection \cite{DBLP:conf/ismar/CailletGN23}.}{Wagner \textit{et al.} \cite{10.1145/3544548.3581423, 10.1145/3530886} designed a Fitts’ Law study to evaluate the efficiency of two different gaze-hand alignment techniques for target selection. Shi \textit{et al.} \cite{10.1145/3591129} utilized eye gaze and hand gestures for region selection in AR. Rodríguez \textit{et al.} \cite{10.1145/3613904.3642758} enabled artists to draw VR sketches using this dual-modal interaction. Bao \textit{et al.} \cite{10108465} explored how hand-eye coordination can facilitate object selection and manipulation in occluded environments. Caillet \textit{et al.} \cite{DBLP:conf/ismar/CailletGN23} designed a two-phase interaction technique with gaze and hand gesture to reduce fatigue in 3D selection.}



\textbf{(b) Gaze + Speech.} Eye gaze and speech are typically integrated to offer hands-free interaction: gaze provides contextual information, while speech functions as a command or query.
\revise{Jing \textit{et al.} visualized shared gaze cues using contextual speech input to enhance the efficiency of XR remote collaboration \cite{DBLP:conf/vr/JingLB22}.}{Jing \textit{et al.} \cite{DBLP:conf/vr/JingLB22} visualized shared gaze cues using contextual speech input to enhance the efficiency of XR remote collaboration.} When the XR system detects keywords such as ``this" or ``that", it guides collaborators to follow the speaker's gaze direction.
\revise{Li \textit{et al.} proposed a method in AR environments that processes multimodal sensory inputs (visual and auditory) from daily activities using LLMs, predicting potential digital follow-up actions \cite{10.1145/3613904.3642068}.}{Li \textit{et al.} \cite{10.1145/3613904.3642068} proposed a method in AR environments that processes multimodal sensory inputs (visual and auditory) from daily activities using LLMs, predicting potential digital follow-up actions.} 
\revise{G-VOILA leveraged gaze data and visual field as contextual information to enhance LLMs' ability to respond to users' daily speech queries \cite{DBLP:journals/imwut/WangSWYYWJXY24}.}{G-VOILA \cite{DBLP:journals/imwut/WangSWYYWJXY24} leveraged gaze data and visual field as contextual information to enhance LLMs' ability to respond to users' daily speech queries.} In addition to serving as inputs, visual and audio modalities can also function as outputs of XR systems to enhance immersion.
\revise{Lee \textit{et al.} designed a multi-modal attention guidance system that utilizes visual cues (lighting effects) and auditory cues (spatial audio) in social VR group conversations, significantly improving users' response times and conversation satisfaction during turn-taking interactions \cite{10462901}.}{Lee \textit{et al.} \cite{10462901} designed a multi-modal attention guidance system that utilizes visual cues (lighting effects) and auditory cues (spatial audio) in VR social group conversations, significantly improving users' response times and conversation satisfaction during turn-taking interaction.}

\textbf{(c) Gesture + Speech.} Gesture and speech are the two most commonly used modalities in delivering presentations. Therefore, they are naturally integrated for augmented presentations. For example,
\revise{Liao \textit{et al.} proposed to augmented live presentations using visuals and animation for expressive storytelling \cite{DBLP:conf/uist/LiaoKJKS22}.To achieve this, they used speech recognition combined with keyword extraction. These keywords are mapped to predefined display contents, such as images and videos. These contents are displayed on the hand in real time.}{Liao \textit{et al.} \cite{DBLP:conf/uist/LiaoKJKS22} proposed augmented live presentations using visuals and animation for expressive storytelling. To achieve this, they used speech recognition combined with keyword extraction. These keywords are mapped to predefined display contents, such as images and videos, which are displayed on the user's hand in real time.} 
However, this method presented inferior visual artifacts due to imperfect system recognition \cite{DBLP:conf/chi/CaoKWAX24}. 
\revise{Cao \textit{et al.} predefined the layouts and effects of graphics and used a set of rules to allow speech, gesture and graphics to be elastically connected for visual quality \cite{DBLP:conf/chi/CaoKWAX24}.}{Cao \textit{et al.} \cite{DBLP:conf/chi/CaoKWAX24} predefined the layouts and effects of graphics and used a set of rules to allow speech, gesture and graphics to be elastically connected for visual quality.}
\revise{Williams \textit{et al.} investigated the effect of referent display methods (text vs. animation) on gesture and speech elicitation \cite{9873984}. In terms of animated referents, they found users elicited interaction proposals with gesture+speech, feeling lower workload than speech only conditions. }{Williams \textit{et al.} \cite{9873984} investigated the effect of referent display methods (text vs. animation) on gesture and speech elicitation. In terms of animated referents, they found users elicited interaction proposals with gesture + speech and suffered lower workload than speech-only interactions. }

\textbf{(d) Gaze + Gesture + Speech.} Several studies have also explored the integration of gaze, gesture, and speech to leverage the benefits of all three modalities.
\revise{Chen \textit{et al.} proposed the Compass+Ring menu that integrates three modalities: gaze is used to point at menu items, speech to confirm selections (avoiding the Midas Touch problem), and gestures to adjust parameters by rotating the wrist or pinching fingers to navigate menu levels. The Ring menu allows for quick adjustments using simple gestures like ``putting on" and ``rotating" a virtual ring \cite{DBLP:conf/ismar/ChenGFCL23}. Lee \textit{et al.} leveraged eye gaze, hand pointing and conversation history to achieve pronoun disambiguation in users' speech. When the user queries using pronouns, the proposed system identifies objects and text within the field-of-view (FoV) to replace the pronouns. This approach enables the voice assistant to provide more accurate responses by incorporating contextual information \cite{DBLP:conf/chi/0005WBCRF24}.}{Chen \textit{et al.} \cite{DBLP:conf/ismar/ChenGFCL23} proposed the Compass+Ring menu that integrates three modalities: gaze to point at menu items, speech to confirm selections, and gestures to adjust parameters by rotating the wrist or pinching fingers to navigate menu levels. The Ring menu allows for quick adjustments using simple gestures like ``putting on" and ``rotating" a virtual ring. Lee \textit{et al.} \cite{DBLP:conf/chi/0005WBCRF24} leveraged eye gaze, hand pointing and conversation history to achieve pronoun disambiguation in users' speech. When the user queries with pronouns, the proposed system replace them with identified objects and text within the field-of-view (FoV). This approach enables the voice assistant to provide more accurate responses by incorporating contextual information.}

\textbf{(e) X + Y.} In the reviewed literature, numerous studies explored multimodal interactions beyond the modalities discussed previously, which here are referred to ``X + Y". Due to space constraints, only three types of them are presented in this section: X + head pose, X + facial expression, and multimodal comparison studies.
\textit{X + head pose.}
Several studies focus on integrating head pose with other modalities \cite{DBLP:journals/imwut/ShenYYS22, DBLP:conf/chi/WeiSYW0YL23, DBLP:conf/chi/HouNSKBG23, marquardt2024selection, DBLP:conf/chi/LeeWSG24}. 
\revise{For example, Clenchclick explored explored the combination of head movements and teeth clenching for target selection in AR \cite{DBLP:journals/imwut/ShenYYS22}. }{For example, Clenchclick \cite{DBLP:journals/imwut/ShenYYS22} explored the combination of head movements and teeth clenching for target selection in AR. }
\textit{X + facial expression.}
Some other studies emphasize combining facial expressions with additional modalities \cite{DBLP:conf/vr/LaiSL24, DBLP:conf/uist/0001QTFLS22}. 
\revise{GazePuffer, for instance, introduced cheek puffing gestures combined with gaze to offer an innovative hands-free interaction \cite{DBLP:conf/vr/LaiSL24}.}{GazePuffer \cite{DBLP:conf/vr/LaiSL24}, for instance, introduced cheek puffing gestures combined with gaze to offer an innovative hands-free interaction.}
\textit{Multimodal Comparison.}
Additionally, some research compares unimodal and multimodal interaction techniques to determine which modalities are better suited for specific tasks or scenarios \cite{10049667, DBLP:conf/ismar/MengXL22, DBLP:conf/ismar/XuMYSL22}.
\revise{In terms of the text selection task in VR, Meng \textit{et al.} designed three hands-free selection mechanisms, including dwell, eye Blinks and voice (hum) to complement head-based pointing \cite{DBLP:conf/ismar/MengXL22}. Yan \textit{et al.} introduced ConeSpeech, a VR-based directional speech interaction method allowing users to selectively communicate with target listeners, minimizing disturbance to bystanders \cite{10049667}.}{In terms of the text selection task in VR, Meng \textit{et al.} \cite{DBLP:conf/ismar/MengXL22} designed three hands-free selection mechanisms, including dwell, eye Blinks and voice (hum) to complement head-based pointing. Yan \textit{et al.} \cite{10049667} introduced ConeSpeech, a VR-based directional speech interaction method allowing users to selectively communicate with target listeners, minimizing disturbance to bystanders.} The authors compare five control modalities, \textit{i.e.} head, gaze, torso, hand, and controller, and ultimately select head orientation for its balance of speed, accuracy, and intuitiveness.

\section{Discussion and Recommendations}

Based on the previous taxonomy and analysis of natural interaction techniques in wearable XR, this section focuses on the challenges these techniques face and offers recommendations for future research directions. Objectively, researchers aim to develop more accurate and efficient natural interaction techniques. Subjectively, the goal is to provide more comfortable and immersive interaction experiences. Additionally, to enhance the usability of these techniques, researchers should apply them in real-world scenarios to improve the overall user experience in XR.

Accordingly, the discussion is organized into three main areas. Section \ref{4.1} presents recommendations for toward more accurate and reliable natural interactions in XR. Section \ref{4.2} discusses recommendations for toward more natural, comfortable, and immersive interactions in XR. Lastly, Section \ref{4.3} offers suggestions for bridging interaction design and practical XR applications.


\subsection{Toward More Accurate and Reliable Natural Interactions in XR}
\label{4.1}
As above section mentioned, current natural interaction such as gaze, gesture and speech, still face the insufficient accuracy during interaction, especially in the complex situations, \textit{e.g.}, outdoors and walking. 
For example, for reducing the effects of Midas touch problem on gaze-only interaction, Section \ref{GazeOnly} mentioned many researchers optimize the virtual menu layouts \cite{DBLP:conf/chi/ChoiSO22, DBLP:conf/vr/0001LC00S22}, or require extra eye movements to confirm the selections \cite{DBLP:conf/vr/OrloskyLSSM24, kim2022lattice}. For the false hand gesture recognition caused by occlusion, researchers usually require users to manipulate objects in certain postures and orientations to avoid occlusion \cite{DBLP:conf/chi/PeiCLZ22, DBLP:journals/tvcg/SongDK23}, or use the wearable device based hand-tracking \cite{DBLP:conf/chi/HeLP22}.
While these methods have shown some success in improving interaction accuracy, there is still significant room for advancement. We propose that the future of achieving more accurate XR interaction lies in three promising research directions: multimodal interaction, error recovery mechanisms, and AI-powered assistance.

\textbf{Multimodal interaction.} Integrating multiple modalities into a unified interaction system significantly enhances the accuracy and reliability of user inputs. For instance, Bao \textit{et al.} addressed the issue of low gaze-pointing accuracy by allowing the hand to refine the pointing direction, thereby facilitating target selection \cite{10108465}. Moreover, combining gaze with hand gesture interaction follows the principle of ``gaze select, hand manipulate" \cite{10.1145/3544548.3581423, 10.1145/3530886}. This approach reduces hand operations and thus decreases the likelihood of hand recognition error.
Multimodal integration can also resolve ambiguities in speech. For instance, Lee \textit{et al.} used gaze or hand pointing to identify specific objects, clarifying vague verbal descriptions such as ``What is this?" \cite{DBLP:conf/chi/0005WBCRF24}.
This highlights the effectiveness of multimodal systems in improving interaction accuracy and reducing ambiguity.

\textbf{Error recovery.} Error recovery mechanisms are crucial for ensuring robust and user-friendly interactions in XR environments. Despite advancements in tracking accuracy, unintended actions or misrecognitions such as clicking the wrong button due to inattention, remain inevitable. 
For example, Sendhilnathan \textit{et al.} categorized gesture-based interaction events into three types: correctly recognized input actions, input recognition errors, and user errors, observing that these categories were consistent across tasks. They then applied a deep learning method to differentiate these events using only eye movement input, achieving promising results \cite{DBLP:conf/uist/SendhilnathanZL22}.
Sidenmark \textit{et al.} designed an error-aware mechanism that adaptively switches to fallback modalities (\textit{e.g.}, head pointing or a controller) when errors or noise occur during gaze interactions. They also adjusted the weighting between the gaze modality and fallback options based on the error ratio \cite{DBLP:journals/tvcg/SidenmarkP0CGWG22}. These approaches highlight the vital role of error recovery in enhancing the reliability and overall usability of XR systems.

\subsection{Toward More Natural, Comfortable and Immersive Interactions in XR}
\label{4.2}
Interaction techniques aim not only for accuracy but also for subjective factors such as comfort, immersion, and usability. In XR environments, the subjective experience of a particular method may be more important to certain users than objective metrics (\textit{e.g.}, temporal performance) \cite{DBLP:conf/vr/HombeckVHDL23}. In the reviewed literature, over 50 studies assess users' subjective experiences with interaction techniques through measures like task load, immersion, preference, and usability. These assessments are typically conducted via post-study questionnaires, such as the NASA Task Load Index (NASA-TLX), System Usability Scale (SUS), and Simulator Sickness Questionnaire (SSQ). This emphasizes the importance of user-friendliness as a critical quality of interaction techniques. We anticipate that future XR interactions will continue to prioritize improvements in: reducing task load, enhancing immersion, and improving subtlety.

\textbf{Reduce task load.}
Physical and cognitive load are critical factors in user’s interactive experience. The NASA-TLX is the most widely used tool for assessing the task load of interaction techniques. 
\revise{As discussed in \ref{Interaction techniques}, each modality presents its own limitations:}{As discussed in \ref{Interaction techniques}, each modality has its own limitations in terms of task load:} hand-gesture-only interactions can lead to arm fatigue \cite{DBLP:conf/vr/QuereMJWW24}; speech-only interactions increase cognitive load due to the requirement of memorizing keywords \cite{DBLP:conf/ismar/ChenGFCL23}; prolonged use of gaze-only interactions can also cause eye fatigue \cite{DBLP:conf/chi/ZhangCSS24}. Our review identified several studies focused on reducing users' burden. One common approach is to effectively integrate multiple modalities. 
\revise{For instance, Bao \textit{et al.} proposed a better method of combining gaze and hand gestures, significantly alleviating arm fatigue \cite{10108465}. Compass+Ring introduced a multimodal menu integrating gaze, speech, and gesture to mitigate eye fatigue \cite{DBLP:conf/ismar/ChenGFCL23}.}{For instance, Bao \textit{et al.} \cite{10108465} proposed a better method of combining gaze and hand gestures, significantly alleviating arm fatigue. Compass+Ring \cite{DBLP:conf/ismar/ChenGFCL23} introduced a multimodal menu integrating gaze, speech and gesture to mitigate eye fatigue.} Additionally, some studies innovate new devices and modalities, offering more effortless interactions such as finger gestures. 
\revise{AO-finger introduced a wristband that recognizes fine-grained finger gestures that require little exertion \cite{DBLP:conf/chi/XuZKN23}.}{AO-finger \cite{DBLP:conf/chi/XuZKN23} introduced a wristband that recognizes fine-grained finger gestures that require little exertion.} We believe that reducing task load will remain a key focus in future interaction design.

\textbf{Enhance immersion.}
\revise{Immersion is a critical component of the XR experience and constitutes one of the 3I of VR \cite{DBLP:books/daglib/0011673}}{Immersion is a critical component of the XR experience and is one of the 3Is of VR \cite{DBLP:books/daglib/0011673}}. Interaction, a bidirectional process, also aims to provide users with an immersive experience. Numerous studies mainly focus on offering feedback with different devices and technologies to enhance the immersion of XR. 
\revise{For instance, Jang \textit{et al.} utilized ultrasonic devices to deliver mid-air haptic feedback, allowing users to touch and explore volume-rendered hologram with their bare hands \cite{DBLP:conf/ismar/JangFP22}. Saint-Aubert \textit{et al.} investigated tactile vibrations speech from users or virtual avatars to enhance persuasiveness, co-presence, and leadership \cite{DBLP:conf/vr/SaintAubertAMPAL23}.}{For instance, Jang \textit{et al.} \cite{DBLP:conf/ismar/JangFP22} utilized ultrasonic devices to deliver mid-air haptic feedback, allowing users to touch and explore volume-rendered hologram with their bare hands. Saint-Aubert \textit{et al.} \cite{DBLP:conf/vr/SaintAubertAMPAL23} investigated tactile vibrations speech from users or virtual avatars to enhance persuasiveness, co-presence, and leadership.} Besides, several studies have proposed novel well-designed interaction techniques to enhance users' immersion. 
\revise{Illumotion is an example to increase presence and reduce cybersickness by designing a hand-gesture-based interaction technique inspired by photo manipulation \cite{DBLP:conf/vr/SinJLLLN24}.}{Illumotion \cite{DBLP:conf/vr/SinJLLLN24} is an example to increase presence and reduce cybersickness by designing a hand-gesture-based interaction technique inspired by photo manipulation.} This suggests that future research will continue to focus on enhancing immersion in XR environments.

\textbf{Improve subtlety.}
For XR devices to be integrated into daily life, interaction techniques must be adaptable across various scenarios without raising concerns. In public environments, conspicuous interactions may lead to social awkwardness and privacy challenges \cite{DBLP:conf/ismar/LiLMHLS22}. Thus, enhancing the subtlety of interaction techniques is crucial for improving social acceptance. Some existing methods, such as hand gestures and speech interactions, have been criticized for their lack of subtlety. Hand gestures are restricted by the FoV of HMD cameras \cite{DBLP:conf/chi/XuZKN23}, while speech interactions may disrupt the experiences of others \cite{DBLP:conf/chi/WangYWJ024}. Recent research has increasingly focused on developing more discreet interaction techniques. Tactile interactions, including micro finger gestures \cite{DBLP:conf/ismar/LiLMHLS22, DBLP:conf/chi/XuZKN23, DBLP:journals/imwut/ChenLYZ22, DBLP:conf/iswc/KitamuraYS23} and wearable devices \cite{DBLP:conf/ismar/DasNH23, DBLP:conf/chi/HeLP22}, are considered to offer greater subtlety and social acceptance compared to hand gestures. Additionally, to enable speech interactions in noisy or speech-unfriendly environments, several studies have proposed methods for silent speech recognition \cite{DBLP:journals/tvcg/CaiML24, DBLP:conf/chi/ZhangLHWLGZ23, DBLP:conf/chi/WangSRZ24}. Therefore, improving the subtlety of interaction techniques will remain a prominent and ongoing topic in XR research.









\subsection{\revise{}{Empowering Multimodal XR Natural Interaction with AI and LLMs}}  
\label{4.3}  

\revise{}{Recent advancements in AI and LLMs have significantly expanded the potential of natural interactions in XR environments. By enhancing input recognition, enabling semantic reasoning, and facilitating multimodal integration, these technologies improve the efficiency, intuitiveness, and usability of XR systems. This section discusses the roles of AI and LLMs, supported by recent research, and explores potential future directions for their application.}

\revise{}{\textbf{AI enhances input recognition and contextual understanding.}  AI plays a pivotal role in addressing challenges related to multimodal input recognition and resolving ambiguities in XR interactions. Machine learning models have significantly improved input accuracy by combining diverse data sources. For example, gaze direction and head orientation are integrated to predict user targets in complex scenes \cite{DBLP:conf/chi/WeiSYW0YL23}. Similarly, error-aware systems powered by deep learning can identify user mistakes or inconsistencies during interactions and adaptively switch to alternative modalities, such as voice commands or hand gestures \cite{DBLP:conf/uist/SendhilnathanZL22, DBLP:journals/tvcg/SidenmarkP0CGWG22}.  Beyond input recognition, AI enhances contextual understanding by integrating sensory data. For instance, fine-grained gesture recognition systems leveraging visual and audio sensors enable precise interactions even in occluded or noisy environments \cite{DBLP:conf/chi/XuZKN23}. In lifelogging applications, AI models can process egocentric video data to automatically segment and tag key events based on temporal and spatial patterns \cite{10484440}. These advancements increase the reliability and robustness of XR systems, making them more responsive and adaptable to user needs.} 

\revise{}{\textbf{LLMs enable semantic reasoning and natural language interaction.}  LLMs bring advanced semantic reasoning capabilities to XR systems, enabling more natural, intuitive, and flexible user interactions. Unlike traditional keyword-based approaches, LLMs can process open-ended user commands and resolve ambiguous references by incorporating contextual cues, such as gaze direction, gestures, or spatial information. For example, recent systems have demonstrated how LLMs can resolve pronoun ambiguities in speech by linking them to objects within the user’s field of view \cite{DBLP:journals/imwut/WangSWYYWJXY24, DBLP:conf/chi/0005WBCRF24}. Beyond disambiguation, LLMs empower non-technical users to modify virtual environments dynamically through natural language commands \cite{DBLP:conf/vr/GiunchiNGS24}. Additionally, they enhance interactivity by supporting intelligent virtual assistants that integrate multimodal inputs, such as speech and gaze, to deliver context-aware and reasoning-driven responses \cite{DBLP:conf/chi/WangYWJ024}. These capabilities make XR systems more intuitive, accessible, and adaptable to diverse user needs. }

\revise{}{\textbf{Future implications.} Looking forward, the integration of AI and LLMs presents transformative opportunities for developing adaptive, user-centric XR systems. A key area of future exploration involves leveraging AI to dynamically model user behavior and preferences, enabling systems to personalize interactions over time. For example, advancements in lifelogging technologies, such as egocentric vision models and wearable AR devices, facilitate the continuous capture and encoding of users’ daily experiences, including what they see, hear, and do \cite{DBLP:conf/chi/0005WBCRF24}. 
By combining AI for event detection with LLMs for semantic reasoning, future systems could enable effortless memory retrieval. Users could query their past experiences with questions such as, ``What did I read during yesterday’s meeting?" or ``Who did I talk to at lunch?” These capabilities align with the broader objective of creating more intuitive, human-centric, and context-aware XR systems. }

\subsection{Bridging Interaction Design and Practical XR Applications}
\label{4.4}

As discussed in Section \ref{Appl.}, nearly 70\% of the research focuses on the design of interaction techniques without specifying their application in concrete scenarios. While these techniques could potentially be adapted to various applications, they often require significant modifications for specific contexts. This gap presents challenges that limit the widespread adoption of natural interaction techniques in XR environments. Therefore, we argue that researchers should focus more on application scenarios when developing natural interaction techniques.

\textbf{Existing applications of natural interaction techniques.} Section \ref{Appl.} highlighted over 20 papers that explore natural interaction techniques in different application contexts, such as sketching\cite{DBLP:journals/imwut/ChenLYZ22, DBLP:journals/tvcg/SongDK23}, virtual meetings \cite{DBLP:conf/vr/SaintAubertAMPAL23, 10462901}, XR navigation\cite{DBLP:conf/vr/QuereMJWW24, DBLP:conf/chi/WangYWJ024}, reading \cite{DBLP:conf/ismar/LeeHM22, DBLP:conf/ismar/MengXL22}, and maintenance \cite{DBLP:conf/vr/QuereMJWW24}. These applications have significant potential for further expansion. For example, in sketching, combining brain-computer interfaces (BCIs) with eye-tracking technology could allow users to control the shape and color of design objects directly through thought, rather than relying solely on gestures or voice commands. In virtual meetings, real-time emotion recognition technology can capture participants' facial expressions, tone of voice, and heart rate to generate personalized avatars.

\textbf{Natural interaction techniques can also be applied to new domains.} Beyond the previously mentioned applications, in healthcare, these techniques offer more convenient care by integrating eye-tracking, gestures, and voice commands. For instance, patients wearing lightweight AR headsets can use eye-tracking to control devices like lighting or the TV, and adjust the bed angle with voice commands.
In entertainment, natural interaction transforms experiences by allowing players to interact with virtual characters through gestures and speech, while the game adjusts its environment or difficulty based on the player's emotions in real time.

\textbf{Application beyond the lab.} Currently, most studies on natural interaction techniques are confined to laboratory or controlled environments. Only a limited number of studies explore their use in more complex, real-world contexts such as shopping malls or city streets \cite{DBLP:conf/chi/0005WBCRF24, 10.1145/3613904.3642068}. To enable the broader adoption of XR technologies, it is crucial to design natural interaction techniques that can address the challenges posed by these complex, uncontrolled environments.

\textbf{Breaking down barriers between applications.} Existing research often develops interaction modalities that are specific to each application, which increases the learning curve for users. If interactions across different XR applications were standardized, similar to the unified operation model of smartphone apps, it would greatly reduce the learning curve for users. This standardization could, in turn, encourage wider adoption of these technologies.














\section{Limitations and Future Work}
While this review of recent papers from top venues offers valuable insights into the latest trends in XR natural interaction techniques, there are several limitations that need to be addressed in future work.

Firstly, due to the vast amount of relevant research, we limited our scope to publications from six major venues between 2022 and 2024.
\revise{However, other venues, such as those within the ACM and IEEE digital libraries, also contain valuable research that was not included.}{However, other venues, such as those within the ACM and IEEE digital libraries, also contain valuable research, which was not included.} Additionally, extending the review to earlier years could provide a clearer understanding of the broader development trends in the field.

Furthermore, in this paper, we categorized the collected literature into four main categories: application context, operation types, performance measures, and interaction modalities. Future work could explore additional categories, such as study types or use cases \cite{DBLP:journals/tvcg/SpittlePC023}, to provide a more detailed understanding.

Finally, our review primarily focused on papers that explicitly mentioned natural interaction techniques for wearable XR. However, many studies on non-wearable natural interaction could potentially be adapted for wearable XR. 
\revise{For example, Ahmad Khan \textit{et al.} explored the synchronization between gaze and speech to implicitly link voice notes with digital text content \cite{10.1145/3491102.3502134}.}{For example, Ahmad Khan \textit{et al.} \cite{10.1145/3491102.3502134} explored the synchronization between gaze and speech to implicitly link voice notes with digital text content.} Although their study was conducted in a desktop environment, this approach could be applied to wearable XR systems as well.


\section{Conclusion}

In this paper, we review 104 research papers on natural interaction techniques for wearable XR, published between 2022 and 2024 in six top venues. We categorize this literature based on application context, operation types, performance measures, and interaction modalities. 
Specifically, we classify operation types into seven categories, distinguishing between active and passive interactions. Interaction modalities are further broken down into nine distinct types. In addition, we present statistical analyses of advanced natural interaction techniques. Building on these insights, we identify key challenges in natural interaction systems and suggest potential avenues for future research.
This review offers valuable insights for researchers aiming to design natural and efficient interaction systems for XR.




\bibliographystyle{fcs}
\bibliography{ref}

\begin{thebibliography}{100}

\bibitem{HoloLens}
Company M.
\newblock Microsoft hololens 2 for precise, efficient hands-free work, 2024.10.
\newblock \url{https://www.microsoft.com/en-us/hololens}

\bibitem{Meta}
Meta .
\newblock Meta quest 3: New mixed reality vr headset - tech specs, 2024.10.
\newblock \url{https://www.meta.com/quest/quest-3/#specs}

\bibitem{DBLP:conf/vr/JingLB22}
Jing A, Lee G~A, Billinghurst M.
\newblock Using speech to visualise shared gaze cues in {MR} remote collaboration.
\newblock In: {IEEE} Conference on Virtual Reality and 3D User Interfaces, {VR} 2022, Christchurch, New Zealand, March 12-16, 2022.
\newblock 2022,  250--259

\bibitem{DBLP:conf/vr/QuereMJWW24}
Quere C, Menin A, Julien R, Wu~H, Winckler M.
\newblock Handynotes: using the hands to create semantic representations of contextually aware real-world objects.
\newblock In: {IEEE} Conference Virtual Reality and 3D User Interfaces, {VR} 2024, Orlando, FL, USA, March 16-21, 2024.
\newblock 2024,  265--275

\bibitem{Barteit2021}
Barteit S, Lanfermann L, B{\"a}rnighausen T, Neuhann F, Beiersmann C.
\newblock Augmented, mixed, and virtual reality-based head-mounted devices for medical education: Systematic review.
\newblock JMIR Serious Games, 2021, 9(3): e29080

\bibitem{an2023arcosmetics}
An~S, Chen J, Zhu Z, Zhou F, Yang Y, Ma~Y, Liu X, Zhu H.
\newblock Arcosmetics: a real-time augmented reality cosmetics try-on system.
\newblock Frontiers of Computer Science, 2023, 17(4): 174706

\bibitem{Apple}
Apple .
\newblock Introducing apple vision pro: Apple’s first spatial computer, 2024.10.
\newblock \url{https://www.apple.com/newsroom/2023/06/introducing-apple-vision-pro/}

\bibitem{spatialcomputing}
Yenduri G, M~R, Maddikunta P~K~R, Gadekallu T~R, Jhaveri R~H, Bandi A, Chen J, Wang W, Shirawalmath A~A, Ravishankar R, Wang W.
\newblock Spatial computing: Concept, applications, challenges and future directions, 2024

\bibitem{hackl2024spatial}
Hackl C, Cronin I.
\newblock Spatial Computing: An AI-Driven Business Revolution.
\newblock John Wiley \& Sons, 2024

\bibitem{10108465}
Bao Y, Wang J, Wang Z, Lu~F.
\newblock Exploring 3d interaction with gaze guidance in augmented reality.
\newblock In: 2023 IEEE Conference Virtual Reality and 3D User Interfaces (VR).
\newblock 2023,  22--32

\bibitem{DBLP:conf/interact/BerardIBEBC09}
B{\'{e}}rard F, Ip~J, Benovoy M, El{-}Shimy D, Blum J~R, Cooperstock J~R.
\newblock Did "minority report" get it wrong? superiority of the mouse over 3d input devices in a 3d placement task.
\newblock In: Human-Computer Interaction - {INTERACT} 2009, 12th {IFIP} {TC} 13 International Conference, Uppsala, Sweden, August 24-28, 2009, Proceedings, Part {II}.
\newblock 2009,  400--414

\bibitem{DBLP:conf/vr/MatthewsTIS22}
Matthews B~J, Thomas B~H, Itzstein G~S~V, Smith R~T.
\newblock Shape aware haptic retargeting for accurate hand interactions.
\newblock In: {IEEE} Conference on Virtual Reality and 3D User Interfaces, {VR} 2022, Christchurch, New Zealand, March 12-16, 2022.
\newblock 2022,  625--634

\bibitem{DBLP:conf/vr/GiunchiNGS24}
Giunchi D, Numan N, Gatti E, Steed A.
\newblock Dreamcodevr: Towards democratizing behavior design in virtual reality with speech-driven programming.
\newblock In: {IEEE} Conference Virtual Reality and 3D User Interfaces, {VR} 2024, Orlando, FL, USA, March 16-21, 2024.
\newblock 2024,  579--589

\bibitem{chai2022speech}
Chai Y, Weng Y, Wang L, Zhou K.
\newblock Speech-driven facial animation with spectral gathering and temporal attention.
\newblock Frontiers of Computer Science, 2022, 16(3): 163703

\bibitem{ding2016survey}
Ding C, Liu L.
\newblock A survey of sketch based modeling systems.
\newblock Frontiers of Computer Science, 2016, 10: 985--999

\bibitem{huang2024matching}
Huang Y, Yang L, Chen G, Zhang H, Lu~F, Sato Y.
\newblock Matching compound prototypes for few-shot action recognition.
\newblock International Journal of Computer Vision, 2024,  1--26

\bibitem{DBLP:conf/ismar/WangGL23}
Wang Z, Gu~X, Lu~F.
\newblock {DEAMP:} dominant-eye-aware foveated rendering with multi-parameter optimization.
\newblock In: Bruder G, Olivier A, Cunningham A, Peng Y~E, Grubert J, Williams I, eds, {IEEE} International Symposium on Mixed and Augmented Reality, {ISMAR} 2023, Sydney, Australia, October 16-20, 2023.
\newblock 2023,  632--641

\bibitem{DBLP:conf/vr/ChaconasH18}
Chaconas N, H{\"{o}}llerer T.
\newblock An evaluation of bimanual gestures on the microsoft hololens.
\newblock In: Proc. IEEE Conf. Virtual Real. 3D User Interfaces.
\newblock Mar. 2018,  33--40

\bibitem{DBLP:conf/chi/Hincapie-RamosGMI14}
Hincapi{\'{e}}{-}Ramos J~D, Guo X, Moghadasian P, Irani P.
\newblock Consumed endurance: a metric to quantify arm fatigue of mid-air interactions.
\newblock In: Proc. Conf. Human Factors Comput. Syst.
\newblock Apr. 2014,  1063--1072

\bibitem{DBLP:conf/chi/CaoKWAX24}
Cao Y, Kazi R~H, Wei L, Aneja D, Xia H.
\newblock Elastica: Adaptive live augmented presentations with elastic mappings across modalities.
\newblock In: Mueller F~F, Kyburz P, Williamson J~R, Sas C, Wilson M~L, Dugas P~O~T, Shklovski I, eds, Proceedings of the {CHI} Conference on Human Factors in Computing Systems, {CHI} 2024, Honolulu, HI, USA, May 11-16, 2024.
\newblock 2024,  599:1--599:19

\bibitem{DBLP:conf/chi/TorreFHBFL24}
Torre F~D~L, Fang C~M, Huang H, Banburski{-}Fahey A, Fernandez J~A, Lanier J.
\newblock {LLMR:} real-time prompting of interactive worlds using large language models.
\newblock In: Mueller F~F, Kyburz P, Williamson J~R, Sas C, Wilson M~L, Dugas P~O~T, Shklovski I, eds, Proceedings of the {CHI} Conference on Human Factors in Computing Systems, {CHI} 2024, Honolulu, HI, USA, May 11-16, 2024.
\newblock 2024,  600:1--600:22

\bibitem{DBLP:journals/tvcg/SidenmarkP0CGWG22}
Sidenmark L, Parent M, Wu~C, Chan J, Glueck M, Wigdor D, Grossman T, Giordano M.
\newblock Weighted pointer: Error-aware gaze-based interaction through fallback modalities.
\newblock {IEEE} Trans. Vis. Comput. Graph., 2022, 28(11): 3585--3595

\bibitem{10.1145/3530886}
Lystb\ae{}k M~N, Rosenberg P, Pfeuffer K, Gr\o{}nb\ae{}k J~E, Gellersen H.
\newblock Gaze-hand alignment: Combining eye gaze and mid-air pointing for interacting with menus in augmented reality.
\newblock Proc. ACM Hum.-Comput. Interact., 2022, 6(ETRA)

\bibitem{10.1145/3613904.3642068}
Li~J~N, Xu~Y, Grossman T, Santosa S, Li~M.
\newblock Omniactions: Predicting digital actions in response to real-world multimodal sensory inputs with llms.
\newblock In: Proceedings of the 2024 CHI Conference on Human Factors in Computing Systems, CHI '24.
\newblock 2024

\bibitem{DBLP:conf/embc/WangDCS15}
Wang H, Dong X, Chen Z, Shi B~E.
\newblock Hybrid gaze/eeg brain computer interface for robot arm control on a pick and place task.
\newblock In: 37th Annual International Conference of the {IEEE} Engineering in Medicine and Biology Society, {EMBC} 2015, Milan, Italy, August 25-29, 2015.
\newblock 2015,  1476--1479

\bibitem{DBLP:journals/tvcg/TranBWBP23}
Tran T~T~M, Brown S, Weidlich O, Billinghurst M, Parker C.
\newblock Wearable augmented reality: Research trends and future directions from three major venues.
\newblock {IEEE} Trans. Vis. Comput. Graph., 2023, 29(11): 4782--4793

\bibitem{DBLP:conf/ismar/HertelKSBSS21}
Hertel J, Karaosmanoglu S, Schmidt S, Br{\"{a}}ker J, Semmann M, Steinicke F.
\newblock A taxonomy of interaction techniques for immersive augmented reality based on an iterative literature review.
\newblock In: {IEEE} International Symposium on Mixed and Augmented Reality, {ISMAR} 2021, Bari, Italy, October 4-8, 2021.
\newblock 2021,  431--440

\bibitem{pirker2021potential}
Pirker J, Dengel A.
\newblock The potential of 360 virtual reality videos and real vr for education—a literature review.
\newblock IEEE computer graphics and applications, 2021, 41(4): 76--89

\bibitem{katona2021review}
Katona J.
\newblock A review of human--computer interaction and virtual reality research fields in cognitive infocommunications.
\newblock Applied Sciences, 2021, 11(6): 2646

\bibitem{DBLP:journals/vi/ZhangWZST23}
Zhang Y, Wang Z, Zhang J, Shan G, Tian D.
\newblock A survey of immersive visualization: Focus on perception and interaction.
\newblock Vis. Informatics, 2023, 7(4): 22--35

\bibitem{DBLP:journals/tvcg/SpittlePC023}
Spittle B, Pascual M~F, Creed C, Williams I.
\newblock A review of interaction techniques for immersive environments.
\newblock {IEEE} Trans. Vis. Comput. Graph., 2023, 29(9): 3900--3921

\bibitem{DBLP:conf/chi/WangYWJ024}
Wang Z, Yuan L, Wang L, Jiang B, Zeng W.
\newblock Virtuwander: Enhancing multi-modal interaction for virtual tour guidance through large language models.
\newblock In: Mueller F~F, Kyburz P, Williamson J~R, Sas C, Wilson M~L, Dugas P~O~T, Shklovski I, eds, Proceedings of the {CHI} Conference on Human Factors in Computing Systems, {CHI} 2024, Honolulu, HI, USA, May 11-16, 2024.
\newblock 2024,  612:1--612:20

\bibitem{DBLP:conf/vr/YangQCSBLL24}
Yang J~J, Qiu L, Corona{-}Moreno E~A, Shi L, Bui H, Lam M~S, Landay J~A.
\newblock {AMMA:} adaptive multimodal assistants through automated state tracking and user model-directed guidance planning.
\newblock In: {IEEE} Conference Virtual Reality and 3D User Interfaces, {VR} 2024, Orlando, FL, USA, March 16-21, 2024.
\newblock 2024,  892--902

\bibitem{DBLP:conf/ismar/GhamandiHNGKTL23}
Ghamandi R, Hmaiti Y, Nguyen T~T, Ghasemaghaei A, Kattoju R~K, II~E~M~T, LaViola J~J.
\newblock What and how together: {A} taxonomy on 30 years of collaborative human-centered {XR} tasks.
\newblock In: Bruder G, Olivier A, Cunningham A, Peng Y~E, Grubert J, Williams I, eds, {IEEE} International Symposium on Mixed and Augmented Reality, {ISMAR} 2023, Sydney, Australia, October 16-20, 2023.
\newblock 2023,  322--335

\bibitem{DBLP:conf/ismar/ChowdhuryUPIH22}
Chowdhury S, Ullah A~K~M~A, Pelmore N~B, Irani P, Hasan K.
\newblock Wriarm: Leveraging wrist movement to design wrist+arm based teleportation in {VR}.
\newblock In: Duh H~B~L, Williams I, Grubert J, Jones J~A, Zheng J, eds, {IEEE} International Symposium on Mixed and Augmented Reality, {ISMAR} 2022, Singapore, October 17-21, 2022.
\newblock 2022,  317--325

\bibitem{DBLP:conf/chi/SchmitzGS022}
Schmitz M, G{\"{u}}nther S, Sch{\"{o}}n D, M{\"{u}}ller F.
\newblock Squeezy-feely: Investigating lateral thumb-index pinching as an input modality.
\newblock In: Barbosa S~D~J, Lampe C, Appert C, Shamma D~A, Drucker S~M, Williamson J~R, Yatani K, eds, {CHI} '22: {CHI} Conference on Human Factors in Computing Systems, New Orleans, LA, USA, 29 April 2022 - 5 May 2022.
\newblock 2022,  61:1--61:15

\bibitem{DBLP:conf/ismar/BanMNK22}
Ban R, Matsumoto K, Narumi T, Kuzuoka H.
\newblock Wormholes in {VR:} teleporting hands for flexible passive haptics.
\newblock In: {IEEE} International Symposium on Mixed and Augmented Reality, {ISMAR} 2022, Singapore, October 17-21, 2022.
\newblock 2022,  748--757

\bibitem{DBLP:conf/ismar/YuZDV022}
Yu~D, Zhou Q, Dingler T, Velloso E, Gon{\c{c}}alves J.
\newblock Blending on-body and mid-air interaction in virtual reality.
\newblock In: Duh H~B~L, Williams I, Grubert J, Jones J~A, Zheng J, eds, {IEEE} International Symposium on Mixed and Augmented Reality, {ISMAR} 2022, Singapore, October 17-21, 2022.
\newblock 2022,  637--646

\bibitem{DBLP:conf/ismar/LeeHM22}
Lee G, Healey J, Manocha D.
\newblock Vrdoc: Gaze-based interactions for {VR} reading experience.
\newblock In: Duh H~B~L, Williams I, Grubert J, Jones J~A, Zheng J, eds, {IEEE} International Symposium on Mixed and Augmented Reality, {ISMAR} 2022, Singapore, October 17-21, 2022.
\newblock 2022,  787--796

\bibitem{DBLP:conf/chi/ChoiSO22}
Choi M, Sakamoto D, Ono T.
\newblock Kuiper belt: Utilizing the "out-of-natural angle" region in the eye-gaze interaction for virtual reality.
\newblock In: Barbosa S~D~J, Lampe C, Appert C, Shamma D~A, Drucker S~M, Williamson J~R, Yatani K, eds, {CHI} '22: {CHI} Conference on Human Factors in Computing Systems, New Orleans, LA, USA, 29 April 2022 - 5 May 2022.
\newblock 2022,  357:1--357:17

\bibitem{kim2022lattice}
Kim T, Ham A, Ahn S, Lee G.
\newblock Lattice menu: A low-error gaze-based marking menu utilizing target-assisted gaze gestures on a lattice of visual anchors.
\newblock In: Proceedings of the 2022 CHI Conference on Human Factors in Computing Systems.
\newblock 2022,  1--12

\bibitem{DBLP:conf/vr/0001LC00S22}
Yi~X, Lu~Y, Cai Z, Wu~Z, Wang Y, Shi Y.
\newblock Gazedock: Gaze-only menu selection in virtual reality using auto-triggering peripheral menu.
\newblock In: {IEEE} Conference on Virtual Reality and 3D User Interfaces, {VR} 2022, Christchurch, New Zealand, March 12-16, 2022.
\newblock 2022,  832--842

\bibitem{DBLP:journals/tvcg/WangZ022}
Wang Z, Zhao Y, Lu~F.
\newblock Gaze-vergence-controlled see-through vision in augmented reality.
\newblock {IEEE} Trans. Vis. Comput. Graph., 2022, 28(11): 3843--3853

\bibitem{DBLP:journals/imwut/ChenLYZ22}
Chen T, Li~T, Yang X, Zhu K.
\newblock Efring: Enabling thumb-to-index-finger microgesture interaction through electric field sensing using single smart ring.
\newblock Proc. {ACM} Interact. Mob. Wearable Ubiquitous Technol., 2022, 6(4): 161:1--161:31

\bibitem{DBLP:conf/uist/SendhilnathanZL22}
Sendhilnathan N, Zhang T, Lafreniere B, Grossman T, Jonker T~R.
\newblock Detecting input recognition errors and user errors using gaze dynamics in virtual reality.
\newblock In: Agrawala M, Wobbrock J~O, Adar E, Setlur V, eds, The 35th Annual {ACM} Symposium on User Interface Software and Technology, {UIST} 2022, Bend, OR, USA, 29 October 2022 - 2 November 2022.
\newblock 2022,  38:1--38:19

\bibitem{DBLP:conf/uist/LiaoKJKS22}
Liao J, Karim A, Jadon S~S, Kazi R~H, Suzuki R.
\newblock Realitytalk: Real-time speech-driven augmented presentation for {AR} live storytelling.
\newblock In: Agrawala M, Wobbrock J~O, Adar E, Setlur V, eds, The 35th Annual {ACM} Symposium on User Interface Software and Technology, {UIST} 2022, Bend, OR, USA, 29 October 2022 - 2 November 2022.
\newblock 2022,  17:1--17:12

\bibitem{DBLP:conf/uist/0001QTFLS22}
Yi~X, Qiu L, Tang W, Fan Y, Li~H, Shi Y.
\newblock {DEEP:} 3d gaze pointing in virtual reality leveraging eyelid movement.
\newblock In: Agrawala M, Wobbrock J~O, Adar E, Setlur V, eds, The 35th Annual {ACM} Symposium on User Interface Software and Technology, {UIST} 2022, Bend, OR, USA, 29 October 2022 - 2 November 2022.
\newblock 2022,  3:1--3:14

\bibitem{DBLP:conf/ismar/XuMYSL22}
Xu~W, Meng X, Yu~K, Sarcar S, Liang H.
\newblock Evaluation of text selection techniques in virtual reality head-mounted displays.
\newblock In: Duh H~B~L, Williams I, Grubert J, Jones J~A, Zheng J, eds, {IEEE} International Symposium on Mixed and Augmented Reality, {ISMAR} 2022, Singapore, October 17-21, 2022.
\newblock 2022,  131--140

\bibitem{DBLP:journals/imwut/ShenYYS22}
Shen X, Yan Y, Yu~C, Shi Y.
\newblock Clenchclick: Hands-free target selection method leveraging teeth-clench for augmented reality.
\newblock Proc. {ACM} Interact. Mob. Wearable Ubiquitous Technol., 2022, 6(3): 139:1--139:26

\bibitem{DBLP:conf/ismar/MengXL22}
Meng X, Xu~W, Liang H.
\newblock An exploration of hands-free text selection for virtual reality head-mounted displays.
\newblock In: Duh H~B~L, Williams I, Grubert J, Jones J~A, Zheng J, eds, {IEEE} International Symposium on Mixed and Augmented Reality, {ISMAR} 2022, Singapore, October 17-21, 2022.
\newblock 2022,  74--81

\bibitem{DBLP:conf/ismar/DasNH23}
Das S, Nasser A, Hasan K.
\newblock Fingerbutton: Enabling controller-free transitions between real and virtual environments.
\newblock In: Bruder G, Olivier A, Cunningham A, Peng Y~E, Grubert J, Williams I, eds, {IEEE} International Symposium on Mixed and Augmented Reality, {ISMAR} 2023, Sydney, Australia, October 16-20, 2023.
\newblock 2023,  533--542

\bibitem{DBLP:conf/ismar/ZhuSSG23}
Zhu F, Sidenmark L, Sousa M, Grossman T.
\newblock Pinchlens: Applying spatial magnification and adaptive control-display gain for precise selection in virtual reality.
\newblock In: Bruder G, Olivier A, Cunningham A, Peng Y~E, Grubert J, Williams I, eds, {IEEE} International Symposium on Mixed and Augmented Reality, {ISMAR} 2023, Sydney, Australia, October 16-20, 2023.
\newblock 2023,  1221--1230

\bibitem{DBLP:journals/tvcg/SongDK23}
Song Z, Dudley J~J, Kristensson P~O.
\newblock Hotgestures: Complementing command selection and use with delimiter-free gesture-based shortcuts in virtual reality.
\newblock {IEEE} Trans. Vis. Comput. Graph., 2023, 29(11): 4600--4610

\bibitem{DBLP:conf/chi/0003HLG23}
Tseng W, Huron S, Lecolinet E, Gugenheimer J.
\newblock Fingermapper: Mapping finger motions onto virtual arms to enable safe virtual reality interaction in confined spaces.
\newblock In: Schmidt A, V{\"{a}}{\"{a}}n{\"{a}}nen K, Goyal T, Kristensson P~O, Peters A, Mueller S, Williamson J~R, Wilson M~L, eds, Proceedings of the 2023 {CHI} Conference on Human Factors in Computing Systems, {CHI} 2023, Hamburg, Germany, April 23-28, 2023.
\newblock 2023,  874:1--874:14

\bibitem{DBLP:conf/ismar/ChenHTHH23}
Chen Y, Hsieh C, Then M~Y~J, Han P, Hung Y.
\newblock Leap to the eye: Implicit gaze-based interaction to reveal invisible objects for virtual environment exploration.
\newblock In: Bruder G, Olivier A, Cunningham A, Peng Y~E, Grubert J, Williams I, eds, {IEEE} International Symposium on Mixed and Augmented Reality, {ISMAR} 2023, Sydney, Australia, October 16-20, 2023.
\newblock 2023,  214--222

\bibitem{DBLP:conf/chi/SidenmarkCNLPG23}
Sidenmark L, Clarke C, Newn J, Lystb{\ae}k M~N, Pfeuffer K, Gellersen H.
\newblock Vergence matching: Inferring attention to objects in 3d environments for gaze-assisted selection.
\newblock In: Schmidt A, V{\"{a}}{\"{a}}n{\"{a}}nen K, Goyal T, Kristensson P~O, Peters A, Mueller S, Williamson J~R, Wilson M~L, eds, Proceedings of the 2023 {CHI} Conference on Human Factors in Computing Systems, {CHI} 2023, Hamburg, Germany, April 23-28, 2023.
\newblock 2023,  257:1--257:15

\bibitem{10.1145/3544548.3581423}
Wagner U, Lystb\ae{}k M~N, Manakhov P, Gr\o{}nb\ae{}k J~E~S, Pfeuffer K, Gellersen H.
\newblock A fitts’ law study of gaze-hand alignment for selection in 3d user interfaces.
\newblock In: Proceedings of the 2023 CHI Conference on Human Factors in Computing Systems, CHI '23.
\newblock 2023

\bibitem{10.1145/3591129}
Shi R, Wei Y, Qin X, Hui P, Liang H~N.
\newblock Exploring gaze-assisted and hand-based region selection in augmented reality.
\newblock Proc. ACM Hum.-Comput. Interact., 2023, 7(ETRA)

\bibitem{DBLP:conf/ismar/CailletGN23}
Caillet A~C, Goguey A, Nigay L.
\newblock 3d selection in mixed reality: Designing a two-phase technique to reduce fatigue.
\newblock In: Bruder G, Olivier A, Cunningham A, Peng Y~E, Grubert J, Williams I, eds, {IEEE} International Symposium on Mixed and Augmented Reality, {ISMAR} 2023, Sydney, Australia, October 16-20, 2023.
\newblock 2023,  800--809

\bibitem{DBLP:conf/ismar/ChenGFCL23}
Chen X, Guo D, Feng L, Chen B, Liu W.
\newblock Compass+ring: {A} multimodal menu to improve interaction performance and comfortability in one-handed scenarios.
\newblock In: Bruder G, Olivier A, Cunningham A, Peng Y~E, Grubert J, Williams I, eds, {IEEE} International Symposium on Mixed and Augmented Reality, {ISMAR} 2023, Sydney, Australia, October 16-20, 2023.
\newblock 2023,  473--482

\bibitem{DBLP:conf/chi/WeiSYW0YL23}
Wei Y, Shi R, Yu~D, Wang Y, Li~Y, Yu~L, Liang H.
\newblock Predicting gaze-based target selection in augmented reality headsets based on eye and head endpoint distributions.
\newblock In: Schmidt A, V{\"{a}}{\"{a}}n{\"{a}}nen K, Goyal T, Kristensson P~O, Peters A, Mueller S, Williamson J~R, Wilson M~L, eds, Proceedings of the 2023 {CHI} Conference on Human Factors in Computing Systems, {CHI} 2023, Hamburg, Germany, April 23-28, 2023.
\newblock 2023,  283:1--283:14

\bibitem{DBLP:conf/chi/HouNSKBG23}
Hou B~J, Newn J, Sidenmark L, Khan A~A, B{\ae}kgaard P, Gellersen H.
\newblock Classifying head movements to separate head-gaze and head gestures as distinct modes of input.
\newblock In: Schmidt A, V{\"{a}}{\"{a}}n{\"{a}}nen K, Goyal T, Kristensson P~O, Peters A, Mueller S, Williamson J~R, Wilson M~L, eds, Proceedings of the 2023 {CHI} Conference on Human Factors in Computing Systems, {CHI} 2023, Hamburg, Germany, April 23-28, 2023.
\newblock 2023,  253:1--253:14

\bibitem{10049667}
Yan Y, Liu H, Shi Y, Wang J, Guo R, Li~Z, Xu~X, Yu~C, Wang Y, Shi Y.
\newblock Conespeech: Exploring directional speech interaction for multi-person remote communication in virtual reality.
\newblock IEEE Transactions on Visualization and Computer Graphics, 2023, 29(5): 2647--2657

\bibitem{DBLP:conf/vr/SindhupathirajaUDH24}
Sindhupathiraja S~R, Ullah A~K~M~A, Delamare W, Hasan K.
\newblock Exploring bi-manual teleportation in virtual reality.
\newblock In: {IEEE} Conference Virtual Reality and 3D User Interfaces, {VR} 2024, Orlando, FL, USA, March 16-21, 2024.
\newblock 2024,  754--764

\bibitem{DBLP:conf/chi/DupreARSP24}
Dupr{\'{e}} C, Appert C, Rey S, Saidi H, Pietriga E.
\newblock Tripad: Touch input in {AR} on ordinary surfaces with hand tracking only.
\newblock In: Mueller F~F, Kyburz P, Williamson J~R, Sas C, Wilson M~L, Dugas P~O~T, Shklovski I, eds, Proceedings of the {CHI} Conference on Human Factors in Computing Systems, {CHI} 2024, Honolulu, HI, USA, May 11-16, 2024.
\newblock 2024,  754:1--754:18

\bibitem{DBLP:conf/vr/OrloskyLSSM24}
Orlosky J, Liu C, Sakamoto K, Sidenmark L, Mansour A.
\newblock Eyeshadows: Peripheral virtual copies for rapid gaze selection and interaction.
\newblock In: {IEEE} Conference Virtual Reality and 3D User Interfaces, {VR} 2024, Orlando, FL, USA, March 16-21, 2024.
\newblock 2024,  681--689

\bibitem{DBLP:conf/chi/ZhangCSS24}
Zhang C, Chen T, Shaffer E, Soltanaghai E.
\newblock Focusflow: 3d gaze-depth interaction in virtual reality leveraging active visual depth manipulation.
\newblock In: Mueller F~F, Kyburz P, Williamson J~R, Sas C, Wilson M~L, Dugas P~O~T, Shklovski I, eds, Proceedings of the {CHI} Conference on Human Factors in Computing Systems, {CHI} 2024, Honolulu, HI, USA, May 11-16, 2024.
\newblock 2024,  372:1--372:18

\bibitem{DBLP:conf/chi/TurkmenGBSASPM24}
T{\"{u}}rkmen R, Gelmez Z~E, Batmaz A~U, Stuerzlinger W, Asente P, Sarac M, Pfeuffer K, Machuca M~D~B.
\newblock Eyeguide {\&} eyeconguide: Gaze-based visual guides to improve 3d sketching systems.
\newblock In: Mueller F~F, Kyburz P, Williamson J~R, Sas C, Wilson M~L, Dugas P~O~T, Shklovski I, eds, Proceedings of the {CHI} Conference on Human Factors in Computing Systems, {CHI} 2024, Honolulu, HI, USA, May 11-16, 2024.
\newblock 2024,  178:1--178:14

\bibitem{DBLP:conf/chi/ZennerKFAK24}
Zenner A, Karr C, Feick M, Ariza O, Kr{\"{u}}ger A.
\newblock Beyond the blink: Investigating combined saccadic {\&} blink-suppressed hand redirection in virtual reality.
\newblock In: Proceedings of the {CHI} Conference on Human Factors in Computing Systems, {CHI} 2024, Honolulu, HI, USA, May 11-16, 2024.
\newblock 2024,  750:1--750:14

\bibitem{10.1145/3613904.3642758}
Rodriguez R, Sullivan B~T, Barrera~Machuca M~D, Batmaz A~U, Tornatzky C, Ortega F~R.
\newblock An artists' perspectives on natural interactions for virtual reality 3d sketching.
\newblock In: Proceedings of the 2024 CHI Conference on Human Factors in Computing Systems, CHI '24.
\newblock 2024

\bibitem{DBLP:conf/vr/LaiSL24}
Lai Y, Sun M, Li~Z.
\newblock Gazepuffer: Hands-free input method leveraging puff cheeks for {VR}.
\newblock In: {IEEE} Conference Virtual Reality and 3D User Interfaces, {VR} 2024, Orlando, FL, USA, March 16-21, 2024.
\newblock 2024,  331--341

\bibitem{marquardt2024selection}
Marquardt A, Steininger M, Trepkowski C, Weier M, Kruijff E.
\newblock Selection performance and reliability of eye and head gaze tracking under varying light conditions.
\newblock In: 2024 IEEE Conference Virtual Reality and 3D User Interfaces (VR).
\newblock 2024,  546--556

\bibitem{DBLP:conf/chi/PeiCLZ22}
Pei S, Chen A, Lee J, Zhang Y.
\newblock Hand interfaces: Using hands to imitate objects in {AR/VR} for expressive interactions.
\newblock In: Barbosa S~D~J, Lampe C, Appert C, Shamma D~A, Drucker S~M, Williamson J~R, Yatani K, eds, {CHI} '22: {CHI} Conference on Human Factors in Computing Systems, New Orleans, LA, USA, 29 April 2022 - 5 May 2022.
\newblock 2022,  429:1--429:16

\bibitem{DBLP:conf/chi/SatriadiSECCLYD22}
Satriadi K~A, Smiley J, Ens B, Cordeil M, Czauderna T, Lee B, Yang Y, Dwyer T, Jenny B.
\newblock Tangible globes for data visualisation in augmented reality.
\newblock In: Barbosa S~D~J, Lampe C, Appert C, Shamma D~A, Drucker S~M, Williamson J~R, Yatani K, eds, {CHI} '22: {CHI} Conference on Human Factors in Computing Systems, New Orleans, LA, USA, 29 April 2022 - 5 May 2022.
\newblock 2022,  505:1--505:16

\bibitem{DBLP:journals/tvcg/XuZSFY23}
Xu~X, Zhou Y, Shao B, Feng G, Yu~C.
\newblock Gesturesurface: {VR} sketching through assembling scaffold surface with non-dominant hand.
\newblock {IEEE} Trans. Vis. Comput. Graph., 2023, 29(5): 2499--2507

\bibitem{DBLP:journals/imwut/ZhanXZGCGLQ23}
Zhan L, Xiong T, Zhang H, Guo S, Chen X, Gong J, Lin J, Qin Y.
\newblock Toucheditor: Interaction design and evaluation of a flexible touchpad for text editing of head-mounted displays in speech-unfriendly environments.
\newblock Proc. {ACM} Interact. Mob. Wearable Ubiquitous Technol., 2023, 7(4): 198:1--198:29

\bibitem{9873984}
Williams A~S, Ortega F~R.
\newblock The impacts of referent display on gesture and speech elicitation.
\newblock IEEE Transactions on Visualization and Computer Graphics, 2022, 28(11): 3885--3895

\bibitem{DBLP:journals/ijhci/DengSZK24}
Deng C, Sun L, Zhou C, Kuai S.
\newblock Dual-gain mode of head-gaze interaction improves the efficiency of object positioning in a 3d virtual environment.
\newblock Int. J. Hum. Comput. Interact., 2024, 40(8): 2067--2082

\bibitem{DBLP:conf/vr/HombeckVHDL23}
Hombeck J~N, Voigt H, Heggemann T, Datta R~R, Lawonn K.
\newblock Tell me where to go: Voice-controlled hands-free locomotion for virtual reality systems.
\newblock In: {IEEE} Conference Virtual Reality and 3D User Interfaces, {VR} 2023, Shanghai, China, March 25-29, 2023.
\newblock 2023,  123--134

\bibitem{DBLP:conf/vr/MortezapoorVVK23}
Mortezapoor S, Vasylevska K, Vonach E, Kaufmann H.
\newblock Cobodeck: {A} large-scale haptic {VR} system using a collaborative mobile robot.
\newblock In: {IEEE} Conference Virtual Reality and 3D User Interfaces, {VR} 2023, Shanghai, China, March 25-29, 2023.
\newblock 2023,  297--307

\bibitem{DBLP:conf/vr/SinJLLLN24}
Sin Z~P~T, Jia Y, Li~R~C, Leong H~V, Li~Q, Ng~P~H~F.
\newblock illumotion: An optical-illusion-based {VR} locomotion technique for long-distance 3d movement.
\newblock In: {IEEE} Conference Virtual Reality and 3D User Interfaces, {VR} 2024, Orlando, FL, USA, March 16-21, 2024.
\newblock 2024,  924--934

\bibitem{10.1145/3613904.3642147}
Kang S, Jeong J, Lee G~A, Kim S~H, Yang H~J, Kim S.
\newblock The rayhand navigation: A virtual navigation method with relative position between hand and gaze-ray.
\newblock In: Proceedings of the 2024 CHI Conference on Human Factors in Computing Systems, CHI '24.
\newblock 2024

\bibitem{DBLP:conf/chi/LeeWSG24}
Lee H~S, Weidner F, Sidenmark L, Gellersen H.
\newblock Snap, pursuit and gain: Virtual reality viewport control by gaze.
\newblock In: Mueller F~F, Kyburz P, Williamson J~R, Sas C, Wilson M~L, Dugas P~O~T, Shklovski I, eds, Proceedings of the {CHI} Conference on Human Factors in Computing Systems, {CHI} 2024, Honolulu, HI, USA, May 11-16, 2024.
\newblock 2024,  375:1--375:14

\bibitem{10.1145/3491102.3517682}
Stemasov E, Wagner T, Gugenheimer J, Rukzio E.
\newblock Shapefindar: Exploring in-situ spatial search for physical artifact retrieval using mixed reality.
\newblock In: Proceedings of the 2022 CHI Conference on Human Factors in Computing Systems, CHI '22.
\newblock 2022

\bibitem{DBLP:conf/ismar/SongDK22}
Song Z, Dudley J~J, Kristensson P~O.
\newblock Efficient special character entry on a virtual keyboard by hand gesture-based mode switching.
\newblock In: {IEEE} International Symposium on Mixed and Augmented Reality, {ISMAR} 2022, Singapore, October 17-21, 2022.
\newblock 2022,  864--871

\bibitem{DBLP:conf/chi/HeLP22}
He~Z, Lutteroth C, Perlin K.
\newblock Tapgazer: Text entry with finger tapping and gaze-directed word selection.
\newblock In: Barbosa S~D~J, Lampe C, Appert C, Shamma D~A, Drucker S~M, Williamson J~R, Yatani K, eds, {CHI} '22: {CHI} Conference on Human Factors in Computing Systems, New Orleans, LA, USA, 29 April 2022 - 5 May 2022.
\newblock 2022,  337:1--337:16

\bibitem{DBLP:journals/tvcg/ShenDK23}
Shen J, Dudley J~J, Kristensson P~O.
\newblock Fast and robust mid-air gesture typing for {AR} headsets using 3d trajectory decoding.
\newblock {IEEE} Trans. Vis. Comput. Graph., 2023, 29(11): 4622--4632

\bibitem{DBLP:conf/iui/ZhaoPTZWJBG23}
Zhao M, Pierce A~M, Tan R, Zhang T, Wang T, Jonker T~R, Benko H, Gupta A.
\newblock Gaze speedup: Eye gaze assisted gesture typing in virtual reality.
\newblock In: Proceedings of the 28th International Conference on Intelligent User Interfaces, {IUI} 2023, Sydney, NSW, Australia, March 27-31, 2023.
\newblock 2023,  595--606

\bibitem{cui2023glancewriter}
Cui W, Liu R, Li~Z, Wang Y, Wang A, Zhao X, Rashidian S, Baig F, Ramakrishnan I, Wang F, others .
\newblock Glancewriter: Writing text by glancing over letters with gaze.
\newblock In: Proceedings of the 2023 CHI Conference on Human Factors in Computing Systems.
\newblock 2023,  1--13

\bibitem{DBLP:conf/chi/ZhangLHWLGZ23}
Zhang R, Li~K, Hao Y, Wang Y, Lai Z, Guimbreti{\`{e}}re F, Zhang C.
\newblock Echospeech: Continuous silent speech recognition on minimally-obtrusive eyewear powered by acoustic sensing.
\newblock In: Schmidt A, V{\"{a}}{\"{a}}n{\"{a}}nen K, Goyal T, Kristensson P~O, Peters A, Mueller S, Williamson J~R, Wilson M~L, eds, Proceedings of the 2023 {CHI} Conference on Human Factors in Computing Systems, {CHI} 2023, Hamburg, Germany, April 23-28, 2023.
\newblock 2023,  852:1--852:18

\bibitem{shen2024ringgesture}
Shen J, Boldu R, Kalla A, Glueck M, Surale H~B, Karlson A.
\newblock Ringgesture: A ring-based mid-air gesture typing system powered by a deep-learning word prediction framework.
\newblock IEEE Transactions on Visualization and Computer Graphics, 2024

\bibitem{DBLP:conf/vr/HuDK24}
Hu~J, Dudley J~J, Kristensson P~O.
\newblock Skimr: Dwell-free eye typing in mixed reality.
\newblock In: {IEEE} Conference Virtual Reality and 3D User Interfaces, {VR} 2024, Orlando, FL, USA, March 16-21, 2024.
\newblock 2024,  439--449

\bibitem{DBLP:journals/tvcg/CaiML24}
Cai Z, Ma~Y, Lu~F.
\newblock Robust dual-modal speech keyword spotting for {XR} headsets.
\newblock {IEEE} Trans. Vis. Comput. Graph., 2024, 30(5): 2507--2516

\bibitem{10474330}
Ren Y, Zhang Y, Liu Z, Xie N.
\newblock Eye-hand typing: Eye gaze assisted finger typing via bayesian processes in ar.
\newblock IEEE Transactions on Visualization and Computer Graphics, 2024, 30(5): 2496--2506

\bibitem{DBLP:journals/corr/abs-2405-18537}
Jadon S~S, Faridan M, Mah E, Vaish R, Willett W, Suzuki R.
\newblock Augmented conversation with embedded speech-driven on-the-fly referencing in {AR}.
\newblock CoRR, 2024, abs/2405.18537

\bibitem{DBLP:journals/imwut/WangSWYYWJXY24}
Wang Z, Shi Y, Wang Y, Yao Y, Yan K, Wang Y, Ji~L, Xu~X, Yu~C.
\newblock {G-VOILA:} gaze-facilitated information querying in daily scenarios.
\newblock Proc. {ACM} Interact. Mob. Wearable Ubiquitous Technol., 2024, 8(2): 78:1--78:33

\bibitem{DBLP:conf/chi/0005WBCRF24}
Lee J, Wang J, Brown E, Chu L, Rodriguez S~S, Froehlich J~E.
\newblock Gazepointar: {A} context-aware multimodal voice assistant for pronoun disambiguation in wearable augmented reality.
\newblock In: Mueller F~F, Kyburz P, Williamson J~R, Sas C, Wilson M~L, Dugas P~O~T, Shklovski I, eds, Proceedings of the {CHI} Conference on Human Factors in Computing Systems, {CHI} 2024, Honolulu, HI, USA, May 11-16, 2024.
\newblock 2024,  408:1--408:20

\bibitem{DBLP:conf/chi/WangLZ24}
Wang X, Lafreniere B, Zhao J.
\newblock Exploring visualizations for precisely guiding bare hand gestures in virtual reality.
\newblock In: Mueller F~F, Kyburz P, Williamson J~R, Sas C, Wilson M~L, Dugas P~O~T, Shklovski I, eds, Proceedings of the {CHI} Conference on Human Factors in Computing Systems, {CHI} 2024, Honolulu, HI, USA, May 11-16, 2024.
\newblock 2024,  636:1--636:19

\bibitem{DBLP:journals/tvcg/ShenDMK22}
Shen J, Dudley J~J, Mo~G~B, Kristensson P~O.
\newblock Gesture spotter: {A} rapid prototyping tool for key gesture spotting in virtual and augmented reality applications.
\newblock {IEEE} Trans. Vis. Comput. Graph., 2022, 28(11): 3618--3628

\bibitem{DBLP:conf/chi/LeeZAYGLKYDLSGZ24}
Lee C, Zhang R, Agarwal D, Yu~T~C, Gunda V, Lopez O, Kim J, Yin S, Dong B, Li~K, Sakashita M, Guimbreti{\`{e}}re F, Zhang C.
\newblock Echowrist: Continuous hand pose tracking and hand-object interaction recognition using low-power active acoustic sensing on a wristband.
\newblock In: Proceedings of the {CHI} Conference on Human Factors in Computing Systems, {CHI} 2024, Honolulu, HI, USA, May 11-16, 2024.
\newblock 2024,  403:1--403:21

\bibitem{DBLP:conf/vr/RuppGBK24}
Rupp D, GrieBer P, B{\"{o}}nsch A, Kuhlen T~W.
\newblock Authentication in immersive virtual environments through gesture-based interaction with a virtual agent.
\newblock In: {IEEE} Conference on Virtual Reality and 3D User Interfaces Abstracts and Workshops, {VR} Workshops 2024, Orlando, FL, USA, March 16-21, 2024.
\newblock 2024,  54--60

\bibitem{10522613}
Liu T, Xiao Y, Hu~M, Sha H, Ma~S, Gao B, Guo S, Liu Y, Song W.
\newblock Audiogest: Gesture-based interaction for virtual reality using audio devices.
\newblock IEEE Transactions on Visualization and Computer Graphics, 2024,  1--13

\bibitem{DBLP:conf/chi/XuZKN23}
Xu~C, Zhou B, Krishnan G, Nayar S~K.
\newblock Ao-finger: Hands-free fine-grained finger gesture recognition via acoustic-optic sensor fusing.
\newblock In: Schmidt A, V{\"{a}}{\"{a}}n{\"{a}}nen K, Goyal T, Kristensson P~O, Peters A, Mueller S, Williamson J~R, Wilson M~L, eds, Proceedings of the 2023 {CHI} Conference on Human Factors in Computing Systems, {CHI} 2023, Hamburg, Germany, April 23-28, 2023.
\newblock 2023,  306:1--306:14

\bibitem{DBLP:conf/iswc/KitamuraYS23}
Kitamura R, Yamamoto T, Sugiura Y.
\newblock Touchlog: Finger micro gesture recognition using photo-reflective sensors.
\newblock In: Tentori M, Weibel N, Laerhoven K~V, Zhou Z, eds, Proceedings of the 2023 International Symposium on Wearable Computers, {ISWC} 2023, Cancun, Quintana Roo, Mexico, October 8-12, 2023.
\newblock 2023,  92--97

\bibitem{DBLP:conf/ismar/LiLMHLS22}
Li~T, Liu Y, Ma~S, Hu~M, Liu T, Song W.
\newblock Nailring: An intelligent ring for recognizing micro-gestures in mixed reality.
\newblock In: Duh H~B~L, Williams I, Grubert J, Jones J~A, Zheng J, eds, {IEEE} International Symposium on Mixed and Augmented Reality, {ISMAR} 2022, Singapore, October 17-21, 2022.
\newblock 2022,  178--186

\bibitem{DBLP:conf/chi/Pohlmann0MMB23}
P{\"{o}}hlmann K~M~T, Li~G, McGill M, Markoff R, Brewster S~A.
\newblock You spin me right round, baby, right round: Examining the impact of multi-sensory self-motion cues on motion sickness during a {VR} reading task.
\newblock In: Schmidt A, V{\"{a}}{\"{a}}n{\"{a}}nen K, Goyal T, Kristensson P~O, Peters A, Mueller S, Williamson J~R, Wilson M~L, eds, Proceedings of the 2023 {CHI} Conference on Human Factors in Computing Systems, {CHI} 2023, Hamburg, Germany, April 23-28, 2023.
\newblock 2023,  712:1--712:16

\bibitem{DBLP:conf/chi/MedlarLG24}
Medlar A, Lehtikari M~T, Glowacka D.
\newblock Behind the scenes: Adapting cinematography and editing concepts to navigation in virtual reality.
\newblock In: Mueller F~F, Kyburz P, Williamson J~R, Sas C, Wilson M~L, Dugas P~O~T, Shklovski I, eds, Proceedings of the {CHI} Conference on Human Factors in Computing Systems, {CHI} 2024, Honolulu, HI, USA, May 11-16, 2024.
\newblock 2024,  545:1--545:12

\bibitem{DBLP:conf/chi/FeickR0K22}
Feick M, Regitz K~P, Tang A, Kr{\"{u}}ger A.
\newblock Designing visuo-haptic illusions with proxies in virtual reality: Exploration of grasp, movement trajectory and object mass.
\newblock In: Barbosa S~D~J, Lampe C, Appert C, Shamma D~A, Drucker S~M, Williamson J~R, Yatani K, eds, {CHI} '22: {CHI} Conference on Human Factors in Computing Systems, New Orleans, LA, USA, 29 April 2022 - 5 May 2022.
\newblock 2022,  635:1--635:15

\bibitem{DBLP:conf/chi/WuQQCRS24}
Wu~G, Qian J, Quispe S~C, Chen S, Rulff J, Silva C~T.
\newblock Artist: Automated text simplification for task guidance in augmented reality.
\newblock In: Mueller F~F, Kyburz P, Williamson J~R, Sas C, Wilson M~L, Dugas P~O~T, Shklovski I, eds, Proceedings of the {CHI} Conference on Human Factors in Computing Systems, {CHI} 2024, Honolulu, HI, USA, May 11-16, 2024.
\newblock 2024,  939:1--939:24

\bibitem{DBLP:conf/chi/ElsharkawyAYAHK24}
Elsharkawy A~I~A~M, Ataya A~A~S, Yeo D, An~E, Hwang S, Kim S.
\newblock {SYNC-VR:} synchronizing your senses to conquer motion sickness for enriching in-vehicle virtual reality.
\newblock In: Mueller F~F, Kyburz P, Williamson J~R, Sas C, Wilson M~L, Dugas P~O~T, Shklovski I, eds, Proceedings of the {CHI} Conference on Human Factors in Computing Systems, {CHI} 2024, Honolulu, HI, USA, May 11-16, 2024.
\newblock 2024,  257:1--257:17

\bibitem{DBLP:conf/vr/LiLYTFX24}
Li~Y, Liu Z, Yuan L, Tang H, Fan Y, Xie N.
\newblock Dynamic scene adjustment mechanism for manipulating user experience in {VR}.
\newblock In: {IEEE} Conference Virtual Reality and 3D User Interfaces, {VR} 2024, Orlando, FL, USA, March 16-21, 2024.
\newblock 2024,  179--188

\bibitem{DBLP:conf/chi/RaschRS023}
Rasch J, Rusakov V~D, Schmitz M, M{\"{u}}ller F.
\newblock Going, going, gone: Exploring intention communication for multi-user locomotion in virtual reality.
\newblock In: Schmidt A, V{\"{a}}{\"{a}}n{\"{a}}nen K, Goyal T, Kristensson P~O, Peters A, Mueller S, Williamson J~R, Wilson M~L, eds, Proceedings of the 2023 {CHI} Conference on Human Factors in Computing Systems, {CHI} 2023, Hamburg, Germany, April 23-28, 2023.
\newblock 2023,  785:1--785:13

\bibitem{DBLP:conf/chi/TanX0SZHH24}
Tan F~F, Xu~P, Ram A, Suen W~Z, Zhao S, Huang Y, Hurter C.
\newblock Audioxtend: Assisted reality visual accompaniments for audiobook storytelling during everyday routine tasks.
\newblock In: Mueller F~F, Kyburz P, Williamson J~R, Sas C, Wilson M~L, Dugas P~O~T, Shklovski I, eds, Proceedings of the {CHI} Conference on Human Factors in Computing Systems, {CHI} 2024, Honolulu, HI, USA, May 11-16, 2024.
\newblock 2024,  83:1--83:22

\bibitem{DBLP:conf/vr/WangZF24}
Wang X, Zhang W, Fu~H.
\newblock {A3RT:} attention-aware {AR} teleconferencing with life-size 2.5d video avatars.
\newblock In: {IEEE} Conference Virtual Reality and 3D User Interfaces, {VR} 2024, Orlando, FL, USA, March 16-21, 2024.
\newblock 2024,  211--221

\bibitem{DBLP:conf/uist/Tao022}
Tao Y, Lopes P.
\newblock Integrating real-world distractions into virtual reality.
\newblock In: Agrawala M, Wobbrock J~O, Adar E, Setlur V, eds, The 35th Annual {ACM} Symposium on User Interface Software and Technology, {UIST} 2022, Bend, OR, USA, 29 October 2022 - 2 November 2022.
\newblock 2022,  5:1--5:16

\bibitem{10462901}
Lee G, Lee D~Y, Su~G~M, Manocha D.
\newblock “may i speak?”: Multi-modal attention guidance in social vr group conversations.
\newblock IEEE Transactions on Visualization and Computer Graphics, 2024, 30(5): 2287--2297

\bibitem{wang2024tasks}
Wang Z, Lu~F.
\newblock Tasks reflected in the eyes: Egocentric gaze-aware visual task type recognition in virtual reality.
\newblock IEEE Transactions on Visualization and Computer Graphics, 2024

\bibitem{DBLP:conf/chi/ShenS022}
Shen V, Shultz C~D, Harrison C.
\newblock Mouth haptics in {VR} using a headset ultrasound phased array.
\newblock In: Barbosa S~D~J, Lampe C, Appert C, Shamma D~A, Drucker S~M, Williamson J~R, Yatani K, eds, {CHI} '22: {CHI} Conference on Human Factors in Computing Systems, New Orleans, LA, USA, 29 April 2022 - 5 May 2022.
\newblock 2022,  275:1--275:14

\bibitem{DBLP:conf/chi/TatzgernDWCEDGH22}
Tatzgern M, Domhardt M, Wolf M, Cenger M, Emsenhuber G, Dinic R, Gerner N, Hartl A.
\newblock Airres mask: {A} precise and robust virtual reality breathing interface utilizing breathing resistance as output modality.
\newblock In: Barbosa S~D~J, Lampe C, Appert C, Shamma D~A, Drucker S~M, Williamson J~R, Yatani K, eds, {CHI} '22: {CHI} Conference on Human Factors in Computing Systems, New Orleans, LA, USA, 29 April 2022 - 5 May 2022.
\newblock 2022,  274:1--274:14

\bibitem{DBLP:conf/chi/0001OPSB24}
Kim M~J, Ofek E, Pahud M, Sinclair M~J, Bianchi A.
\newblock Big or small, it's all in your head: Visuo-haptic illusion of size-change using finger-repositioning.
\newblock In: Mueller F~F, Kyburz P, Williamson J~R, Sas C, Wilson M~L, Dugas P~O~T, Shklovski I, eds, Proceedings of the {CHI} Conference on Human Factors in Computing Systems, {CHI} 2024, Honolulu, HI, USA, May 11-16, 2024.
\newblock 2024,  751:1--751:15

\bibitem{DBLP:conf/vr/YamazakiH23}
Yamazaki Y, Hasegawa S.
\newblock Providing 3d guidance and improving the music-listening experience in virtual reality shooting games using musical vibrotactile feedback.
\newblock In: {IEEE} Conference Virtual Reality and 3D User Interfaces, {VR} 2023, Shanghai, China, March 25-29, 2023.
\newblock 2023,  276--285

\bibitem{DBLP:conf/uist/ShenRM0S23}
Shen V, Rae{-}Grant T, Mullenbach J, Harrison C, Shultz C~D.
\newblock Fluid reality: High-resolution, untethered haptic gloves using electroosmotic pump arrays.
\newblock In: Follmer S, Han J, Steimle J, Riche N~H, eds, Proceedings of the 36th Annual {ACM} Symposium on User Interface Software and Technology, {UIST} 2023, San Francisco, CA, USA, 29 October 2023- 1 November 2023.
\newblock 2023,  8:1--8:20

\bibitem{DBLP:conf/uist/JinguWS23}
Jingu A, Withana A, Steimle J.
\newblock Double-sided tactile interactions for grasping in virtual reality.
\newblock In: Follmer S, Han J, Steimle J, Riche N~H, eds, Proceedings of the 36th Annual {ACM} Symposium on User Interface Software and Technology, {UIST} 2023, San Francisco, CA, USA, 29 October 2023- 1 November 2023.
\newblock 2023,  9:1--9:11

\bibitem{DBLP:conf/vr/SaintAubertAMPAL23}
Saint{-}Aubert J, Argelaguet F, Mac{\'{e}} M~J, Pacchierotti C, Amedi A, L{\'{e}}cuyer A.
\newblock Persuasive vibrations: Effects of speech-based vibrations on persuasion, leadership, and co-presence during verbal communication in {VR}.
\newblock In: {IEEE} Conference Virtual Reality and 3D User Interfaces, {VR} 2023, Shanghai, China, March 25-29, 2023.
\newblock 2023,  552--560

\bibitem{DBLP:conf/chi/Hedeshy0MS21}
Hedeshy R, Kumar C, Menges R, Staab S.
\newblock Hummer: Text entry by gaze and hum.
\newblock In: Kitamura Y, Quigley A, Isbister K, Igarashi T, Bj{\o}rn P, Drucker S~M, eds, {CHI} '21: {CHI} Conference on Human Factors in Computing Systems, Virtual Event / Yokohama, Japan, May 8-13, 2021.
\newblock 2021,  741:1--741:11

\bibitem{10.1145/3485279.3485283}
Schneider D, Biener V, Otte A, Gesslein T, Gagel P, Campos C, Pucihar v~K, Kljun M, Ofek E, Pahud M, Kristensson P~O, Grubert J.
\newblock Accuracy evaluation of touch tasks in commodity virtual and augmented reality head-mounted displays.
\newblock In: Proceedings of the 2021 ACM Symposium on Spatial User Interaction, SUI '21.
\newblock 2021

\bibitem{cai2020generalizing}
Cai M, Lu~F, Sato Y.
\newblock Generalizing hand segmentation in egocentric videos with uncertainty-guided model adaptation.
\newblock In: Proceedings of the ieee/cvf conference on computer vision and pattern recognition.
\newblock 2020,  14392--14401

\bibitem{cai2018desktop}
Cai M, Lu~F, Gao Y.
\newblock Desktop action recognition from first-person point-of-view.
\newblock IEEE transactions on cybernetics, 2018, 49(5): 1616--1628

\bibitem{cheng2024appearance}
Cheng Y, Wang H, Bao Y, Lu~F.
\newblock Appearance-based gaze estimation with deep learning: A review and benchmark.
\newblock IEEE Transactions on Pattern Analysis and Machine Intelligence, 2024

\bibitem{DBLP:conf/ismar/WangZLL21}
Wang Z, Zhao Y, Liu Y, Lu~F.
\newblock Edge-guided near-eye image analysis for head mounted displays.
\newblock In: {IEEE} International Symposium on Mixed and Augmented Reality, {ISMAR} 2021, Bari, Italy, October 4-8, 2021.
\newblock 2021,  11--20

\bibitem{DBLP:journals/tbe/GuestrinE06}
Guestrin E~D, Eizenman M.
\newblock General theory of remote gaze estimation using the pupil center and corneal reflections.
\newblock {IEEE} Trans. Biomed. Eng., 2006, 53(6): 1124--1133

\bibitem{DBLP:journals/corr/abs-2003-08806}
Wu~Z, Rajendran S, As~v~T, Zimmermann J, Badrinarayanan V, Rabinovich A.
\newblock Magiceyes: {A} large scale eye gaze estimation dataset for mixed reality.
\newblock CoRR, 2020, abs/2003.08806

\bibitem{DBLP:conf/etra/SantiniNK19}
Santini T, Niehorster D~C, Kasneci E.
\newblock Get a grip: slippage-robust and glint-free gaze estimation for real-time pervasive head-mounted eye tracking.
\newblock In: Krejtz K, Sharif B, eds, Proceedings of the 11th {ACM} Symposium on Eye Tracking Research {\&} Applications, {ETRA} 2019, Denver , CO, USA, June 25-28, 2019.
\newblock 2019,  17:1--17:10

\bibitem{6949395}
Narcizo F~B, Queiroz R.~d J~E, Gomes H~M.
\newblock Remote eye tracking systems: Technologies and applications.
\newblock In: 2013 26th Conference on Graphics, Patterns and Images Tutorials.
\newblock 2013,  15--22

\bibitem{DBLP:conf/etra/DierkesKB19}
Dierkes K, Kassner M, Bulling A.
\newblock A fast approach to refraction-aware eye-model fitting and gaze prediction.
\newblock In: Krejtz K, Sharif B, eds, Proceedings of the 11th {ACM} Symposium on Eye Tracking Research {\&} Applications, {ETRA} 2019, Denver , CO, USA, June 25-28, 2019.
\newblock 2019,  23:1--23:9

\bibitem{DBLP:conf/chi/KytoEPLB18}
Kyt{\"{o}} M, Ens B, Piumsomboon T, Lee G~A, Billinghurst M.
\newblock Pinpointing: Precise head- and eye-based target selection for augmented reality.
\newblock In: Mandryk R~L, Hancock M, Perry M, Cox A~L, eds, Proceedings of the 2018 {CHI} Conference on Human Factors in Computing Systems, {CHI} 2018, Montreal, QC, Canada, April 21-26, 2018.
\newblock 2018, ~81

\bibitem{DBLP:conf/chi/WangSRZ24}
Wang X, Su~Z, Rekimoto J, Zhang Y.
\newblock Watch your mouth: Silent speech recognition with depth sensing.
\newblock In: Mueller F~F, Kyburz P, Williamson J~R, Sas C, Wilson M~L, Dugas P~O~T, Shklovski I, eds, Proceedings of the {CHI} Conference on Human Factors in Computing Systems, {CHI} 2024, Honolulu, HI, USA, May 11-16, 2024.
\newblock 2024,  323:1--323:15

\bibitem{DBLP:conf/ismar/WangSHRS024}
Wang Z, Sun J, Hu~M, Rao M, Song W, Lu~F.
\newblock Gazering: Enhancing hand-eye coordination with pressure ring in augmented reality.
\newblock In: Eck U, Sra M, Stefanucci J~K, Sugimoto M, Tatzgern M, Williams I, eds, {IEEE} International Symposium on Mixed and Augmented Reality, {ISMAR} 2024, Bellevue, WA, USA, October 21-25, 2024.
\newblock 2024,  534--543

\bibitem{DBLP:journals/thms/WangWYL21}
Wang Z, Wang H, Yu~H, Lu~F.
\newblock Interaction with gaze, gesture, and speech in a flexibly configurable augmented reality system.
\newblock {IEEE} Trans. Hum. Mach. Syst., 2021, 51(5): 524--534

\bibitem{DBLP:books/daglib/0011673}
Burdea G~C, Coiffet P.
\newblock Virtual reality technology {(2.} ed.).
\newblock Wiley, 2003

\bibitem{DBLP:conf/ismar/JangFP22}
Jang J, Frier W, Park J.
\newblock Multimodal volume data exploration through mid-air haptics.
\newblock In: Duh H~B~L, Williams I, Grubert J, Jones J~A, Zheng J, eds, {IEEE} International Symposium on Mixed and Augmented Reality, {ISMAR} 2022, Singapore, October 17-21, 2022.
\newblock 2022,  243--251

\bibitem{10484440}
Shiota T, Takagi M, Kumagai K, Seshimo H, Aono Y.
\newblock { Egocentric Action Recognition by Capturing Hand-Object Contact and Object State }.
\newblock In: 2024 IEEE/CVF Winter Conference on Applications of Computer Vision (WACV).
\newblock January 2024,  6527--6537

\bibitem{10.1145/3491102.3502134}
Khan A~A, Newn J, Bailey J, Velloso E.
\newblock Integrating gaze and speech for enabling implicit interactions.
\newblock In: Proceedings of the 2022 CHI Conference on Human Factors in Computing Systems, CHI '22.
\newblock 2022

\end{thebibliography}


\end{document}